\pdfoutput=1  
\documentclass[
  aps,
  prab,
  reprint,
  superscriptaddress,
  amsmath,amssymb,
  floatfix,
  nofootinbib,
]{revtex4-2}

\usepackage{graphicx}
\usepackage{amsmath}
\usepackage{booktabs}

\usepackage{array}
\makeatletter
\@ifpackagelater{array}{2022/01/01}{\let\switch@array\relax}{}
\makeatother

\usepackage{siunitx}
\usepackage{tikz}
\usetikzlibrary{arrows.meta,positioning,fit,calc}
\usepackage[colorlinks=true,allcolors=blue,bookmarks=true,
  bookmarksnumbered=true,bookmarksopen=true]{hyperref}

\usepackage{xcolor}

\begin{document}

\title{HELIX: a hybrid envelope--multiparticle linac code with
       differentiable space-charge optimization}

\author{Abhishek Pathak}
\email{abhishek@fnal.gov}
\affiliation{Fermi National Accelerator Laboratory, Batavia, Illinois 60510, USA}

\date{\today}

\begin{abstract}
HELIX (Hybrid Envelope-multiparticle LInac eXplorer) is a Python
hadron-linac framework developed at Fermilab for the Proton Improvement
Plan~II (PIP-II) superconducting $\mathrm{H}^-$ linac.  Differentiable
beam dynamics and card-driven envelope--multiparticle design codes exist
separately; to our knowledge no published framework has combined
exact-gradient matching through a self-consistent nonlinear
particle-in-cell (PIC) space-charge solve with the lattice-card workflow
used for machine design.  HELIX does.  One TraceWin-format lattice
drives an rms-envelope solver with three-dimensional linearized space
charge, a multiparticle tracker with a 3-D FFT PIC solve, and linear
matrix analysis over one data model, and the constraint cards that steer
six classical matching algorithms also steer a differentiable seventh.
That seventh re-expresses a documented fixed-energy subset of the
tracking-plus-PIC map as a single PyTorch graph, differentiated through
the whole deposit-solve-gather-kick cycle and the transport between
kicks; for a fixed computational branch it returns reverse-mode
Jacobians exact to floating-point precision, at a cost per Jacobian set
by the number of constraints rather than the number of knobs.  On a
six-quadrupole manufactured verification problem both matchers land far
inside the imposed engineering tolerances, so convergence depth measures
the optimizer, not usability: the exact-Jacobian run reaches a
tolerance-normalized residual of $3.2\times10^{-10}$, some 200 times
below the finite-difference endpoint at essentially the same cumulative
forward-equivalent budget, bit-reproducible on a fixed FP64 CPU
configuration.  What carries engineering weight is that the Jacobian
cost is flat in the knob count: finite differences win at six knobs,
measured wall times bracket the crossover between ten and twelve knobs,
reverse mode is $1.45\times$ faster at sixteen, and peak memory stays
within $2.6$--$3.4~\mathrm{GB}$.  Under a finer model across independent
bunch seeds the remaining mismatch is set by coarse-model bias, not by
the optimizer.  Verification is hierarchical, from analytic and
independently integrated envelope references through
$10^{-9}$--$10^{-13}$ cross-implementation PIC parity across the FP64
backends to mode-matched TraceWin benchmarks on the PIP-II
low-energy, medium-energy and accelerating lines.
With space charge on, the four envelope rms moments along the PIP-II
medium-energy beam transport and half-wave-resonator line agree within
0.7\% at every recorded point and 0.15\% at exit.  Over the 186-m
accelerating linac the energy agrees within 0.03\% throughout, and at
the common endpoint HELIX transmits 97.14\% against partran's 97.29\%,
inside the two ensembles' combined binomial uncertainty; multiparticle
rms sizes agree at the few-percent level.  HELIX also provides
scope-guarded machine-learned field-map surrogates for envelope
transport, multi-objective Pareto exploration, element- and beam-error
Monte Carlo with orbit correction, failure-compensation analysis, and
RFQ transport inside the same 3-D FFT PIC solve.
Outside the differentiable path the PIC kernels run under NumPy/SciPy or
C++/OpenMP with optional CUDA and Metal Poisson solvers; the front end
also reads a MAD-X subset, and exchanges TraceWin particle files
and openPMD-layout HDF5.
The differentiable path is CPU-only within its documented fixed-energy
scope and its matching residual uses a coarse mesh; cross-code
validation is concentrated on PIP-II-family lattices.
\end{abstract}

\maketitle

\section{Introduction}
\label{sec:intro}

The Proton Improvement Plan~II (PIP-II), under construction at
Fermilab, centers on an 800-MeV superconducting H$^-$ linac built
from continuous-wave-capable structures and cryomodules to support
high-intensity proton delivery to the
Booster~\cite{pathak_pip2design}.  From a 30-keV ion source the
beam passes through a radio-frequency quadrupole (RFQ), which
bunches it and accelerates it to
2.1~MeV, and a chain of superconducting cavity families then carries
it to full energy.  The RFQ delivers a nominal 5-mA beam; a
bunch-by-bunch chopper in the medium-energy beam transport (MEBT)
then removes bunches to form the programmed pulse pattern, giving
the machine's 2-mA average design
current~\cite{pathak_pip2design,pip2fdr}.  Except where stated, the
space-charge calculations of this paper use the unchopped
\SI{5}{\milli\ampere}, which is the current that fixes the charge per
bunch, $Q = I/f_\mathrm{bunch}$; the multiparticle runs track a single
bunch, so the chopped average does not enter.  The beam is dc in the
low-energy beam transport (LEBT) and bunched
downstream.\footnote{The transfer-line envelope study of
Sec.~\ref{sec:app_foil} uses \SI{4.84}{\milli\ampere}; its paired
foil-kick particle runs are space-charge-off.}  This
space-charge-dominated regime sets the terms of the design:
space-charge nonlinearities and
parametric resonances can grow the emittance, exchange it between
planes, and raise beam loss; the lattice counters these effects
through adiabatic phase-advance variation, zero-current phase
advances kept below $90^\circ$ per period, and operating points
placed in resonance-free regions of the Hofmann stability
chart~\cite{pathak_pip2design}.

Designing, commissioning, and
operating such a machine leans on simulation at two very different
fidelities.  Lattice design, matching, and tolerance work want
thousands of cheap evaluations, which rms-envelope models with
linearized space charge provide in seconds at section scale; loss
budgets, halo, and
transmission questions want self-consistent multiparticle tracking
with a three-dimensional particle-in-cell (PIC) space-charge solve,
which costs minutes per evaluation.  In practice the two fidelities
live in one workflow: designs are iterated at envelope speed and
confirmed at PIC fidelity, with matching, error tolerancing, and
failure studies wrapped around both.

The tools that support this workflow today sit largely at two
poles.  The
established hadron-linac workhorses (TraceWin~\cite{tracewin},
IMPACT-Z~\cite{impactz}, and TRACK~\cite{track_code}) cover the
physics and, in TraceWin's case, define the de facto lattice language
of this community, but TraceWin and TRACK are distributed as
precompiled binaries, and none of the three is Python-native.  The
modern open
frameworks (Bmad~\cite{bmad}, Xsuite~\cite{xsuite},
ImpactX~\cite{impactx}, and the machine-learning-oriented
Cheetah~\cite{cheetah}) are architecturally contemporary but are not
organized around the envelope-first, lattice-card-driven linac design
workflow.  Closest to that workflow among recent codes is
AVAS~\cite{avas}, a newer arrival on the workhorse side of this
divide, which pairs envelope
and multiparticle modes behind built-in parameter matching, error
analysis and a graphical interface; its paper documents
neither TraceWin-format lattice input nor a differentiable path, and
specifies no space-charge coupling in its envelope mode.
Section~\ref{sec:discussion} surveys this landscape in detail.

Meanwhile, differentiable beam-dynamics simulation has
moved quickly: Qiang demonstrated gradient-based design optimization
through a differentiable self-consistent space-charge
model~\cite{qiang2023differentiable}, and automatic differentiation
has since reached Julia tracking with space-charge
sensitivities~\cite{jutrack} and the ImpactX envelope model, where
compiler-level differentiation supports a gradient-matched FODO
demonstration~\cite{huebl2025impactx}.  Gradient-based Twiss
matching has been shown with an auto-differentiable parallel PIC
code~\cite{qiang2025admodule}, a reverse-mode
integrated-Green-function space-charge kick has been added to
Cheetah~\cite{dhamrait2026}, and differentiable tracking has been
benchmarked systematically for optimization~\cite{huhn2025}.  At
Fermilab itself, the author's PIP-II digital-twin framework couples
differentiable symplectic transport to neural-network surrogates
trained on PIC simulations for many-parameter operational
tuning~\cite{pathak_dt}; it differentiates through the learned
surrogates rather than through a self-consistent space-charge
solve.

The two lines of development have so far stayed separate: the
differentiable engines are
not built around the card-driven design workflow, and the
design workhorses document no gradients through their space-charge
solvers.  To our knowledge, no published
framework yet combines exact-gradient matching through a
self-consistent
nonlinear PIC solve with a card-driven hybrid
envelope-multiparticle hadron-linac workflow of the kind used for
machine design.

This paper describes HELIX (Hybrid Envelope-multiparticle LInac
eXplorer), a Python toolkit developed at Fermilab for PIP-II that
supplies this combination: a single lattice description drives both
envelope-speed design iteration and PIC-fidelity confirmation, and
the constraint cards that steer its classical matching algorithms
also steer exact-gradient matching, within a documented scope,
through a differentiable
re-implementation of the PIC solve.  Machine learning is folded
into the same fabric: beyond the PyTorch-based differentiable
solver, scope-guarded neural surrogates, trained against each
field-map element's own transfer-matrix ground truth, can stand in
for the costliest elements of a zero-current envelope pass and, through
an experimental fast path, of a multiparticle one
(Sec.~\ref{sec:surrogates}).  HELIX matured from the
Linac\_Gen preview reported in Ref.~\cite{pathak_linacgen} into a
tested, documented package with a reference manual and an automated
test suite, and it is publicly released under the GNU General Public
License v3.0, with the release approved by Fermilab and the
U.S.\ Department of Energy (see the
data availability statement).  It sits within a
broader Fermilab effort on simulation-driven operations, alongside
virtual-accelerator controls~\cite{miceli_twinac} and beam-based
measurements on Fermilab's existing linac~\cite{pathak_hla}.  Its
individual ingredients each have published antecedents; the
contribution is their integration into one tool, in which the
matching engine dispatches the differentiable solver as a peer of the
classical algorithms rather than as a separate research code, and
the production envelope and multiparticle solvers are benchmarked
quantitatively against TraceWin, the design code of record for
PIP-II, on the machine's design lattices.

The specific contributions reported here are as follows.
\begin{enumerate}
\item A hybrid architecture uses a lattice front end that reads
  established lattice formats: a documented subset of the TraceWin
  language and a MAD-X subset, backed by an I/O layer for TraceWin
  particle files and openPMD-layout HDF5 results.  Under one data
  model, GUI, and batch CLI, it drives three interchangeable solver
  modes: rms-envelope with three-dimensional linearized space charge;
  multiparticle tracking with a 3-D FFT PIC solve (NumPy/SciPy or
  C++/OpenMP kernels, with GPU-accelerated Poisson solves), seeded
  from a built-in initial-distribution generator or an imported
  particle file; and linear matrix analysis.  The same architecture
  includes H$^-$-specific loss machinery (a stripper-foil element and
  Lorentz- and intra-beam-stripping analyzers) and a backtracking
  engine that transports distributions or envelopes backward through
  inverses of the element maps, with documented fallbacks and
  exclusions (Secs.~\ref{sec:physics}
  and~\ref{sec:methods}).
\item A differentiable PyTorch re-implementation covers a documented
  fixed-energy subset of the tracking-plus-PIC forward map.  It is held
  to kernel-level parity with the primary solvers at the $10^{-13}$
  level or better and, on the bend-free FODO space-charge benchmark,
  to machine-precision tracked-beam parity.  It supplies reverse-mode
  Jacobians exact to floating-point precision for a fixed computational
  branch and enters the matching engine as a seventh selectable
  optimization algorithm
  (Secs.~\ref{sec:diffpic} and~\ref{sec:matching}; tracked-beam
  parity quantified in Sec.~\ref{sec:diffpic}, kernel parity and the
  derivative checks in Sec.~\ref{sec:validation}).
\item A matching and design-study suite provides TraceWin-style
  constraint cards, seven single-objective algorithms, multi-objective
  Pareto exploration, machine-learned element surrogates, element- and
  beam-error Monte Carlo with singular-value-decomposition (SVD)
  orbit correction, element-failure criticality and compensation
  analysis, and parallel parameter scans
  (Secs.~\ref{sec:matching}, \ref{sec:surrogates},
  and~\ref{sec:applications}).
\item Hierarchical verification comprises analytic envelope benchmarks
  at the $10^{-8}$ level, cross-implementation PIC parity at
  $10^{-9}$--$10^{-13}$ across the FP64 backends, and mode-matched
  cross-code benchmarks against TraceWin on PIP-II-family lattices.
  In those benchmarks, the envelope mode with space charge follows the
  reference in every rms moment to better than 0.7\% at every recorded
  point along the MEBT+HWR (half-wave resonator) line.  Multiparticle
  rms sizes agree at the few-percent level in the mean, with localized
  waist deviations quantified in Sec.~\ref{sec:validation}.
\item On a six-quadrupole manufactured verification problem,
  gradient-based matching through nonlinear PIC space charge drives
  the tolerance-normalized matching residual to
  $3.2\times10^{-10}$, with every physical mismatch far below its
  imposed engineering tolerance.  The result is bit-reproducible on a
  fixed FP64 CPU configuration and is reached at essentially the same
  cumulative forward-equivalent budget at which a finite-difference
  baseline stops a factor of ${\approx}200$ short.  Finite differences
  won the wall clock at this knob count by $1.74\times$.  In the
  accompanying knob-count scaling study, the reverse-mode Jacobian
  cost is flat in the number of knobs, and reverse-mode peak memory
  stays within \SIrange{2.6}{3.4}{\giga\byte} through sixteen knobs.
  Measured end-to-end wall clocks cross finite differences between the
  ten- and twelve-knob cases, with a $1.45\times$ reverse-mode
  advantage at sixteen
  (Sec.~\ref{sec:applications}; the differentiable path's documented
  scope is collected in Table~\ref{tab:gradscope}).
\end{enumerate}

Section~\ref{sec:physics} presents the physics models,
Sec.~\ref{sec:methods} the numerical methods and architecture,
Sec.~\ref{sec:validation} the verification and benchmarks,
Sec.~\ref{sec:applications} the case studies,
Sec.~\ref{sec:performance} performance and scaling, and
Sec.~\ref{sec:discussion} the relation to existing codes and the
limitations of the present release.

\section{Physics models}
\label{sec:physics}

\begin{figure*}[t]
\begin{tikzpicture}[
  >={Stealth[length=2mm,width=1.7mm]},
  font=\footnotesize,
  box/.style={draw, semithick, rounded corners=2pt, align=center,
              inner sep=3pt, fill=white},
  statebox/.style={box, fill=black!8},
  solverbox/.style={box, line width=0.9pt, fill=white},
  scbox/.style={box, fill=black!4},
  diagbox/.style={box, fill=black!8},
  auxbox/.style={box, densely dashed, fill=black!4},
  flow/.style={->, semithick},
  feed/.style={->, semithick, densely dashed},
  note/.style={font=\scriptsize\itshape},
]
\node[note] at (1.7,3.95)  {beam state and conventions};
\node[note] at (6.9,3.95)  {solver-mode transport};
\node[note] at (13.9,3.95) {diagnostics};
\node[statebox, text width=42mm] (a1) at (1.7,2.6)
  {6-D state $(x, x', y, y', \Delta\phi, \Delta W)$\\
   in (mm, mrad, mm, mrad, deg, MeV);\\
   kinetic $x' \!=\! p_x/p_z$, $y' \!=\! p_y/p_z$};
\node[statebox, text width=42mm] (a2) at (1.7,0.7)
  {centroid $\mu$, covariance $\Sigma$;\\
   $\Sigma \rightarrow D_f\,\Sigma\,D_f^{\mathsf{T}}$ at frequency
   jumps;\\
   $\zeta = -\beta\lambda_{\mathrm{rf}}\Delta\phi/360^\circ$};
\node[statebox, text width=42mm] (a3) at (1.7,-1.35)
  {generated distributions\\
   (Gaussian, waterbag, KV,\\
   parabolic, uniform, thermal-halo)\\
   or \texttt{.dst} import};
\node[solverbox, text width=40mm] (bmat) at (6.9,2.5)
  {\textbf{matrix}: composed linear\\ maps (no space charge)};
\node[solverbox, text width=40mm] (benv) at (6.9,1.15)
  {\textbf{envelope}: $\Sigma \rightarrow M\Sigma M^{\mathsf{T}}$,\\
   self-consistent\\ linearized SC lens};
\node[solverbox, text width=42mm] (bmp) at (6.9,-0.85)
  {\textbf{multiparticle}: exact matrix\\
   maps; field-map KD / Strang-like\\
   DKD; outer
   $T_{\Lambda/2} \circ S_\Lambda \circ T_{\Lambda/2}$\\
   \textit{(order-unassigned field-map push)}};
\node[diagbox, text width=34mm] (d1) at (13.9,0.8)
  {env + multiparticle moments:\\
   projected $\varepsilon$,
   Twiss; $\varepsilon_n$ (transverse); $\varepsilon_{4D}$;\\
   mode values $\hat\varepsilon_i$
   \textit{(fixed convention)}\\[2pt]
   matrix: periodic Twiss,\\ phase advance\\[2pt]
   multiparticle: transmission,\\ losses, halo};
\node[scbox, text width=32mm] (sc1) at (3.7,-3.45)
  {rms-equivalent ellipsoid\\ (analytic, $R_D$) [envelope]\\
   \textit{axis-aligned, projected sizes}};
\node[scbox, text width=30mm] (sc2) at (7.5,-3.45)
  {3-D FFT PIC (CIC/TSC,\\ IGF kernel) [multiparticle]};
\node[scbox, text width=32mm] (sc3) at (11.3,-3.45)
  {dc: uniform cylinder [both];\\
   Bassetti--Erskine, 2-D PIC\\ {[multiparticle only]}};
\node[note, text width=58mm, align=center] at (9.0,-4.75)
  {all scaled by $1-f$
   (\texttt{SPACE\_CHARGE\_COMP}\\ neutralisation fraction $f$)};
\node[auxbox, text width=126mm, font=\scriptsize] (aux) at (7.6,-5.65)
  {auxiliary physics: 1-D steady-state CSR (bends, multiparticle)
   $\cdot$ foil interactions (per-particle kicks; envelope
   $\Sigma \rightarrow \Sigma + D$; no charge-state conversion)
   $\cdot$ H$^-$ loss analyzers (post-tracking: Lorentz, intra-beam)
   $\cdot$ static element- and beam-error models};
\draw[flow] (a1.east) -- (bmat.west);
\draw[flow] (a1.east) -- ($(benv.west)+(0,0.3)$);
\draw[flow] (a1.east) -- ($(bmp.west)+(0,0.45)$);
\draw[flow] (a2.east) -- (benv.west);
\draw[flow] (a3.east) -- ($(bmp.west)+(0,-0.3)$);
\draw[feed] (bmat.east) -- ($(d1.west)+(0,0.7)$);
\draw[flow] (benv.east) -- (d1.west);
\draw[flow] (bmp.east)  -- ($(d1.west)+(0,-0.7)$);
\draw[flow] (sc1.north) |- ($(benv.west)+(0,-0.32)$);
\draw[flow] (sc2.north) -- (sc2.north |- bmp.south);
\draw[flow] (sc3.north) |- ($(bmp.east)+(0,-0.35)$);
\draw[flow] ($(sc3.north)+(0.5,0)$) |- ($(benv.east)+(0,-0.32)$);
\end{tikzpicture}
\caption{\label{fig:physmodels}%
Physics-model map of this section.  The shared conventions feed all
three solver modes; the covariance path feeds the envelope solver
and generated or imported distributions feed the multiparticle
tracker.  The lower band identifies the collective-field models and
the modes each serves (bracketed tags; arrows drawn per assignment).
Auxiliary physics comprises CSR, foil interactions, post-tracking
H$^-$ loss analyzers, and static element and beam errors, each
acting where its entry states.  The envelope and multiparticle
solvers report shared moment
diagnostics; the matrix mode returns periodic Twiss and phase
advance directly (dashed), and transmission, loss, and halo
diagnostics are specific
to multiparticle tracking.  Italics restate the caveats of the
text.}
\end{figure*}
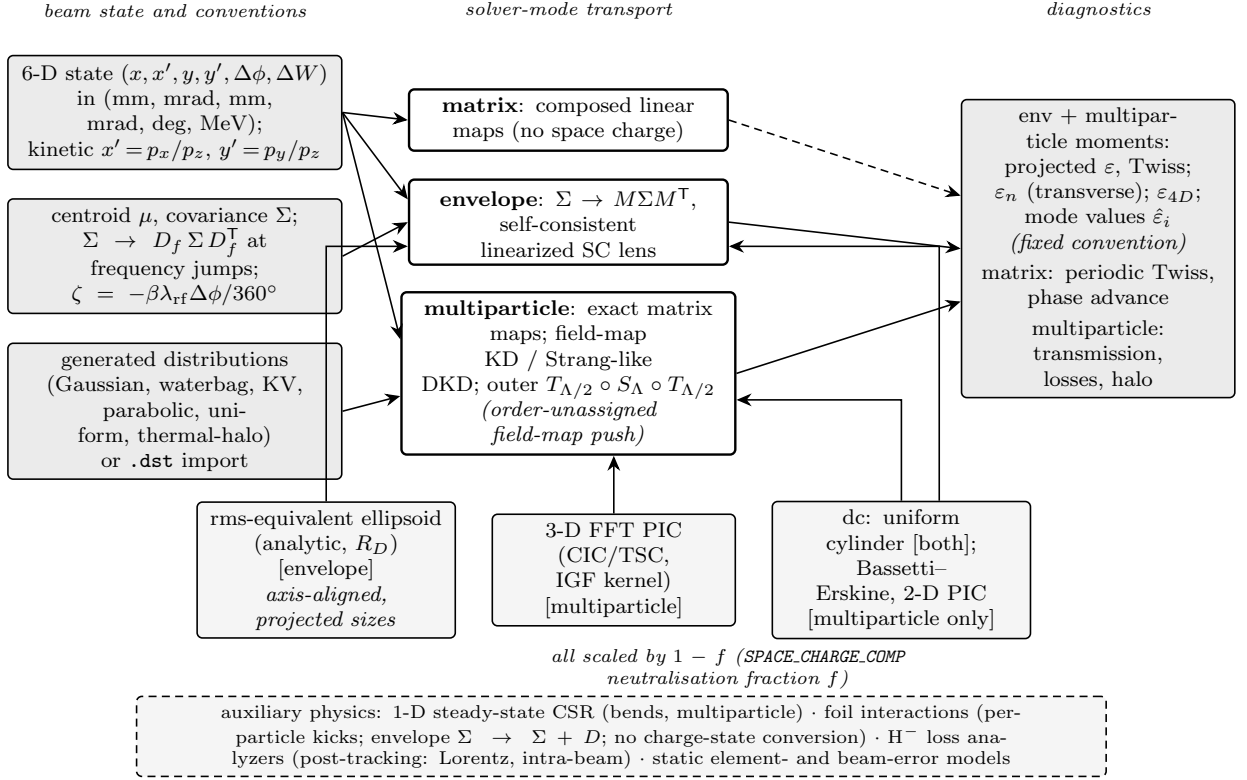

HELIX provides three complementary simulation modes that share a single
lattice description, reference-particle model, and 6-D phase-space
convention: (i) a deterministic rms-envelope solver that propagates the
full $6\times6$ second-moment matrix with an analytic, axis-aligned
three-dimensional
space-charge kick, (ii) a multiparticle tracker with particle-in-cell
(PIC) space charge, and (iii) linear matrix analysis of the composed
transfer maps (Sec.~\ref{sec:methods}); the envelope and
multiparticle paths additionally carry dedicated continuous-beam (dc)
variants for unbunched transport upstream of the first rf structure.  The
coordinates, element parametrization, and sign conventions follow
TraceWin~\cite{tracewin_manual}, so that lattices, field maps, and particle
files can be exchanged between the two codes within the supported
card set (Sec.~\ref{sec:methods});
the numerical models behind those conventions are described below,
and Fig.~\ref{fig:physmodels} maps them onto the three modes.

\subsection{Beam representation and envelope model}
\label{sec:envelope}

\emph{Coordinates.}  Each macroparticle carries the 6-D state
$(x, x', y, y', \Delta\phi, \Delta W)$ in units of
(mm, mrad, mm, mrad, deg, MeV).  The transverse divergences are
kinetic, $x' = p_x/p_z$, rather than canonical $p_x/p_0$; the
longitudinal pair is the rf-phase and kinetic-energy deviation from a
synchronous reference particle that carries the species (rest mass
$m$, signed charge $q_s$), kinetic
energy $W$, local rf frequency $f$, synchronous phase $\phi_s$, and
the relativistic factors $\beta$ and $\gamma$.  The independent
variable is the path length $s$; time is not tracked.  The local rf
wavelength is $\lambda_{\mathrm{rf}} = c/f$, and the lab-frame
longitudinal offset equivalent to a phase deviation is
$\zeta = -\beta\lambda_{\mathrm{rf}}\,\Delta\phi/360^\circ$,
positive toward the bunch head.  Kick calculations do not share a
single unit system: each element folds the required unit factors and
physical constants into its own coefficients (an rf gap takes its peak
voltage in MV and returns an energy kick directly in MeV), while the
collective-field kicks convert positions to metres, evaluate the field
in V/m, and convert the resulting angles back to mrad.  Phase
deviations are expressed in degrees of the \emph{local} rf frequency:
at a frequency jump (e.g., the \SI{162.5}{\mega\hertz} to
\SI{325}{\mega\hertz} transition in PIP-II) the multiparticle tracker
rescales every particle's $\Delta\phi$ by the frequency ratio, which
transforms all moments involving $\Delta\phi$ consistently, so that
the physical bunch length and the physical (mm$\cdot$mrad)
longitudinal emittance are preserved, while the native deg$\cdot$MeV
emittance value itself rescales by the same frequency ratio.  In
envelope mode the code applies the same coordinate change at the
moment level, $\Sigma \rightarrow D_f\,\Sigma\,D_f^{\mathsf{T}}$ with
$D_f = \mathrm{diag}(1,1,1,1,f_\text{new}/f_\text{old},1)$: the entire
phase row and column of the moment matrix is rescaled
($\langle\Delta\phi^2\rangle$ by the squared ratio;
$\langle\Delta\phi\,\Delta W\rangle$ and every phase--transverse
cross moment $\langle x\,\Delta\phi\rangle$, etc., by the ratio), so
the envelope jump is exact for arbitrary transverse--longitudinal
correlations.  For the uncoupled beams of the benchmarks in this
paper the cross moments at the jumps are at the
$10^{-3}$-correlation level, and the full transformation is
numerically indistinguishable from rescaling the longitudinal block
alone.  Collective kicks act
only on the deviation
coordinates; the reference particle is never modified by space
charge.

\emph{Moments, Twiss parameters, and emittances.}  The beam is
summarized by the centroid and the centered (central-moment)
$6\times6$ covariance matrix,
\begin{equation*}
  \mu = \langle u \rangle,
  \qquad
  \Sigma = \bigl\langle (u-\mu)(u-\mu)^{\mathsf{T}} \bigr\rangle,
\end{equation*}
over the particle coordinates
$u = (x, x', y, y', \Delta\phi, \Delta W)$: the moment diagnostics
subtract the measured centroid before forming $\Sigma$, so beam
offsets do not contaminate the second moments.  For each plane, with
$(\xi, \xi')$ the corresponding centered coordinate pair, the rms
emittance and Twiss functions are obtained from the $2\times2$ block,
\begin{equation}
  \varepsilon_\xi = \sqrt{\langle \xi^2\rangle\langle \xi'^2\rangle
                        - \langle \xi \xi'\rangle^2},
  \quad
  \beta_\xi = \frac{\langle \xi^2\rangle}{\varepsilon_\xi},
  \quad
  \alpha_\xi = -\frac{\langle \xi \xi'\rangle}{\varepsilon_\xi}.
  \label{eq:twiss}
\end{equation}
Normalized transverse emittances are
$\varepsilon_n = \beta\gamma\,\varepsilon$.
The longitudinal emittance is native in deg$\cdot$MeV and is converted
to the geometric mm$\cdot$mrad of the equivalent $(z, \delta)$ pair by
dividing by the product of two Jacobian factors,
$\varepsilon_z[\mathrm{mm\,mrad}] =
\varepsilon_z[\mathrm{deg\,MeV}]/(k_\phi k_W)$: the phase-to-length
factor $k_\phi = 360^\circ/(\beta\lambda_{\mathrm{rf}})$ (deg per mm,
with $\lambda_{\mathrm{rf}}$ in mm)
and the energy-to-slope factor
$k_W = \beta^2\gamma\, mc^2 \times 10^{-3}$ (MeV per mrad), which
implements $\Delta W = \beta^2\gamma\, mc^2\,\delta$ with the momentum
deviation $\delta = \Delta p/p$ expressed in mrad.  Initial
distributions are generated from user Twiss parameters (Gaussian with
adjustable cutoff, waterbag, KV, parabolic, uniform, or bi-Gaussian
thermal-halo), optionally seeded with input dispersion, or loaded from
TraceWin \texttt{.dst} files.

\emph{Envelope transport.}  The envelope solver propagates $\Sigma$
element by element through the linear map of each element,
\begin{equation}
  \Sigma \;\rightarrow\; M\,\Sigma\,M^{\mathsf{T}},
  \label{eq:sigmatransport}
\end{equation}
where $M$ is either the analytic $6\times6$ transfer matrix (drifts,
quadrupoles, solenoids, bends, rf gaps) or a numerically linearized
map of the field-map integrator described in
Sec.~\ref{sec:integrators}.  Space charge enters as a thin linear
defocusing lens applied midway through each transport sub-step, with
the kick strength recomputed self-consistently from the current
$\Sigma$.

For bunched beams the kick is built from the uniform triaxial
ellipsoid model of Lapostolle and
Wangler~\cite{lapostolle,wangler}.  The bunch of charge magnitude
$Q = I/f_b$
(with $I$ the beam-current magnitude and $f_b$ the bunch repetition
frequency, which is fixed at
injection and deliberately \emph{not} changed at cavity frequency
jumps) is replaced by the rms-equivalent uniform ellipsoid with
rest-frame semi-axes
$(a, b, c) = \sqrt{5}\,(\sigma_x, \sigma_y, \gamma\sigma_z)$, where
$\sigma_{x,y}$ are the lab-frame transverse sizes (invariant under
the longitudinal boost) and
$\sigma_z = \beta\lambda_{\mathrm{rf}}\sigma_\phi/360^\circ$.  The
self-field is therefore axis-aligned and built from the projected rms
sizes: the full $6\times6$ $\Sigma$, including its cross-plane
correlations, is transported, but coupled (tilted-ellipsoid)
self-field terms are not included; the model therefore targets beams
whose principal axes remain close to the laboratory axes.  The
interior field of the ellipsoid is linear,
\begin{equation}
  E_i(r_i) = \frac{3Q}{4\pi\epsilon_0}\,\frac{M_i\, r_i}{abc},
  \qquad i = x, y, z,
  \label{eq:ellipsoidfield}
\end{equation}
where $r_i$ denotes the field-point coordinate inside the bunch and
$M_i$ are Maxwell depolarization factors that depend only on the
semi-axes,
\begin{equation}
\begin{aligned}
  M_x &= \frac{abc}{2}\!\int_0^\infty\!\!
        \frac{dt}{(t + a^2)^{3/2}\sqrt{(t + b^2)(t + c^2)}} \\
      &= \frac{abc}{3}\,R_D\!\left(b^2, c^2, a^2\right)
\end{aligned}
  \label{eq:depol}
\end{equation}
(and cyclically for $M_y$ and $M_z$),
which satisfy $M_x + M_y + M_z = 1$ and are evaluated exactly through
Carlson's symmetric elliptic integral $R_D$~\cite{carlson}, as in ellipsoidal-bunch
envelope treatments~\cite{ferrario}, rather than through interpolated
form-factor tables.  Over a kick spacing $\Delta s$ the resulting
thin-lens kicks in all three planes are
\begin{align}
  \Delta x' &= \frac{3\, |q_s|\, Q\, M_x\, \Delta s}
                    {4\pi\epsilon_0\, abc\, \beta^2\gamma^2 mc^2}\; x,
  \qquad \text{(and analogously in } y\text{)},
  \label{eq:sckickperp}\\
  \Delta W &\;\rightarrow\; \Delta W + |q_s|\,E_z(z_r)\,\Delta s,
  \qquad
  z_r = \gamma\,\zeta ,
  \label{eq:sckicklong}
\end{align}
where $z_r$ is the rest-frame longitudinal offset of a particle at
phase deviation $\Delta\phi$ (in degrees); positive $\Delta\phi$
(late arrival) maps to negative $z_r$, toward the bunch tail.
Throughout this paper
$q_s$ denotes the signed charge of the single tracked species
($q_s = -e$ for H$^-$; HELIX transports one species per run):
external-field kicks carry the signed $q_s$, while all space-charge
kick expressions carry the magnitude $|q_s|$, matching the code's
unsigned charge deposition (see the discussion
after Eq.~(\ref{eq:pickick}) below).  The $\beta^2\gamma^2$
factor in Eq.~(\ref{eq:sckickperp}) combines the $1/\gamma$ net
transverse Lorentz force in the lab frame, the interaction time
$\Delta t = \Delta s/(\beta c)$ over the kick spacing, and the
$p_z = \beta\gamma\,mc$ momentum used to convert $\Delta p_\perp$
into an angle, while $E_z$
is invariant under the longitudinal boost and produces a direct
energy kick.  Including the longitudinal plane in the linearized kick
matrix lets the envelope solver track rf--space-charge equilibria
that purely transverse formulations miss.

\emph{Continuous beams.}  For dc beams the longitudinal block of
$\Sigma$ is zeroed and the transverse kick reverts to the linear field
of a uniform elliptical cylinder [Eq.~(\ref{eq:dcuniform}) below with
semi-axes $2\sigma_x$, $2\sigma_y$].  HELIX additionally ships a
standalone continuous-beam envelope integrator for unbunched beams,
which solves the rms envelope equations~\cite{sacherer}
\begin{equation}
\begin{aligned}
  \sigma_x'' + \kappa_x(s)\,\sigma_x
   - \frac{\varepsilon_x^2}{\sigma_x^3}
   - \frac{K}{2\,(\sigma_x + \sigma_y)} &= 0, \\
  K &= \frac{|q_s|\, I}{2\pi\epsilon_0\, m c^3 (\beta\gamma)^3},
\end{aligned}
  \label{eq:sacherer}
\end{equation}
(and the analogous equation in $y$) in rms size $\sigma$ and rms
emittance $\varepsilon$ [the equivalent KV edge-radius form
$X'' + \kappa X - \varepsilon_X^2/X^3 - 2K/(X+Y) = 0$ follows with
$X = 2\sigma$ and $\varepsilon_X = 4\varepsilon$], integrated with an
adaptive Runge--Kutta scheme; the focusing function $\kappa(s)$ is
read element by element from hard-edge quadrupoles or from the
on-axis $B_z(s)$ of 1-D/3-D solenoid maps via
$\kappa = [B_z/(2 B\rho)]^2$ (the scalar Larmor-frame focusing
strength; the scalar equations assume negligible canonical angular
momentum), and $\beta$, $\gamma$ are held constant
(no acceleration), matching the low-energy-transport regime the
solver targets.  Its space-charge normalization is consistent with
the dc envelope kick used in production transport
[Eq.~(\ref{eq:dcuniform}) applied to $\Sigma$ with semi-axes
$a = 2\sigma_x$, $b = 2\sigma_y$] in the nonrelativistic regime where
both operate: Eq.~(\ref{eq:sacherer}) carries the full
$(\beta\gamma)^{-3}$ with the $1/\gamma^{2}$ magnetic cancellation,
whereas the production kick is electrostatic and scales as
$(\beta^{3}\gamma)^{-1}$, a $\gamma^{2}$ difference that is
negligible at LEBT energies ($\gamma - 1 \sim 3\times10^{-5}$; see
the discussion below Eq.~(\ref{eq:dcuniform})).  The two paths provide
independent cross-checks of the continuous-beam envelope physics
(Sec.~\ref{sec:validation}).

\emph{Coupling diagnostics and transverse invariants.}  The
projected emittances of
Eq.~(\ref{eq:twiss}) oscillate under solenoidal (transverse $x$--$y$)
coupling and exchange with the longitudinal plane under dispersive
coupling.  HELIX therefore records at every step the 4-D emittance
$\varepsilon_{4D} = \sqrt{\det\Sigma_{4D}}$ [an emittance-squared
quantity, (mm$\cdot$mrad)$^2$], which is invariant under
any linear transport that is symplectic within the transverse
$4\times4$ block (solenoid rotations in particular) but not, in
general, under transverse--longitudinal coupling, which moves
correlations across the block boundary; and, to cover the coupled
6-D case, three dimensionless spectral mode values
$\hat\varepsilon_i$ (the
symplectic eigenvalues of the unit-scaled covariance matrix defined
below, relative to the fixed matrix $J_0$), computed from the trace
invariants of Balandin \emph{et al.}~\cite{balandin}, which in these
mixed-unit internal coordinates are fixed-convention coupling
diagnostics rather than canonical invariants, as made
precise
below.  These
quantities are evaluated on the dimensionless matrix
$\widehat{\Sigma} = S\,\Sigma\,S^{\mathsf{T}}$, with
$S$ the diagonal matrix of inverse internal unit scales
(mm, mrad, deg, MeV) in
the coordinates $(x, x', y, y', \Delta\phi, \Delta W)$, so the
expressions below are algebraically well defined; with
$J_0 = \mathrm{diag}(J_2, J_2, J_2)$,
$J_2 = \bigl(\begin{smallmatrix} 0 & 1 \\ -1 & 0
\end{smallmatrix}\bigr)$, the fixed block-skew matrix used to define
the diagnostic,
\begin{equation}
\begin{aligned}
  I_2 &= -\tfrac{1}{2}\,\mathrm{tr}\!\left[(\widehat{\Sigma} J_0)^2\right],\quad
  I_4 = +\tfrac{1}{2}\,\mathrm{tr}\!\left[(\widehat{\Sigma} J_0)^4\right], \\
  I_6 &= -\tfrac{1}{2}\,\mathrm{tr}\!\left[(\widehat{\Sigma} J_0)^6\right],
\end{aligned}
  \label{eq:invariants}
\end{equation}
the squared mode values $\hat\varepsilon^2$ are the roots of the cubic
\begin{equation}
  \hat\varepsilon^6 - I_2\,\hat\varepsilon^4
  + \frac{I_2^2 - I_4}{2}\,\hat\varepsilon^2
  - \left(\frac{I_2^3}{6} - \frac{I_2 I_4}{2}
          + \frac{I_6}{3}\right) = 0 .
  \label{eq:eigencubic}
\end{equation}
This trace formulation avoids explicit identification and pairing of
the conjugate eigenvalues of $\widehat{\Sigma} J_0$ when the mode values
differ by orders of magnitude, as they do for long proton bunches.
In uncoupled transport the roots reduce exactly to the numeric
values of the projected
emittances $(\varepsilon_x, \varepsilon_y, \varepsilon_z)$ in their
internal units (mm$\cdot$mrad transversely, deg$\cdot$MeV
longitudinally).  The internal coordinates are not canonical pairs:
$x' = p_x/p_z$ is kinetic and $(\Delta\phi, \Delta W)$ carry
deg/MeV scalings, so the traces of Eq.~(\ref{eq:invariants}) are
constants of motion only where the transport is symplectic with
respect to $J_0$ in \emph{these} coordinates.  That holds demonstrably
for fixed-energy linear transport that mixes only coordinate pairs
sharing a common scale factor: each pair
differs from a canonical pair by a constant per-pair factor
(e.g.\ $x'$ from $p_x/p_0$ by the common reference momentum, to
paraxial order), so coupling confined to the transverse
pairs (solenoid rotations in particular, whose two pairs share one
scale factor) leaves the mode values constant while the
projections oscillate; maps that couple the transverse and
longitudinal pairs (dispersive bends) mix pairs of \emph{different}
scale factor and are not symplectic with respect to this fixed $J_0$
even at fixed energy, so there the roots are diagnostics from the
outset.  Because the per-pair scale factors differ between the
transverse and longitudinal pairs, the mode values
under transverse--longitudinal coupling depend on
this mm/mrad/deg/MeV convention (converting phases to radians, for
instance, would change the coupled roots).  Through acceleration, rf kicks,
and frequency jumps
those scale factors change and the maps are not symplectic with
respect to $J_0$; the $\hat\varepsilon_i$ are therefore retained
only as fixed-convention diagnostics of
coupling and phase-space dilution, with no invariance implied.

\subsection{Multiparticle tracking and integrators}
\label{sec:integrators}

The multiparticle tracker dispatches on element type: matrix elements
(drifts, quadrupoles, bends, solenoids) apply their exact linear maps,
thin elements apply momentum kicks, and field-map elements are
integrated numerically in sub-steps whose density follows the
TraceWin-style step configuration (separate cadences for integration
and for space-charge kicks), with automatic refinement for
magnetic-only solenoid maps and for cavities whose field-map grids are
finer than the requested step.  Aperture losses are checked once per
space-charge bundle and after any trailing sub-steps, and per sub-step in
the two-sub-step matrix elements (quadrupoles, bends, hard-edge
solenoids).  Static misalignments $(dx, dy,
\theta_{\mathrm{tilt}})$ are realized as exact coordinate transforms
into and out of each element frame (in envelope mode the tilt acts as
a rotation of $\Sigma$; rigid offsets leave second moments
invariant).

Through a field map, one integration slice of length $h$ is a
split-operator map composed of a kick $K_h$ (the impulse of the
interpolated $(\mathbf{E}, \mathbf{B})$ fields sampled at the slice
midpoint in $z$ with each particle's rf phase) and a drift $D_h$.
Two compositions are available for standalone 3-D field maps,
selected globally,
\begin{equation}
  \Psi^{\mathrm{KD}}_h = D_h \circ K_h,
  \qquad
  \Psi^{\mathrm{DKD}}_h = D_{h/2} \circ K_h \circ D_{h/2},
  \label{eq:splitting}
\end{equation}
a first-order kick--drift scheme (the default, retained for TraceWin
parity) and a Strang-like symmetric drift--kick--drift
composition~\cite{strang} in which transverse positions are advanced by half a
slice before the fields are sampled.  The DKD layout centers the
transverse position used for field sampling, reducing the leading
position-sampling asymmetry of the kick--drift composition; because
the
kick acts on the
kinetic coordinates $(x, x')$ rather than on canonical pairs,
contains first-order phase and momentum-rescaling sub-updates, and
is not treated as an exact split flow, no
formal convergence-order or symplecticity claim is made for the
complete rf/Lorentz
map in these
variables.  The kick itself applies the per-slice energy
gain $\delta W = q_s E_z\,h$ evaluated at each particle's
phase (the reference particle is advanced through the same on-axis
field at the synchronous phase, so the deviation coordinate
$\Delta W$ updates by the difference), the transverse Lorentz impulse
\begin{equation}
\begin{aligned}
  \Delta x' &= \frac{q_s\,h}{\gamma\beta^2 mc^2}
              \left[E_x + \beta c\,(y' B_z - B_y)\right], \\
  \Delta y' &= \frac{q_s\,h}{\gamma\beta^2 mc^2}
              \left[E_y + \beta c\,(B_x - x' B_z)\right],
\end{aligned}
  \label{eq:lorentzkick}
\end{equation}
the adiabatic damping of both divergences by
$[1 + \delta W/(\beta^2\gamma\, mc^2)]^{-1}$ that keeps
$x' = p_x/p_z$ consistent as $p_z$ grows, and the first-order phase
slip $\Delta\phi \rightarrow \Delta\phi - [360^\circ
h/(\beta^3\gamma^3 mc^2 \lambda_{\mathrm{rf}})]\,\Delta W$ of an
off-energy particle, with $\Delta W$ here the particle's accumulated
energy deviation.

Space-charge kicks are inserted by a second, outer symmetric
transport--kick--transport composition.
Sub-steps are grouped into bundles of length $\Lambda$ set by the
space-charge cadence, and each bundle is advanced as
\begin{equation}
  \Phi_\Lambda = T_{\Lambda/2} \circ S_\Lambda \circ T_{\Lambda/2},
  \label{eq:strangsc}
\end{equation}
where $T$ is the space-charge-free transport (the composition of
element slices) and $S_\Lambda$ is the collective kick accumulated
over $\Lambda$, evaluated at the bundle midpoint.  When $T$ and
$S_\Lambda$
are exact or symmetric second-order submaps this is the usual Strang
split~\cite{strang}; where $T$ contains the order-unassigned
field-map push above, no additional formal order claim is made for
the complete map.  The same bundling
is used for drifts and field maps in both the
multiparticle and envelope solvers, which see identical
kick placement there: both derive the bundle from the same
integration- and space-charge-step densities.  The two-sub-step
hard-edge elements are the exception---quadrupoles, bends, and
hard-edge solenoids take one kick per sub-step, so two per element, in
multiparticle mode, against the configured space-charge cadence in
envelope mode.  In envelope mode the bundle transport matrices for
field maps are obtained by numerically linearizing the single-particle
push: central-difference Jacobians of the map are evaluated about the
reference orbit, with the reference particle advanced to the bundle
midpoint so accelerating cavities use the locally correct $\beta$ and
$\gamma$.  Because these Jacobians are comparatively expensive,
element matrices can be memoized in an opt-in cache keyed on the
element parameters and a fingerprint of the reference state.  The
cached path serves repeated transfer-matrix work (phase-advance
analysis and the matching tab's linearizations) at seconds-scale
re-evaluation cost; the envelope solver itself does not consume this
cache and re-tracks its probes on every run
(Sec.~\ref{sec:costhierarchy}).

A continuous beam transitions to bunched tracking automatically at
the first rf bunching element (an rf gap or rf field map with nonzero
accelerating field, or an RFQ cell).  In multiparticle mode the
particles already carry longitudinal coordinates sampled uniformly
over one rf period, so the cavity bunches them physically with no
coordinate remapping; in envelope mode the longitudinal block of
$\Sigma$ is seeded with the uniform-distribution equivalent
$\sigma_\phi = 180^\circ/\sqrt{3}$ and the configured dc energy
spread.  This seeding is a moment-level bookkeeping of the uniform
phase distribution, not a capture model: the linearly propagated
$\Sigma$ cannot represent nonlinear rf capture, bunching, or the
associated loss, so envelope work downstream of a buncher should
start from an already-bunched covariance (multiparticle mode
captures the beam physically).

\subsection{Space-charge models}
\label{sec:spacecharge}

\emph{3-D PIC for bunched beams.}  The production space-charge model
is an electrostatic PIC solve in the bunch frame.  Each kick performs
the following cycle.  (1)~Particle positions are boosted
longitudinally to the beam rest frame, $z_r = \gamma\zeta$, with
transverse coordinates unchanged; the rest-frame charge density is
automatically diluted by $1/\gamma$ because the grid is built on the
boosted coordinates.  (2)~A Cartesian grid (default $96^3$ nodes) is
centered on the bunch centroid with half-extent $5\,\sigma$ per axis
by default (the mesh study of Sec.~\ref{sec:validation} quantifies
the sensitivity of this choice; the bunched-beam cross-code
benchmarks there use
$\pm6\sigma$);
the grid is frozen after the first kick or re-fitted every kick in
adaptive mode.  Particles are never discarded for leaving the box:
both the cell index and the in-cell fraction are clipped, in
deposition and gather alike, so an out-of-box particle deposits its
full charge in the nearest boundary cell, is gathered from that cell,
and charge is conserved exactly.  What is one-sided at the box faces
is the field-gradient stencil of step~(5); that one-sided difference,
not the box fit, sets the momentum-balance cost measured in the
field-accuracy test of Sec.~\ref{sec:validation}, where a single tail
particle in the outermost cell dominates the residual net kick.  (3)~Macroparticle charges $q_m = Q/N = I/(f_b N)$,
with $N$ the initial macroparticle count so that losses deplete the
bunch charge, are deposited with either the cloud-in-cell (CIC) or
the triangular-shaped-cloud (TSC) B-spline shape
function~\cite{hockney_eastwood}.  CIC spreads each particle
trilinearly over the $2^3$ surrounding nodes with per-axis weights
$\{1-\delta, \delta\}$, where $\delta \in [0,1)$ is the fractional
position within the cell; TSC spreads it quadratically over $3^3$
nodes with per-axis weights
\begin{equation}
  w_{-1} = \tfrac{1}{2}\left(\tfrac{1}{2}-\delta_c\right)^{2},\quad
  w_{0}  = \tfrac{3}{4}-\delta_c^{2},\quad
  w_{+1} = \tfrac{1}{2}\left(\tfrac{1}{2}+\delta_c\right)^{2},
  \label{eq:tsc}
\end{equation}
with $\delta_c \in [-\tfrac12,\tfrac12)$ the offset from the center of
the containing cell (so that $w_0$ weights the lower of the two nodes
bracketing the particle), reducing grid-noise emittance growth at about $3.4\times$ the
deposition cost.  The stencil is thus anchored to the containing cell
rather than to the nearest node; the gather applies the same
convention, which amounts to a half-cell relabeling of the mesh and
does not affect the space-charge kick.  (4)~The open-boundary Poisson equation is solved by
Hockney's method~\cite{hockney_eastwood}: the density is zero-padded
onto a doubled grid and the potential is obtained as the discrete
free-space convolution
\begin{equation}
  \phi_{\mathbf{i}} = V_c \sum_{\mathbf{j}}
      G_{\mathbf{i}-\mathbf{j}}\, \rho_{\mathbf{j}},
  \label{eq:hockney}
\end{equation}
evaluated with FFTs, where $V_c = \Delta x\,\Delta y\,\Delta z$ and
the Green's-function FFT is precomputed once per grid.  Two kernels
are implemented.  The default is the integrated Green's function
(IGF)~\cite{qiang2006,qiang2024igf}, the exact average of the
Coulomb kernel over the node-centered cell,
\begin{equation}
\begin{aligned}
  G_{ijk} &= \frac{1}{4\pi\epsilon_0 V_c}\,
     F(x,y,z)\,\Big|_{x_-}^{x_+}\Big|_{y_-}^{y_+}\Big|_{z_-}^{z_+},\\
  x_\pm &= \Bigl(i \pm \tfrac{1}{2}\Bigr)\Delta x, \;\ldots
\end{aligned}
  \label{eq:igf}
\end{equation}
computed by eight-corner inclusion--exclusion of the closed-form
antiderivative $\partial^3 F/\partial x\,\partial y\,\partial z
= 1/r$,
\begin{equation}
\begin{split}
  F(x,y,z) ={}& yz\,\sinh^{-1}\!\frac{x}{\sqrt{y^2+z^2}}
             + zx\,\sinh^{-1}\!\frac{y}{\sqrt{z^2+x^2}} \\
            &{}+ xy\,\sinh^{-1}\!\frac{z}{\sqrt{x^2+y^2}}
             - \frac{x^2}{2}\tan^{-1}\!\frac{yz}{xr} \\
            &{}- \frac{y^2}{2}\tan^{-1}\!\frac{zx}{yr}
             - \frac{z^2}{2}\tan^{-1}\!\frac{xy}{zr},
\end{split}
  \label{eq:igfantideriv}
\end{equation}
with $r = \sqrt{x^2+y^2+z^2}$.  This $\sinh^{-1}$/$\arctan$ form of
the antiderivative, also used by Cheetah's space-charge
extension~\cite{dhamrait2026}, is odd under
coordinate inversion and therefore yields a reflection-symmetric
kernel on the doubled grid, avoiding the branch shifts of the
equivalent $\mathrm{atan2}$/$\log$ form.  The alternative
\texttt{point} kernel samples $G = 1/(4\pi\epsilon_0 r)$ at cell
centers with the self-cell regularized to
$r_0 = \tfrac{1}{2}\sqrt{\Delta x^2 + \Delta y^2 + \Delta z^2}$; it
is retained only for legacy regression baselines, since the IGF
removes the point-sampled kernel's near-source bias.  (5)~The electric field
$\mathbf{E} = -\nabla\phi$ is formed by second-order central
differences (one-sided at the box faces) and gathered to the
particles with the same shape function used for deposition.
(6)~Finally, the rest-frame fields are converted directly into
lab-frame kicks,
\begin{equation}
\begin{aligned}
  \Delta x' &= \frac{|q_s|\,E^{(r)}_x\,\Delta s}{\beta^2\gamma^2\,mc^2}, \\
  \Delta y' &= \frac{|q_s|\,E^{(r)}_y\,\Delta s}{\beta^2\gamma^2\,mc^2}, \\
  \Delta W  &\;\rightarrow\; \Delta W + |q_s|\,E^{(r)}_z\,\Delta s,
\end{aligned}
  \label{eq:pickick}
\end{equation}
with no explicit inverse boost: as in
Eq.~(\ref{eq:sckickperp}), the frame transformation and the
$\mathbf{E} + \mathbf{v}\times\mathbf{B}$ cancellation are folded
into the $\beta^2\gamma^2$ factor, and $E_z$ is boost-invariant.
Charges are deposited unsigned: the deposited density represents the
magnitude of the bunch charge, so the computed field points outward
from that positive representation, and multiplying by the charge
magnitude $|q_s|$
[Eqs.~(\ref{eq:sckickperp}), (\ref{eq:sckicklong}),
and~(\ref{eq:pickick}), as well as the dc kicks below] yields the
repulsive self-force for either sign of the single tracked species
(H$^-$ included, with no sign flip in the kick); external-field
kicks retain the signed $q_s$,
Eq.~(\ref{eq:lorentzkick}).

\emph{2-D models for continuous beams.}  When the beam is unbunched
the longitudinal force vanishes and three transverse kick models of
increasing fidelity are available.  The default is the linear field
of a uniform-density elliptical cylinder carrying current $I$,
\begin{equation}
  E_x = \frac{I\,x}{\pi\epsilon_0\,\beta c\; a\,(a+b)},
  \qquad
  E_y = \frac{I\,y}{\pi\epsilon_0\,\beta c\; b\,(a+b)},
  \label{eq:dcuniform}
\end{equation}
with rms-equivalent semi-axes $a = 2\sigma_x$ and $b = 2\sigma_y$,
applied as $\Delta u' = |q_s| E_u\,\Delta s/(\beta^2\gamma\,mc^2)$, a
kick factor shared by all three models of this paragraph.
This is explicitly a nonrelativistic electrostatic approximation:
the kick keeps only the electric self-field and omits the
$1/\gamma^2$ magnetic self-field cancellation (which would make the
denominator $\beta^2\gamma^3\,mc^2$), a difference that is negligible
in the $\gamma \simeq 1$ low-energy transport where dc beams arise
(for the PIP-II LEBT, $\gamma - 1 \sim 3\times10^{-5}$).  The form
matches the TraceWin continuous-beam
formula~\cite{tracewin_manual} and, applied to $\Sigma$, reproduces the dc
envelope kick exactly.  The second model keeps the measured
$(\sigma_x, \sigma_y)$ rigid but applies the nonlinear per-particle
field of a 2-D Gaussian charge density in the closed form of Bassetti
and Erskine~\cite{bassetti_erskine},
\begin{equation}
  E_y + i E_x = \frac{\lambda}{2\epsilon_0\sqrt{\pi}\,\sigma_\Delta}
  \left[\,w(z_1) -
  e^{-\frac{x^2}{2\sigma_x^2}-\frac{y^2}{2\sigma_y^2}}\,
  w(z_2)\right],
  \label{eq:bassettierskine}
\end{equation}
with line density $\lambda = I/\beta c$,
$\sigma_\Delta = \sqrt{2(\sigma_x^2 - \sigma_y^2)}$ for
$\sigma_x > \sigma_y$ (the axes are swapped internally in the
opposite case),
$z_1 = (x + iy)/\sigma_\Delta$,
$z_2 = (x\sigma_y/\sigma_x + i y\sigma_x/\sigma_y)/\sigma_\Delta$,
and $w(z) = e^{-z^{2}}\,\mathrm{erfc}(-iz)$ the Faddeeva function;
for
$|\sigma_x - \sigma_y| < 0.05\,\max(\sigma_x,\sigma_y)$ the round-beam
limit
$E_r = \lambda\,[1 - e^{-r^2/2\sigma^2}]/(2\pi\epsilon_0 r)$ with
$\sigma^2 = (\sigma_x^2 + \sigma_y^2)/2$ is used instead.  That
threshold is a conservative guard against the $\sigma_\Delta \to 0$
singularity rather than a floating-point necessity: the difference of
the two Faddeeva terms is a finite, asymmetry-independent fraction of
either one, and the double-precision pair still reproduces the
round-beam limit to $5\times10^{-15}$ at relative asymmetries of
$10^{-14}$.  At the threshold itself the substituted round-beam field
departs from the exact elliptical result by up to 2.6\% within
$2\sigma$.  The third
model is a full 2-D PIC solve of the actual particle distribution in
the $(x, y)$ plane, serving the
same role as TraceWin's PICNIC-2D solver~\cite{tracewin}: each of the
surviving macroparticles carries line charge
$\lambda/N_{\mathrm{live}}$, and the cycle is CIC deposition,
doubled-grid Hockney convolution [the
cell-area-weighted 2-D analogue of Eq.~(\ref{eq:hockney})]
with the logarithmic kernel
$G(\mathbf{r}) = -\ln(r/r_\ast)/(2\pi\epsilon_0)$,
central-difference gradient, and CIC gather.  The reference
$r_\ast$ is arbitrary, since only $-\nabla\phi$ is used, and the
self-cell is regularized at
the equal-area radius $r_{\mathrm{eff}} = \sqrt{\Delta x\,\Delta
y/\pi}$.  In either
solver mode, \texttt{SPACE\_CHARGE\_COMP} cards scale the applied
space-charge current by $1-f$, with $f$ the neutralisation fraction,
so successive sections of a line can run partially or fully
neutralized; the PIP-II LEBT benchmark deck of
Sec.~\ref{sec:validation} uses such sectioned compensation.

\emph{CSR wake.}  Beyond these core models, the tracker includes a
1-D steady-state coherent-synchrotron-radiation (CSR) wake applied
per sub-step inside bends~\cite{saldin_csr}.  With $\zeta$ the
bunch coordinate of Sec.~\ref{sec:envelope} (positive toward the
bunch head),
$n(\zeta)$ the number line density, and $R$ the bend radius,
the energy change per unit path of the singly charged tracked
species is
\begin{equation}
  \frac{dW}{ds}(\zeta) = -\frac{2\,e^{2}}
       {4\pi\epsilon_0\, 3^{1/3} R^{2/3}}
  \int_{-\infty}^{\zeta}
     \frac{n'(\zeta')}{(\zeta-\zeta')^{1/3}}\, d\zeta' ,
  \label{eq:csr}
\end{equation}
with the integrable kernel singularity integrated analytically over
each density bin; the density is a per-kick histogram of the live
particles (200 bins by default, Gaussian-smoothed over 1.5 bins
before differentiation), renormalized so that its integral is the
\emph{design} bunch population set by the beam current, so the CSR
source is not depleted by losses; in this it follows the continuous-beam
line charge rather than the bunched macrocharge, which is
the one collective source that scraping does reduce.  This is the free-space, ultrarelativistic, steady-state
model: entrance and exit transients and shielding are not included,
the kick changes only particle energies (transverse effects emerge
downstream through dispersion), and it acts only in multiparticle
tracking.  It contributes to none of the studies in this paper,
whose lattices are bend-free except for the transfer line of
Sec.~\ref{sec:app_foil}, where the particle runs carry zero bunch
charge.

\emph{Foil interactions.}  A zero-length foil element combines
Highland multiple scattering~\cite{highland,pdg} with a
minimum-ionizing mean energy loss (a fixed tabulated mass stopping
power, in \si{\mega\electronvolt\square\centi\meter\per\gram},
multiplied by the areal density; the $\beta$-dependent
Bethe rise is not modeled, so the fixed value is representative near
the stopping-power minimum and increasingly understates the mean
loss toward low energies) and regime-dispatched
straggling~\cite{vavilov,pdg}: a mean-pinned truncated-Landau sample
in the thin-absorber regime and a Gaussian for thick absorbers, with
the intermediate Vavilov regime assigned to the Gaussian branch.
Tracking samples the angular and energy-loss kicks per particle; in
envelope mode the covariance-diffusion update
$\Sigma \to \Sigma + D$ adds the Highland $\theta_0^2$ and a Gaussian
straggling $\sigma_E^2$ to the moments while the mean loss shifts the reference
energy.  Charge-state conversion is not simulated, so for H$^-$ the
element is a material-interaction model that is physically
meaningful only up to the foil itself; in the application of
Sec.~\ref{sec:app_foil} the foil terminates the tracked
line.  The dispatch
thresholds, the thin-target extrapolation of the Highland fit, and
the regime placement of the demonstration foil are quantified in
Sec.~\ref{sec:app_foil}.

\emph{H$^-$ loss analyzers.}  Two post-tracking analyzers estimate
stripping losses on the recorded $s$-grid: one for magnetic
(Lorentz) stripping, which evaluates the empirical fractional-loss
rate of Folsom \emph{et al.}~\cite{folsom_stripping} on the design
fields (bends at their design field; quadrupoles as $|G|\,r$ at the
combined radius $r = 2\sqrt{\sigma_x^2 + \sigma_y^2}$; field maps at
their per-element peak field, an axial component entering through
the rms divergence), yielding a screening estimate along the lattice
rather than a particle-resolved loss map; and one for
intra-beam stripping, which evaluates the closed-form rate of
Lebedev \emph{et al.}~\cite{lebedev_ibst} on the recorded rms beam
moments.  Residual-gas and blackbody-radiation stripping are not
modeled.

\emph{Error models.}  Statistical element- and beam-error models
complete the set: geometric misalignments reuse the exact frame
transforms above ($\theta_{\mathrm{tilt}}$ is a roll about the beam
axis), field-amplitude and rf phase errors perturb the element
parameters directly, and beam errors perturb the generated input
distribution (centroid, emittance scale, Twiss mismatch, current);
the present semantics are static and uncoupled
(Sec.~\ref{sec:limitations}).  Of these auxiliary models, the foil
and the error models are exercised in the PIP-II case studies of
Sec.~\ref{sec:applications}.

\section{Numerical methods and software architecture}
\label{sec:methods}

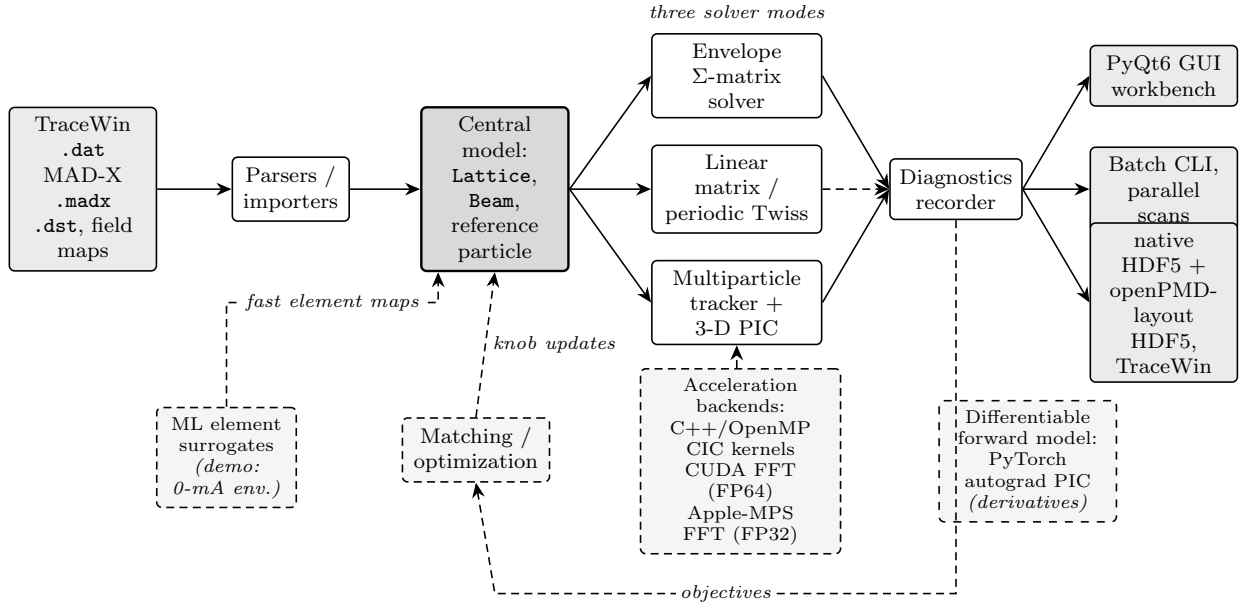
\begin{figure*}[t]
\centering
\begin{tikzpicture}[
  >={Stealth[length=2mm,width=1.7mm]},
  font=\footnotesize,
  box/.style={draw, semithick, rounded corners=2pt, align=center,
              inner sep=3.5pt, minimum height=8mm, fill=white},
  inputbox/.style={box, fill=black!8},
  corebox/.style={box, line width=0.9pt, fill=black!15},
  solverbox/.style={box, fill=white},
  outbox/.style={box, fill=black!8},
  sidebox/.style={box, densely dashed, fill=black!4},
  flow/.style={->, semithick},
  feed/.style={->, semithick, densely dashed},
  lab/.style={font=\scriptsize\itshape, inner sep=1.5pt, fill=white},
]
\node[inputbox, text width=17mm] (inputs) at (0,0)
  {TraceWin \texttt{.dat}\\ MAD-X \texttt{.madx}\\ \texttt{.dst}, field maps};
\node[box, text width=13mm] (parser) at (2.75,0)
  {Parsers /\\ importers};
\node[corebox, text width=17mm] (model) at (5.45,0)
  {Central model:\\ \texttt{Lattice}, \texttt{Beam},\\ reference particle};
\node[solverbox, text width=20mm] (env) at (8.65,1.5)
  {Envelope\\ $\Sigma$-matrix solver};
\node[solverbox, text width=20mm] (mat) at (8.65,0)
  {Linear matrix /\\ periodic Twiss};
\node[solverbox, text width=20mm] (mp)  at (8.65,-1.5)
  {Multiparticle\\ tracker + 3-D PIC};
\node[font=\scriptsize\itshape] at (8.65,2.35) {three solver modes};
\node[box, text width=15mm] (diag) at (11.55,0)
  {Diagnostics\\ recorder};
\node[outbox, text width=17mm] (gui)   at (14.3,1.5)
  {PyQt6 GUI\\ workbench};
\node[outbox, text width=17mm] (cli)   at (14.3,0)
  {Batch CLI,\\ parallel scans};
\node[outbox, text width=17mm] (files) at (14.3,-1.5)
  {native HDF5 +\\ openPMD-layout\\ HDF5, TraceWin};
\node[sidebox, text width=16mm, font=\scriptsize] (surr) at (1.9,-3.55)
  {ML element\\ surrogates\\ \textit{(demo: 0-mA env.)}};
\node[sidebox, text width=17mm] (match) at (5.2,-3.45)
  {Matching /\\ optimization};
\node[sidebox, text width=24mm, font=\scriptsize] (accel) at (8.7,-3.6)
  {Acceleration backends:\\ C++/OpenMP CIC kernels\\ CUDA FFT (FP64)\\
   Apple-MPS FFT (FP32)};
\node[sidebox, dashed, text width=21mm, font=\scriptsize] (diff) at (12.5,-3.6)
  {Differentiable forward model:\\ PyTorch autograd PIC\\
   \textit{(derivatives)}};
\draw[flow] (inputs) -- (parser);
\draw[flow] (parser) -- (model);
\draw[flow] (model.east) -- (env.west);
\draw[flow] (model.east) -- (mat.west);
\draw[flow] (model.east) -- (mp.west);
\draw[flow] (env.east) -- (diag.west);
\draw[feed] (mat.east) -- (diag.west);
\draw[flow] (mp.east)  -- (diag.west);
\draw[flow] (diag.east) -- (gui.west);
\draw[flow] (diag.east) -- (cli.west);
\draw[flow] (diag.east) -- (files.west);
\draw[feed] (accel.north -| mp.south) -- (mp.south);
\draw[feed] (diag.south) -- (11.55,-5.35) -- (5.45,-5.35)
  node[lab, midway] {objectives} -- (match.south);
\draw[feed] (match.north) -- (model.south)
  node[lab, midway, right=1pt] {knob updates};
\draw[feed] (surr.north) -- (1.9,-1.5) -- (4.7,-1.5)
  node[lab, midway] {fast element maps} -- (4.7,-1.5 |- model.south);
\end{tikzpicture}
\caption{HELIX architecture and dataflow.  Lattice descriptions
(TraceWin \texttt{.dat} or MAD-X files, plus \texttt{.dst}
distributions and tabulated field maps) are parsed into a central
model consisting of a \texttt{Lattice}, a \texttt{Beam}, and a
reference particle.  A run dispatches this model to one of three
solver modes (envelope $\Sigma$-matrix, linear matrix/periodic Twiss,
or multiparticle tracking with 3-D PIC space charge); the envelope
and multiparticle solvers
populate a common diagnostics recorder feeding the PyQt6 workbench,
the batch CLI with process-parallel scans, and native HDF5,
openPMD-layout HDF5, and
TraceWin-format file outputs (the matrix path, dashed at the
recorder, returns its optics directly).  Dashed elements are the layers built
around this pipeline: acceleration backends (C++/OpenMP deposit/gather
kernels, CUDA and Apple-MPS FFTs) and the differentiable
PyTorch forward model of
Sec.~\ref{sec:diffpic} (a derivative provider rather than an
accelerator), serving the multiparticle solver, the matching
and optimization engine (Sec.~\ref{sec:matching}) that consumes
recorded diagnostics as objectives and writes updated knob values back
into the lattice, and machine-learned element surrogates
(Sec.~\ref{sec:surrogates}) that substitute fast maps for expensive
field-map elements.}
\label{fig:architecture}
\end{figure*}

HELIX, currently distributed under the Python package name
\texttt{linac\_gen} \cite{pathak_linacgen} (HELIX is the public name
at release; the package import will remain \texttt{linac\_gen}), is
implemented in Python.  Its architecture, sketched in
Fig.~\ref{fig:architecture}, is organized around a deliberately small
data model: a \texttt{Lattice} (an ordered element list plus step and
error configuration), a \texttt{ReferenceParticle} (species, kinetic
energy, rf frequency, and derived kinematics), and a \texttt{Beam} (an
$N\times 6$ macroparticle array with per-particle loss flags and beam
current).  Every simulation is a function of these objects: a front
end reading established lattice formats (a documented subset of
the TraceWin lattice
format \cite{tracewin} as its primary dialect, plus a MAD-X subset,
with the \texttt{.dst}/openPMD-layout HDF5 interchange surface
detailed
below) builds the central model,
a run dispatches it to one of three interchangeable solver modes, and
the envelope and multiparticle modes stream their results into a
common diagnostics recorder that
feeds the graphical workbench, the batch command-line interface (CLI),
and the file outputs (the matrix mode returns its map and derived
optics directly).  The core is pure Python on
NumPy/SciPy~\cite{numpy,scipy}, in the
spirit of recent accelerator toolkits such as Xsuite \cite{xsuite} and
Cheetah \cite{cheetah}; compiled and GPU acceleration is added only in
the two hot spots of the particle-in-cell (PIC) loop, charge
deposit/field gather and the Poisson FFT, where profiling identifies
potential benefit at sufficient workload (measured crossovers in
Sec.~\ref{sec:performance}).  This section describes the lattice front end, the solver
dispatch, the acceleration layers, the input/output (I/O) surface, and
the user interfaces; the differentiable-PIC, matching, and surrogate
engines that close the feedback loops in Fig.~\ref{fig:architecture}
are described in Secs.~\ref{sec:diffpic}--\ref{sec:surrogates}.

\subsection{Lattice language and element library}
\label{sec:lattice}

The primary lattice front end is a parser for TraceWin-format
\texttt{.dat} files including the element cards (\texttt{DRIFT},
\texttt{QUAD}, \texttt{SOLENOID}, \texttt{BEND}, \texttt{EDGE},
\texttt{GAP}, \texttt{NCELLS}, \texttt{FIELD\_MAP}, \texttt{RFQ\_CELL},
\texttt{THIN\_STEERING}, \texttt{APERTURE}, \texttt{MARKER},
\texttt{SPACE\_CHARGE\_COMP}), the overlapping-field
\texttt{SUPERPOSE\_MAP} clusters described below, the control
directives (\texttt{FREQ},
\texttt{PARTRAN\_STEP} step densities, \texttt{LATTICE} subsections,
\texttt{FIELD\_MAP\_PATH}), the \texttt{ERROR\_*} statistical-error
directives, and the \texttt{SET\_*}/\texttt{ADJUST\_*} matching
language.  Deck features a solver cannot honor exactly
are never silently dropped: each is reported through the
parser's downgrade channel, which appends an explicit warning in
permissive mode and raises in strict mode.  The cases are graded
rather than uniform.  Dynamic
\texttt{ERROR\_*} forms and coupled error groups still take effect
with degraded semantics---reduced to static per-seed perturbations and
to independent per-element draws respectively; the
longitudinal-shift, pitch, and yaw error components are inert; and
several further directives are recognized and then apply nothing at
all.  Section~\ref{sec:limitations} grades the full set against the
manual's known-limitations list.  None of these downgraded forms bears
on the benchmark results of
Sec.~\ref{sec:validation}: the one \texttt{ERROR\_*} directive present
in those decks is inert in both codes' nominal runs, as the
configuration table of Sec.~\ref{sec:tracewin} records.  The parser runs in a permissive mode that skips unknown
cards and records them as warnings in the returned metadata, or in a
strict mode that raises on any unrecognized input.  Matching
directives are not discarded: each \texttt{SET}/\texttt{ADJUST}
matching card
becomes a first-class lattice element that carries its typed
arguments, is walked over harmlessly by the trackers, is collected
into variables and constraints by the matching engine
(Sec.~\ref{sec:matching}), and round-trips semantically within the
supported card schema through the
\texttt{.dat} writer; runtime control cards (\texttt{FREQ},
synchronous-phase and beam-reference settings) instead update the
tracking state as the solvers pass them.  HELIX interprets and writes all \texttt{.dat}
lengths in millimeters, which is also the documented convention of
the TraceWin \texttt{.dat} language itself~\cite{tracewin_manual};
within the supported card set, files therefore carry the same units in
both programs, and portability is limited by the coverage gaps
discussed in Sec.~\ref{sec:limitations} rather than by units.

The element library organizes its concrete element classes
under four base classes that define how an element acts on the beam:
\emph{transfer-map elements} with closed-form $6\times6$ matrices
(drift, hard-edge quadrupole, hard-edge solenoid, sector/rectangular
dipole); \emph{thin-kick elements} (rf gap, steerer, thin multipole,
stripping/scattering foil, ideal thin lens); \emph{field-map elements}
integrated substep-by-substep through their field models (1-D/2-D and
3-D tabulated field maps, the multi-gap \texttt{NCELLS} cavity
expanded into per-cell thin gaps, single RFQ cells, and a whole-RFQ
element); and
\emph{control, diagnostic, and externally applied elements} that
carry no through-transport map of their own but can still act on the
beam (markers; aperture masks, which act by removing particles;
space-charge-compensation tags, which rescale the collective force;
and the dipole edge,
whose linear fringe kick is applied separately).  Quadrupoles,
solenoids, dipoles, rf gaps, and field maps
additionally carry misalignment and field-error mixins that
the Monte~Carlo error-study machinery perturbs (the applied
geometric errors are the transverse offsets and the roll;
longitudinal shift, pitch, and yaw are parsed but not applied,
Sec.~\ref{sec:limitations}).

Field maps follow the TraceWin conventions.  The five-digit
\texttt{FIELD\_MAP} geometry code,
$\mathrm{geom} = \mathrm{aper}\cdot 10^4 + \mathrm{rf}_B\cdot 10^3 +
\mathrm{rf}_E\cdot 10^2 + \mathrm{stat}_B\cdot 10 + \mathrm{stat}_E$,
is decoded digit-by-digit into up to four field channels (static and
rf, electric and magnetic), each with its own 1-D, 2-D cylindrical, or
3-D Cartesian geometry; a negative code requests TraceWin's
second-order off-axis expansion, which matters only for on-axis
channels (1-D and $G(z)$ maps) where the release applies a first-order
paraxial reconstruction instead---for those a negative code is
reported through the downgrade channel (a warning in permissive mode,
an error in strict mode), whereas 2-D and 3-D channels are sampled
directly at full fidelity and a negative code over them is accepted
unchanged.  A factory then
instantiates either
the 1-D/2-D or the 3-D field-map element, and per-component readers
load the associated TraceWin electric- and magnetic-component field
files (header dimensions in meters,
converted to the internal millimeter grids), with a per-file cache so
that lattices referencing the same cavity template many times parse
the file once.  During tracking, 1-D maps are sampled by linear
interpolation of the on-axis profile combined with the first-order
off-axis expansion, 2-D cylindrical maps by bilinear
interpolation in $(r, z)$, and 3-D Cartesian maps by trilinear
interpolation by default (an opt-in tricubic mode is provided for
sharp-gradient solenoid maps).  For radio-frequency quadrupoles, HELIX provides both a
per-cell element using the two-term Crandall potential and a
whole-RFQ element constructed by replacing a parsed
\texttt{RFQ\_CELL} sequence with the help of a \texttt{.vane}
geometry file;
the latter exposes five vane-field models (\texttt{2term},
\texttt{8term}, \texttt{8term\_full}, \texttt{laplace2d},
\texttt{laplace3d}), of which the default two-term model is the
production path and the higher-order models are retained as documented
research and diagnostic options.  Both RFQ elements are integrated
substep-by-substep inside the multiparticle tracker, so the 3-D FFT PIC
space-charge solve of Sec.~\ref{sec:spacecharge} acts through the RFQ on
the same cadence as the rest of the line.  The established RFQ codes are
closed, export-controlled, or separately licensed, and the openly
available implementations descend from PARMTEQ and compute space charge
on an $r$--$z$ mesh; to our knowledge no other openly licensed code
tracks an RFQ inside a self-consistent 3-D space-charge solve.  TraceWin's \texttt{SUPERPOSE\_MAP}
cards, which place several field maps at explicit longitudinal offsets
so their fields overlap, are parsed into a single container element:
it spans from the cluster entrance to $\max_i(z_{0,i}+L_i)$ (a negative
member offset is TraceWin's convention for entering partway into a
map~\cite{tracewin_manual}, so the portion ahead of the entrance lies
outside the tracked span by design; a regression test pins this
behavior against the equivalent explicitly shifted map) and
integrates once through that span while vector-summing the field of
every member covering each longitudinal position, so overlapping
solenoid and cavity fields add rather than acting in sequence.  All
radio-frequency members of one cluster must share a single frequency.
The container round-trips back to its \texttt{SUPERPOSE\_MAP} cards on
write.

A separate importer reads a practical subset of MAD-X: variable
assignments with arithmetic expressions, element definitions,
\texttt{SEQUENCE}/\texttt{ENDSEQUENCE} blocks with automatic drift
fill between placed elements, and the \texttt{BEAM} and \texttt{USE}
commands.  Elements are mapped onto the HELIX classes (e.g.,
\texttt{SBEND}/\texttt{RBEND} become edge--dipole--edge triplets;
quadrupole $k_1$ and solenoid $k_s$ are converted to field quantities
via the beam rigidity of the \texttt{BEAM} reference energy, which
is unambiguous for the fixed-rigidity transport lines the importer
targets; a sequence that accelerates would need per-element
rigidities the importer does not infer), lengths are converted from meters to
millimeters, and unsupported constructs are reported as warnings
rather than silent omissions.

\subsection{Solver modes and dispatch}
\label{sec:solvers}

A \texttt{Simulation} facade wires the model to the solvers: given
$(\texttt{Lattice}, \texttt{Beam}, \texttt{SpaceChargeConfig})$,
\texttt{run()} executes the multiparticle tracker and
\texttt{run\_envelope()} the envelope solver, while the linear-matrix
path is a stateless function of the lattice and reference particle.
The batch CLI exposes the same choice as
\texttt{run -{}-mode envelope|mp|matrix}, and the GUI as envelope and
multiparticle run actions plus a transfer-matrix viewer, all sharing
the same core implementations.

The \emph{envelope} solver propagates the $6\times6$ beam
$\Sigma$-matrix element by element, applying space charge as thin
linear defocusing kicks derived from the rms-equivalent uniform
ellipsoid model of Sec.~\ref{sec:physics}; for unbunched (continuous)
beams the same $\Sigma$-matrix solver switches its space-charge kick
to the two-dimensional continuous-beam model of
Sec.~\ref{sec:physics}, the configuration behind the LEBT
benchmark of Sec.~\ref{sec:validation}, while the standalone
Sacherer-type envelope ODE integrator of Sec.~\ref{sec:envelope} is a
separately selectable solver kind (at fixed kinematics and
over drift, quadrupole, and solenoid transport only, per its
constant-$\beta,\gamma$ scope), exercised against the closed-form
benchmarks of Sec.~\ref{sec:validation} rather than on the
TraceWin-comparison paths.  Tabulated field-map elements enter
the envelope solver through numerically linearized maps: twelve probe
particles are tracked
through the map and assembled into a central-difference Jacobian (the
RFQ-cell, whole-RFQ, and \texttt{NCELLS} classes are the
exception---they compose their matrices analytically from the same
drift/kick sequence their tracker uses, with no probe particles).  The
1-D/2-D class, the superposed-cluster class, and the 3-D class when it
is sliced for
space-charge transport, use displacements of $\pm 0.01$ in each of
the first five internal
coordinates [mm, mrad, deg] and $\pm 1$~keV in energy; the 3-D
class's full-element matrix uses its own tighter per-coordinate
steps ($\pm 0.001$ transversely, $\pm 0.01$ in phase, $\pm 0.1$~keV
in energy).
The envelope solver returns rms sizes,
transverse projected Twiss functions, and emittances versus $s$ in the
diagnostics recorder, and an aperture-profile helper assembles the
piecewise beam-pipe half-widths along $s$ (including per-slice
\texttt{.ouv} profiles) for envelope--aperture overlay plots.
Because the envelope solver is deterministic and free of sampling
noise, it
is the default engine for matching and design scans.

The \emph{multiparticle} tracker propagates the macroparticle ensemble
through each element's \texttt{track} method: thin elements apply
kicks, thick linear elements apply their transfer matrices, and
field-map elements integrate substep by substep (a first-order
kick--drift composition with fields sampled at the slice midpoint
for 1-D/2-D maps; kick--drift or Strang-like
drift--kick--drift for 3-D maps,
Sec.~\ref{sec:integrators}), with 3-D PIC space-charge kicks
interleaved at a configurable cadence, in the lineage of multiparticle
space-charge codes such as IMPACT-Z \cite{impactz}.  It returns the
same envelope-level diagnostics plus optional full phase-space
snapshots (recorded where snapshot locations, cadence, or markers
are configured),
per-particle loss records, transverse kurtosis-based halo
parameters, and transmission: the
quantities envelope
tracking cannot provide.

The \emph{matrix} mode composes the
$6\times6$ linear one-turn/one-pass map; its default public
operation returns decoupled per-plane periodic
Twiss parameters and phase advances, interpreting the composed map
as a stable repeating cell, and raises on transverse coupling.
Coupled optics are handled instead by a separate \texttt{twiss}
helper, whose eigenvector-based branch
resolves transversely coupled cells and which returns matched
solutions on
either the whole lattice (periodic systems) or a single focusing
cell, back-propagated to the entrance as the
input match of a transfer line, including the periodic dispersion of
bending cells, which the beam configuration's dispersion terms can
imprint on generated bunches.  A further fixed-point helper adds
space charge and dispersion to the matched-input search by forward
shooting, iterating the input Twiss to a self-consistent fixed point;
it is not exercised by the benchmarks of this paper.  The
multiparticle tracker and the zero-current envelope solver also run
\emph{backward}: a backtracking engine transports a
distribution or envelope from a downstream plane to the entrance
through inverses of the element maps.  For particle distributions
the inverses are exact algebraic inverses of
the matrix elements and exact reverse integration of the supported
field-map integrators, with documented fallbacks and exclusions.
\texttt{NCELLS}, RFQ cells, and surrogate maps fall back to
linearized inverses, and foil kicks are refused; a forward CSR kick has
no backward model, so backtracking through one is refused unless the
caller explicitly opts into an approximate reconstruction, in which
case the CSR kick is skipped.
3-D PIC kicks from the standard solver are undone exactly for
adaptive forward grids and,
for fixed grids, when the forward solver's frozen grid is reused,
approximately otherwise, while the analytic dc kernels are stateless
and exact.  (An experimental machine-learned coarse-grid
space-charge backend, used in no result of this paper, is excluded:
it has no reverse replay, and rebuilding its kicks with the standard
solver during reversal requires the caller's explicit approximate-mode
authorization, backtracking raising otherwise.)
Zero-current envelope backtracking instead inverts each
element's linear matrix (analytic transfer and kick matrices where
defined, fitted matrices for field-map elements), so additive
effects are not reconstructed: neither the foil's covariance
diffusion nor its envelope-mode reference-energy loss is inverted.  Thus a
measured or simulated exit distribution (a
\texttt{.dst} file) or a desired exit state can be pulled back to the
injection plane, with a validation mode that checks forward--backward
closure; aperture losses are inherently non-invertible, so
reconstruction applies to the surviving beam, and with space charge
that reconstruction is only approximate for a beam that lost
particles, since the survivors were forward-tracked in the collective
field of particles later scraped away and that field is not
reconstructed from the survivors alone.  The qualitative cost ordering spans
orders of magnitude---matrix composition is a chain of matrix
products once each element's linearized matrix is in hand, though on a
field-map deck building those matrices dominates it (the twelve-probe
linearization costs about \SI{0.3}{\second} per tabulated map, some
\SI{6}{\second} for the twenty maps of the MEBT+HWR line),
envelope runs complete in seconds on section-scale lattices and in
minutes on the full 256-m linac (seconds-scale \emph{repeated}
evaluation belongs to the opt-in transfer-matrix cache,
Sec.~\ref{sec:costhierarchy}), and multiparticle runs with space
charge take from minutes to tens of minutes---which is what
makes the hybrid workflow effective: iterate a design at envelope
speed, then confirm with multiparticle physics.  Measured timings are
reported in Sec.~\ref{sec:performance}.

\subsection{Acceleration layers}
\label{sec:acceleration}

The PIC hot loop is accelerated at two levels, and re-implemented at a
third for a different purpose.  First, the
cloud-in-cell (CIC) charge-deposit and field-gather kernels are
implemented in C++ and bound with pybind11~\cite{pybind11}, compiled with OpenMP when
available.  The parallel deposit avoids data races by accumulating
into per-thread private buffers that are reduced in a fixed
thread-index order, and multithreaded results are bitwise
reproducible
run to run for a fixed build and thread count---a property pinned by
repeated-run regression gates and made contractual by an explicit
\texttt{schedule(static)} clause on the parallel deposit loop, so the
mapping of particles to threads no longer rests on the OpenMP
runtime's default work assignment.  If the extension cannot be built or imported, HELIX falls
back transparently to a pure NumPy implementation of the same kernels;
the higher-order triangular-shaped-cloud (TSC) deposit/gather pair
currently exists only in the Python implementation.

Second, the FFTs of the Hockney--Eastwood Poisson solve can execute on
a GPU through a backend-dispatch layer.  Two backends are supported: a
CuPy/CUDA backend running in FP64, which reproduces the CPU reference
to better than $10^{-9}$ relative (quantified in
Sec.~\ref{sec:validation}), and a PyTorch \cite{pytorch} Metal (MPS)
backend for
Apple-silicon GPUs, which runs in FP32 because the Metal hardware does
not support FP64 (in the configurations studied here the FP32
backend differences stay well below the finite-macroparticle
variation of the reported rms moments; the backend is documented and
excluded from the reported cross-code baselines, but is covered by its
own CPU-parity regression tests).  Backend resolution is explicit: the
space-charge configuration accepts
\texttt{auto}/\texttt{cpu}/\texttt{gpu}/\texttt{cuda}/\texttt{mps},
with \texttt{auto} preferring CUDA and otherwise the multithreaded
\texttt{scipy.fft} CPU path (the FP32 MPS backend is never
auto-selected and must be requested explicitly), and the environment variable
\texttt{LINAC\_GEN\_USE\_GPU} overrides the configured value for
headless deployment.  Whether the GPU pays depends on grid size and
host--device transfer cost; measurements are given in
Sec.~\ref{sec:performance}.  Third, a separate PyTorch implementation
of the full PIC cycle provides end-to-end differentiability rather
than raw throughput; it is the subject of Sec.~\ref{sec:diffpic}.

\subsection{I/O and interoperability}
\label{sec:io}

HELIX reads and writes the TraceWin binary \texttt{.dst} particle
format, whose on-disk records store $(x,x',y,y',\phi,W)$ in cm, rad,
and absolute MeV; the loader converts to the internal
$(\mathrm{mm},\mathrm{mrad},\deg,\mathrm{MeV})$ deviation coordinates
using the bunch centroid as the reference (a \texttt{.dst} record
carries no separate synchronous particle, so an imported centroid
offset is absorbed into the reference rather than retained as a beam
centroid) and exposes the raw centroid
so callers can restore absolute values, and the writer inverts the
transformation given the stored centroid (an API-level round trip;
the standard CLI export re-references the phase), enabling
distribution interchange with TraceWin-based
workflows.  On the results side, two TraceWin-compatible ASCII writers
emit the per-element ``partran''-style output table and the 26-column
envelope export format, reproducing TraceWin's three-digit-exponent
number formatting so that files can be compared column by column
against reference outputs.  Native results are written to HDF5 with
envelope arrays, reference-particle evolution, optional full
phase-space snapshots, and the beam configuration stored as
attributes.  A provenance group records the code commit and package
version; the lattice file's SHA-256 hash and the referenced field-map
file hashes (\texttt{SUPERPOSE} cluster members included); the input
beam's SHA-256 hash; the random seed; the space-charge configuration
together with the integration and space-charge cadences (steps per
metre); and the parser's downgrade ledger.  It also records an
identity manifest for any engaged surrogate, the effective compute
backend, the tracker floating-point precision, whether the compiled
C++/OpenMP kernels or the NumPy fallback ran, and the OpenMP thread
count, schedule, and dynamic-teams setting.  The file, not just the
project configuration, therefore records the principal inputs and
settings of the run performed.

Coverage reaches past the field-map data itself:
because the hash covers every file sharing an element's field-file
prefix, the derived aperture (\texttt{.ouv}) and
space-charge-compensation (\texttt{.scc}) side files are captured with
it, and the \texttt{APERTURE} card takes numeric operands only, so no
aperture input reachable from the deck is left unhashed.  The weights
checksum is likewise recomputed from the \texttt{weights.pt} bytes on
every load-from-disk route: command line, workbench, and Python API
alike.  This guarantee has one gap: a surrogate handed to the registry
by a training call in the same process, with neither a reload nor a
recorded weights directory, contributes an identity-only manifest line
and no weights checksum.
A parallel writer emits openPMD-layout HDF5 files
following the 1.1 record structure~\cite{openpmd}, with float64
particle records in SI units whenever phase-space snapshots are
enabled (the default CLI multiparticle run records none), plus a
HELIX-specific envelope group that standard readers simply ignore
and that HELIX's own loader reads back; full semantic conformance
(the time-record and algorithm-metadata attributes) is release
work, particle import remains the
\texttt{.dst} route, and neither this writer nor the TraceWin-format
writers above carries the provenance group.
Together with the \texttt{.dat} parser/writer, the field-map readers,
and the MAD-X importer of Sec.~\ref{sec:lattice}, this covers the
interchange surface; project files (\texttt{.lgproj}) reference the
lattice file and persist beam,
convergence, and numerics settings shared between the GUI and
the CLI, and persisted lattice, distribution, and field-map paths
are written in portable relative form where a usable common root
exists, so projects and exports typically relocate cleanly between
machines.

\subsection{User interfaces}
\label{sec:interfaces}

The graphical workbench is a PyQt6 application with eight tabs (Beam,
Lattice, Matching, Numerics, Surrogates, Error Study, Failure Study,
and Results) covering beam and distribution setup, lattice editing
and \texttt{.dat} import, interactive \texttt{SET}/\texttt{ADJUST}
matching, space-charge grid/kernel/backend selection with convergence
scans, surrogate training and deployment, Monte~Carlo tolerance and
element-failure studies, and a post-run diagnostics dashboard, with
envelope and multiparticle runs launched from the toolbar and a
backtrack-distribution dialog for the backward transport above.  The
built-in convergence scans sweep any of the five numerical
knobs (grid size, grid extent, either step density, or macroparticle
count) over user-chosen values on the loaded lattice, plot and
tabulate the moment response per point, and write a recommended
setting
back into the run configuration (singly, or for all four PIC/step
axes in sequence).  That recommendation is a heuristic, not a proof
of convergence: it is the smallest setting
whose exit $\varepsilon_x$ lies within 1\% of the finest scanned
point's, tested on a single observable against a single reference
level, and the user can override it; establishing convergence in the
stronger sense---across observables, over a physical $s$-range, and
against the two finest levels rather than one---remains the analyst's
responsibility.  Scans of exactly this kind, run at production
scale, underpin the convergence analysis of
Sec.~\ref{sec:validation}, and a project file persists the resulting
numerics so the chosen configuration travels with its lattice.

The same engines run headless through \texttt{python -m linac\_gen}
with the subcommands \texttt{run} (one simulation in envelope,
multiparticle, or matrix mode), \texttt{scan} (parameter sweeps to CSV
via repeated \texttt{-{}-vary} ranges), \texttt{batch} (multi-run
campaigns from a JSON job file), \texttt{twiss} (whole-lattice or
cell-based matched Twiss), \texttt{mo} (multi-objective Pareto
optimization over \texttt{ADJUST} knobs), \texttt{failures} (element
failure impact and recovery), \texttt{backtrack} (backward transport
from a downstream \texttt{.dst} or design target, with the
forward--backward closure validator), and \texttt{match} (the matching
engine).  Inputs are bare lattice files or \texttt{.lgproj}
projects (\texttt{batch} takes its JSON job file),
with command-line overrides for beam scalars, individual
element parameters, and space-charge settings resolved in the order
command line, project, default.  Because scan points are independent
simulations, \texttt{scan -{}-parallel~$N$} distributes them over a
process pool; each worker defaults its internal FFT threading to one
thread (existing environment settings take precedence) so that
parallelism comes from the pool rather than
oversubscribed threading, and results are collected asynchronously as
points complete.

A third front end drives the same run configuration through
natural language.  It is outside the scope of this paper and was not
used for any result reported here.

\subsection{Differentiable PIC and gradient-based matching}
\label{sec:diffpic}

The engines described so far compute beam observables; this one also
computes their derivatives.  HELIX re-implements its multiparticle
forward map (element transport \emph{and} the self-consistent
space-charge solve) as a single PyTorch~\cite{pytorch} computation
graph, so that sensitivities of any tracked quantity with
respect to lattice parameters follow from reverse-mode automatic
differentiation rather than finite differences.  These Jacobians are
exact to floating-point precision for a fixed computational branch of
the discretized map: the integer cell assignments of the
deposit/gather stencils are detached from the graph (gradients flow
only through the continuous shape-function weights), so the
differentiated map is piecewise smooth, nonsmooth only at discrete
stencil and clamp branch changes, principally
cell-boundary crossings.  Differentiable
self-consistent space charge is established prior art: Qiang
demonstrated gradient-based design optimization through a
differentiable space-charge model~\cite{qiang2023differentiable},
Cheetah~\cite{cheetah} has been extended with a reverse-mode
integrated-Green-function space-charge kick~\cite{dhamrait2026}, and
ImpactX differentiates its envelope model with space charge at the
compiler level~\cite{huebl2025impactx}.  HELIX's contribution is not
the idea but its integration: the differentiable path sits inside the
production matching engine of Sec.~\ref{sec:matching} as one of its
selectable algorithms, is driven by the same
\texttt{SET}/\texttt{ADJUST} lattice cards as every other optimizer,
and is held to numerical parity with the production solvers it
mirrors.

\emph{Differentiable PIC.}  The torch space-charge kick mirrors the
production cycle of Sec.~\ref{sec:spacecharge} stage by
stage (rest-frame boost, grid fit, CIC/TSC deposit, doubled-grid
Hockney convolution with the IGF kernel, central-difference field
gradient, gather, and momentum kick), with every stage re-expressed in
autograd-capable torch operations.  Each stage has its own
differentiability requirement.  Deposition becomes an autograd-safe
scatter-add in which the integer cell assignment is detached from the
graph and the gradient flows through the continuous trilinear (CIC) or
quadratic (TSC) shape-function weights, the standard construction for
a differentiable PIC deposit (the quantitative derivative
verification of Sec.~\ref{sec:validation} exercises the CIC path);
the gather reuses the identical
stencil, so the deposit--gather pair is a volume-weighted adjoint
pair by construction,
which suppresses spurious self-forces up to the field-gradient
discretization (the cancellation is not exact on the discrete
gradient stencil, particularly near the grid faces where the
differences are one-sided).  The Poisson stage differentiates through torch's complex
FFTs, and $\mathbf{E} = -\nabla\phi$ is assembled from out-of-place
slice differences that reproduce the numpy solver's boundary treatment
exactly.  Because the torch kick refits its grid to the instantaneous
distribution at every kick, the IGF and point Green's-function
builders are themselves differentiable with respect to the cell
spacing, so gradients propagate through the grid geometry rather than
treating the mesh as frozen.  The implementation runs in
\texttt{torch.float64} on the CPU only (FP64 is required for parity
with the production path and for clean gradients), and its parity is
pinned by the test suite and quantified in
Sec.~\ref{sec:validation}: deposit and gather match the numpy kernels
at the round-off floor and the full kick cycle to $\sim 10^{-13}$
relative on the
kick increment.  A
thin wrapper lets the ordinary numpy tracker call the torch kick
(\texttt{sc\_backend="torch"}), with its
result detached from autograd; the wrapper rebuilds its grid
adaptively at every kick and does not reproduce fixed-grid
production configurations, so it serves parity testing rather than
as a general substitute, and end-to-end differentiability requires
the tracker
below.

\emph{Differentiable tracking.}  Gradients reach lattice parameters
only if transport lives in the same graph.  For lattices without space
charge the torch path precomposes the per-element $6\times6$ maps into
one lattice matrix; that shortcut is incompatible with space charge,
whose kick depends on the evolving distribution.  A stepwise tracker
therefore advances the $(N,6)$ beam tensor element by element through
the torch transfer maps of the five linear element types (drift,
quadrupole, solenoid, dipole, edge) and interleaves the torch
space-charge kick on the production tracker's bundling cadence
(Sec.~\ref{sec:integrators}); because the linear submaps are exact
and the space-charge kick is impulsive, this composition is a
genuine second-order Strang split here: drifts split into
transport--kick--transport bundles on the configured step grid,
quadrupoles and solenoids into two sub-steps with midpoint
kicks (exact, since their maps satisfy $M(L) = M(L/n)^n$), and
dipoles into a symmetric kick--map--kick split (matching the
production placement for drifts, quadrupoles, and solenoids; the
dipole split is a distinct symmetric composition).  Tunable parameters (quadrupole
gradient, solenoid field, dipole angle) enter as override tensors
keyed to their elements, so lattice objects are never mutated.  With
space charge the tracked beam reproduces the production numpy tracker
to machine precision on a bend-free FODO benchmark: every coordinate
column of the final $(N,6)$ particle tensor agrees at the
$10^{-14}$ level, each column normalized by its own peak
magnitude so that no unit convention enters the metric (adaptive
$32^3$ grid, CPU FP64).  The accumulated space-charge increment
itself, which reaches $\sim 10$~mrad in $x'$ and $20^\circ$ in phase
over the one-metre cell, agrees at the same level under
the same per-column normalization against a zero-current reference
run.  The level, not the digit, is the claim: repeated runs of this
fixed-seed benchmark scatter over
$1.6$--$1.7\times10^{-14}$, the difference between the two
implementations being itself at the scale of their run-to-run
variability.  This end-to-end parity is consistent with the per-kernel checks of
Sec.~\ref{sec:validation}: the space-charge kick reproduces its numpy
counterpart at the $10^{-13}$ level and the deposit/gather kernels
theirs at the round-off floor, and the drift and quadrupole transfer
maps, computed
independently in the two codes rather than shared, agree at the same
round-off level, so the tracked chain accumulates no error beyond floating-point
round-off.  Reverse-mode differentiation of
such a chain ordinarily stores every intermediate grid, so autograd
memory grows linearly with the number of space-charge kicks, the
scaling confronted also by the differentiable Cheetah
extension~\cite{dhamrait2026}.  HELIX bounds it with per-element
gradient checkpointing: the forward pass stores only each element's
small $(N,6)$ input and the backward pass recomputes the space-charge
grids on the fly, removing the dominant per-kick grid storage at
roughly twice the forward cost.  Peak
grid memory drops from all kicks' grids to a
single element's working set, while checkpoint inputs and the
surrounding graph are still retained.  The
recomputation is exact because the kick contains no internal
randomness.

\begin{table}[b]
\caption{\label{tab:gradscope}%
Documented scope of the differentiable (gradient) matching path.
Every exclusion listed here is enforced mechanically: a pre-run
check raises before any optimization begins, naming the offending
knob, element, or card.  Implemented and enforced scope is broader
than what is quantitatively demonstrated in this paper: the
derivative benchmarks of
Sec.~\ref{sec:validation} and the end-to-end parity check above tune
quadrupole gradients on a bend-free
lattice, so solenoid, dipole, and edge support and the TSC deposit
are exercised only by unit and smoke tests, while the CIC deposit
carries the full Taylor-remainder verification.}
\begin{ruledtabular}
\begin{tabular}{lp{5.9cm}}
Knobs & quadrupole gradient, solenoid field, dipole angle; rf
        phase/amplitude and energy knobs excluded \\
Elements & drift, quadrupole, solenoid, dipole, edge; passive
        matching and marker cards;
        trained surrogate cavities at frozen reference kinematics
        (zero-net-gain elements).
        Rf gaps, native field-map cavities, and control cards that
        change the reference kinematics mid-lattice refused \\
Constraints & transverse \texttt{SET\_TWISS}/\texttt{SET\_SIZE} exit
        equalities;
        centroid (\texttt{SET\_POSITION}, \texttt{DIAG\_POSITION}),
        longitudinal \texttt{SET\_TWISS} flags, one-sided size,
        phase-advance, energy, emittance, and
        transmission cards refused \\
Physics & fixed synchronous energy---no matching \emph{through}
        acceleration; 3-D PIC space charge (CIC quantitatively
        verified, TSC smoke-tested) \\
Benchmark defaults & $32^3$ grid, adaptive $\pm4\sigma$ extent, CPU
        FP64,
        1500-particle bunch, fixed seed (grid, extent and particle
        count all overridable through the multiparticle configuration;
        bit-reproducible on a fixed hardware/software
        configuration) \\
\end{tabular}
\end{ruledtabular}
\end{table}

\emph{Gradient matching.}  The matcher consumes this machinery as
\texttt{match(algorithm="gradient")}, optionally with
\texttt{space\_charge=True}.  Table~\ref{tab:gradscope} states the
documented scope up front.  A scope check runs before any
optimization and rejects knobs and elements
outside the differentiable path's documented scope, raising an error
that names the offending card or element and directs the user to
\texttt{least\_squares}.  Variables must tune
a quadrupole gradient, solenoid field, or dipole angle, and every
element
must be one of the five linear types, a passive command card,
or an autograd-capable surrogate field map evaluated at the matcher's
fixed reference kinematics; rf gaps and native field-map
cavities are refused.  The torch path does not advance the
synchronous energy, so matching \emph{through} acceleration is outside
the present scope, and a surrogate is appropriate only for
zero-net-energy-gain elements such as the demonstrated bunchers: a
surrogate of an accelerating cavity would hand downstream maps stale
reference kinematics.  A surrogate whose input leaves its trained
scope during optimization raises a hard error rather than silently
substituting an identity map (Sec.~\ref{sec:surrogates}).  The
same check rejects every excluded constraint card of
Table~\ref{tab:gradscope} by name: one-sided size, phase-advance,
energy, emittance, and transmission cards, the centroid
(\texttt{SET\_POSITION}, \texttt{DIAG\_POSITION}) cards the
$\Sigma$-only torch mirror cannot reproduce, and
\texttt{SET\_TWISS} cards carrying longitudinal flags, which the
torch mirror does not implement (the general matcher evaluates them
from the recorded longitudinal Twiss).  Because the torch mirror composes every
command card as an identity, cards that mutate the reference
kinematics mid-lattice (\texttt{FREQ} frequency changes,
\texttt{SET\_BEAM\_ENERGY}, \texttt{SET\_BEAM\_E0\_P0}) are refused
as well; the ubiquitous header \texttt{FREQ} whose value equals the
incoming frequency is exempt, since its jump ratio is unity and the
identity composition is then exact.  The design rule throughout is
that anything the differentiable path cannot represent stops the
run rather than quietly altering the optimized problem.  Within that scope, the no-space-charge residual
propagates the input $\Sigma$ through the precomposed torch matrix,
and the engine verifies at the starting point that it reproduces the
numpy matcher's residual vector to $10^{-6}$ relative, refusing to run
otherwise: a starting-point consistency check that the gradient
algorithm is posed on the same residual the other algorithms see,
rather than a proof of equality across the whole bounded domain.  With space charge, the residual
tracks a macroparticle bunch (1500 by default) sampled once from a
whitened Gaussian colored so that its sample covariance reproduces
the input $\Sigma$ to the accuracy of the $10^{-12}$ diagonal
regularization used in the whitening and coloring factors, with a
fixed seed that
makes residual and
Jacobian deterministic and repeated matches bit-reproducible; tracking uses the
checkpointed stepwise tracker with an internal PIC configuration
defaulting to a $32^3$ grid with adaptive $\pm4\sigma$ extent on the
CPU, and the
exit sample covariance feeds the same constraint evaluators.  Because
this forward model is the nonlinear PIC rather than the envelope
space-charge model behind the other algorithms, the engine reports the
final residuals from the PIC model itself: with space charge the
gradient algorithm optimizes its own, fully specified PIC objective,
and independent-seed or higher-fidelity revalidation of the matched
point (as performed in Sec.~\ref{sec:app_gradmatch}) is the
recommended closing step.  The outer loop is the same
bound-constrained trust-region least-squares solver as the default
algorithm, with the Jacobian supplied by
\texttt{torch.autograd.functional.jacobian} (exact to floating-point
precision for the computational branch fixed by the forward
pass) using vectorized
vector--Jacobian products: all rows are assembled from one
graph-recording forward pass, so the Jacobian overhead on top of
that pass scales with the
number of residual components rather than with the number of
knobs, the reverse-mode economics that favor many-knob matching.
Section~\ref{sec:app_gradmatch} demonstrates this machinery recovering
a known matched solution through nonlinear space charge to far below
its imposed engineering tolerances, and
Sec.~\ref{sec:diffcost} quantifies what the differentiability
costs---and where the exact Jacobian repays it.

\subsection{Matching and optimization suite}
\label{sec:matching}

HELIX includes a lattice-driven matching engine modeled on the
TraceWin \texttt{SET}/\texttt{ADJUST} language~\cite{tracewin}:
optimization variables and constraints are declared as cards inside the
lattice file itself, so a matching problem travels with the lattice it
describes.  Each \texttt{ADJUST} card promotes one element attribute (a
quadrupole gradient, solenoid field, cavity phase or field amplitude,
drift length, dipole angle, or steerer kick) to an optimizer degree of
freedom with optional bounds, and the \texttt{ADJUST\_BEAM\_*} variants
expose the input beam's Twiss parameters, emittances, centroid, and
current.  Variables sharing a nonzero link group collapse into a single
optimizer column and move in lockstep, modeling ganged power supplies.
The well-posed form of a mixed-quantity least-squares match is a
tolerance-normalized residual,
\begin{equation}
  r_i = w_i\,\frac{q_i-q_i^{\star}}{\tau_i},
  \qquad
  C = \tfrac{1}{2}\sum_i r_i^{2},
  \label{eq:matchcost}
\end{equation}
where $\tau_i$ carries the same physical units as the target $q_i$ and
represents an engineering tolerance, so each $r_i$ is dimensionless
and $C$ is invariant under a change of unit, while $w_i$ is a purely
dimensionless priority factor; this is the convention the gradient
benchmark of Sec.~\ref{sec:app_gradmatch} uses.  The general matcher
retains the TraceWin form as a legacy compatibility convention: each
\texttt{SET\_*}/\texttt{MIN\_*}
card carries a single weight $k$ applied to its residual in native
units (mm, mrad, deg, MeV); that is, $k$ plays the role of
$1/\tau_i$ in a fixed native-unit system.  The user must therefore
set $k$ to encode each target's inverse tolerance for the scalar cost
to be meaningful, and the numerical problem is not invariant under a
change of unit unless the weights are transformed to match.
\texttt{SET\_TWISS} is a further subtlety: its $k$ operands act as
per-parameter on/off axis
selectors rather than weights, and its residual concatenates the
dimensionless $\alpha$ with the dimensional $\beta$, so a single such
card already mixes differently scaled quantities under the native-unit
convention.  The engine concatenates
all
residuals into $\mathbf{r}(\mathbf{x})$ and minimizes
$C = \tfrac{1}{2}\lVert\mathbf{r}\rVert^{2}$.  A single call collects the
variables and constraints, dispatches the selected optimizer, and writes
the matched values back into the lattice and beam configuration in
place; a baseline evaluation at the starting point is always recorded so
the improvement over the unmatched lattice is quantified.

Several constraint families are implemented.
\texttt{SET\_TWISS} imposes equality on
selected Twiss parameters at the lattice exit, chosen by
per-parameter flags: the transverse $\alpha_{x,y}$, $\beta_{x,y}$,
and the longitudinal pair evaluated from the $\alpha_z$/$\beta_z$
record both tracking modes carry.  A baseline whose longitudinal
record is absent or degenerate (a dc exit, or a vanishing
$\varepsilon_z$) is refused by a companion gate on that baseline
evaluation unless \texttt{allow\_inert\_constraints} is set; a trial point that
turns degenerate mid-match fills its longitudinal slots with a fixed
penalty of $10^{3}$ in the card's native residual units, so the
residual dimension never changes.
\texttt{SET\_SIZE} does the same for the rms sizes
$\sigma_x$, $\sigma_y$ and a longitudinal size, evaluated like
\texttt{SET\_TWISS} at the lattice exit.  The sign of the TraceWin
longitudinal operand selects the target: a positive value is the rms
phase $\sigma_\phi$ in degrees, a negative one the rms bunch length
$\sigma_z$ in mm, converted through the local $\beta\lambda$.  The
separate transverse centroid-inclusion flag is not modeled and is
rejected by the pre-run constraint audit unless the
\texttt{allow\_inert\_constraints} override is set, and sizes are always
evaluated about the centroid.
\texttt{SET\_SIZE\_MAX} and
\texttt{SET\_SIZE\_MIN} set one-sided bounds on the worst-case $\sigma$
over the span of the $N_e$ elements following the card, resolved to the
recorder rows bracketing that element span so the physical window is
fixed by the lattice and does not move with the recording cadence; the
sampled extremum itself remains subject to ordinary recorder-grid
convergence.
\texttt{SET\_BEAM\_PHASE\_ADV} targets the per-plane phase advances $\mu_x$,
$\mu_y$, $\mu_z$ over the span of elements following the card,
obtained by integrating $1/\beta(s)$, a linear-optics measure whose
identification with the true betatron phase advance holds for the
uncoupled, non-accelerating transport it is intended for.  Its
longitudinal advance
substitutes an effective $(z,z')$ beta
$\tilde\beta_z=\gamma^2(k_W/k_\phi)\,\sigma_{\phi\phi}/\varepsilon_{\phi W}$
in mm/mrad for the native deg/MeV one, with $k_\phi$, $k_W$ the
per-record deg/mm and MeV/mrad conversion factors and $\gamma^2$ the
slip-coordinate drift coefficient: a HELIX-defined effective metric
rather than a canonical invariant.
\texttt{SET\_POSITION} is a four-component
centroid equality at the card's location, evaluated from the
propagated centroid under both cost solvers.  The related
per-monitor \texttt{DIAG\_POSITION} residuals are generated from
targeted beam-position-monitor markers rather than from a
\texttt{SET} card, and are active when the marker's diagnostic family is
tied to an \texttt{ADJUST} family or the marker is ungrouped:
per-plane centroid-minus-target differences weighted by the
monitor's stated accuracy, evaluated from the same propagated
centroid under both cost solvers.
\texttt{SET\_KE\_OUT\_MIN} is a one-sided floor
on the exit kinetic energy, which prevents the optimizer from buying
emittance reduction by detuning cavities off crest.  The remaining
three are HELIX-specific one-sided emittance and loss constraints.
\texttt{MIN\_EMIT\_GROWTH} penalizes per-plane emittance growth:
normalized transversely and native deg$\cdot$MeV longitudinally.
\texttt{MIN\_EMIT\_4D\_GROWTH} penalizes growth of the normalized
coupled four-dimensional transverse invariant
$\varepsilon_{4\mathrm{D},n}=(\beta\gamma)^2\sqrt{\det\Sigma_{4\times4}}$,
the $(\beta\gamma)^2$-scaled counterpart of the geometric
$\varepsilon_{4D}$ of Sec.~\ref{sec:envelope}.  It also penalizes a
$\beta\gamma$-scaled longitudinal-emittance metric, a HELIX-defined
heuristic scaling used as an optimization objective and not claimed
as a physical invariant.  By contrast,
\texttt{MIN\_EMIT\_GROWTH} compares the native deg$\cdot$MeV
longitudinal emittance between entrance and exit.  It clock-anchors
the entrance emittance to the exit rf frequency (rescaling by
$f_\mathrm{out}/f_\mathrm{in}$; Sec.~\ref{sec:physics}), so the metric
is frequency-invariant across an rf-frequency jump: a beam preserved
through a frequency change reads near-zero growth, while genuine
growth is still detected, matching the normalized transverse planes.
Each has a multiplicative tolerance, so $x$--$y$ emittance exchange
through solenoid channels is not penalized.  Finally,
\texttt{MIN\_TRANSMISSION} imposes a transmission floor that closes
the loss-gaming loophole of alive-particle emittance estimators.  A few further TraceWin cards
(\texttt{SET\_ACHROMAT}, \ldots) are parsed for
file round-trip but are not yet active residuals; a constraint audit
in \texttt{match()} refuses to run whenever such a stub card is
present, so a stub condition is not silently
ignored (an explicit
\texttt{allow\_inert\_constraints} override, exposed on the API, the
CLI, and the GUI alike, permits the inert zero-residual
cards for legacy decks).

The residuals are computed from a forward simulation whose fidelity is
selectable through the \texttt{cost\_solver} option.  The default
envelope mode runs the rms envelope solver of Sec.~\ref{sec:envelope}
once per evaluation (space charge optional) and is fast enough for
interactive matching; the multiparticle mode runs the full tracker at a
fixed random seed, a deterministic but single-realization
objective for which independent-seed revalidation of the matched
point is the recommended closing step.  It is substantially slower per
evaluation whenever space charge is active, strongly lattice- and
settings-dependent and set mainly by macroparticle count: on the
MEBT+HWR line the measured cost is $1.9\times$ the envelope pass at
the matcher's default $10^{3}$ macroparticles and $8.5\times$ at
$10^{4}$ (Sec.~\ref{sec:applications}).  It
captures nonlinear space charge, halo formation, and actual
transmission, and is required for \texttt{MIN\_TRANSMISSION}:
requesting that card with the envelope cost solver is refused by the
same pre-run audit rather than silently returning a zero
residual.  Both paths populate the same
diagnostic interface, so most constraint evaluators are
solver-agnostic, with solver-specific observables guarded by the
same pre-run audit.  (The one exception is the \texttt{gradient}
algorithm, whose \emph{optimized} residual is not the
\texttt{cost\_solver}'s: it constructs its own differentiable
residual, as documented in
Sec.~\ref{sec:diffpic}: a
precomposed matrix transport of the input $\Sigma$ without space
charge, or a 1500-particle bunch on a $32^3$ grid by default with
space
charge (both overridable through the multiparticle configuration
options).  The starting-point baseline is still
evaluated with the selected \texttt{cost\_solver}, so with space
charge the reported baseline and final costs come from different
models.)

Seven single-objective algorithms sit behind a common dispatch
(Table~\ref{tab:matching}).  \texttt{least\_squares}, the default, is a
bound-constrained trust-region least-squares solver (SciPy's
\texttt{trf} method) with a finite-difference Jacobian; it is local,
cheap, and accepts open-ended bounds.  \texttt{gradient} runs the same
least-squares outer loop but with the exact Jacobian obtained by
automatic differentiation through the differentiable tracking and
space-charge models of Sec.~\ref{sec:diffpic}.
\texttt{differential\_evolution} and \texttt{dual\_annealing} are
SciPy's global box searches.  Their population and annealing moves use
no derivatives, but neither is derivative-free end to end as invoked
here: \texttt{differential\_evolution} finishes with SciPy's
quasi-Newton polish of the best point, and \texttt{dual\_annealing}
couples the same local search to its annealing chains; both
finite-difference the scalar cost.  \texttt{cmaes} wraps the
covariance-matrix-adaptation evolution strategy~\cite{cmaes} with
per-variable step sizes scaled to the bound widths and a default
population of $4+\lfloor 3\ln d\rfloor$, with $d$ the number of
independent optimizer columns; its population can be evaluated
in parallel across worker processes, and the workers honor the
requested cost solver, so a parallel multiparticle CMA population
scores the multiparticle objective rather than an envelope
stand-in.  When parallel evaluation is enabled a separate
\texttt{cmaes\_search\_solver} option can
deliberately score the global search with the cheap envelope
objective while the reported cost and any least-squares polish stay
on the requested solver, and the transmission floor, which reads
structurally zero without particle data, is refused in that
configuration.  The algorithm is intended for
multimodal landscapes and for one-sided constraints whose zero-residual
plateaus defeat local least squares.  \texttt{bayesopt} performs
Gaussian-process Bayesian optimization with BoTorch~\cite{botorch}:
variables are normalized to the unit cube, a Sobol initial design
[default $\min(2d+2,12)$ points] plus the starting point, which is
always evaluated first, seeds a single-task Gaussian process, and a
log-expected-improvement
acquisition~\cite{logei} proposes one candidate per iteration, minimizing the number
of expensive evaluations.  Its optional physics-informed warm start,
available when
the multiparticle cost is selected, first scores a small Sobol batch
(eight points) with the cheap envelope objective and folds the
lowest-cost \emph{locations} into the Gaussian process's initial
design; every point entering the Gaussian process is then evaluated
with the requested multiparticle objective, so envelope values never
masquerade as multiparticle observations.  This is
warm-start seeding of the initial design, not a learned prior mean, and
it concentrates the early expensive evaluations where the envelope
model predicts good solutions.  Finally, \texttt{sequential\_scan} is a
physics-guided coordinate descent that codifies a manual tuning recipe:
it walks the elements in lattice order, bracket-scans each adjustable
parameter in steps of a fixed fraction of its bound width, reverses the
scan direction when the exit normalized emittances grow past a
configurable reference (by default only when both the horizontal and
the longitudinal metrics grow, so that pure plane exchange is not
punished),
and can reject outright any step that induces beam loss.  The global
searches require finite bounds on every variable and refuse otherwise
with an error naming the offending card; they run with fixed seeds so
matches are reproducible.  \texttt{cmaes} and \texttt{bayesopt}
optionally chain a short least-squares polish from their best point.

\begin{table}[b]
\caption{\label{tab:matching}%
Single-objective matching algorithms dispatched by the HELIX matcher
(the \texttt{gradient} algorithm is detailed in
Sec.~\ref{sec:diffpic}).
``Bounds'' states whether finite bounds are required on every
\texttt{ADJUST} variable (enforced with a diagnostic error naming the
unbounded card).}
\begin{ruledtabular}
\footnotesize
\begin{tabular}{llll}
Algorithm & Type & Derivatives & Bounds \\
\colrule
\texttt{least\_squares} & local & finite diff. & optional \\
\texttt{gradient} & local & auto.\ diff.\ (fixed branch) & optional \\
\texttt{differential\_evolution} & global & none (FD polish) & finite \\
\texttt{dual\_annealing} & global & none (FD polish) & finite \\
\texttt{cmaes}~\cite{cmaes} & global & none & finite \\
\texttt{bayesopt}~\cite{botorch} & global & none (GP model) & finite \\
\texttt{sequential\_scan} & heuristic & none & finite \\
\end{tabular}
\end{ruledtabular}
\end{table}

For genuinely competing objectives, HELIX complements the scalar
matcher with a Pareto explorer.  It accepts two or more objectives from
a built-in library (per-plane and four-dimensional emittance growth,
transmission loss, exit energy, and exit or peak rms sizes) over the
same \texttt{ADJUST} decision variables and returns the non-dominated
front, using either the NSGA-II genetic algorithm from
pymoo~\cite{pymoo} or, for expensive multiparticle evaluations, the
sample-efficient Bayesian qNEHVI acquisition~\cite{qnehvi} from
BoTorch~\cite{botorch}.  The scalar matcher and the Pareto
explorer are both reachable from the Python API, the command line, and
the GUI.

\subsection{Machine-learning element surrogates}
\label{sec:surrogates}

The surrogate acceleration validated in this work is direct
substitution of one field-map transfer matrix in zero-current
envelope transport.  Registry-based sliced space-charge engagement
presently delegates to the native integrator, and the multiparticle
linear-map path is experimental; neither is counted as validated
acceleration.  The remainder of this subsection describes the model,
the several engagement routes, and their present limitations against
that scope.

\begin{figure*}[t]
\includegraphics[width=\textwidth]{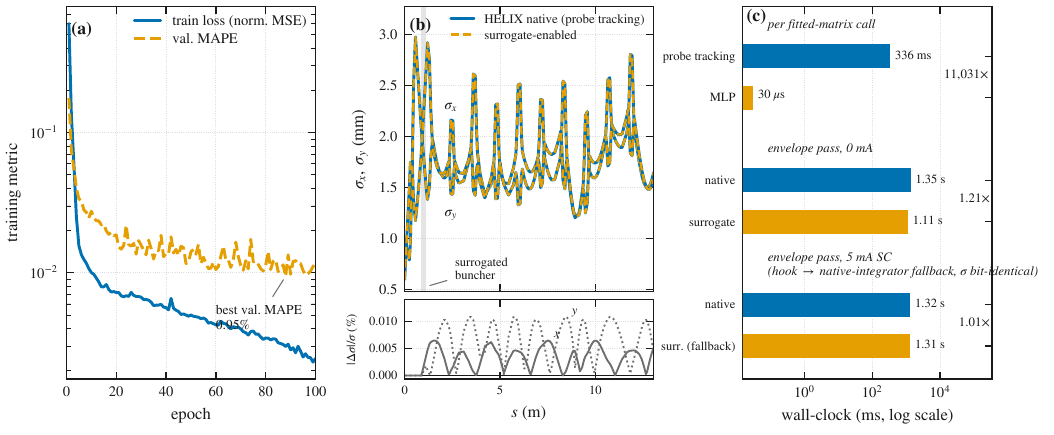}
\caption{\label{fig:surrogate}%
Surrogate workflow demonstrated end to end on the first PIP-II MEBT
buncher (240-mm 3-D field map), trained for this paper at the
moderate level (2000 Latin-hypercube samples $\times$ 100 epochs,
fixed seed; \SI{264}{\second} total, of which \SI{262}{\second} is
12-worker data generation).  (a)~Training history; best validation
mean absolute percentage error 0.95\%; the persisted model is
that best-epoch checkpoint, and its held-out test MAPE on 400
independent post-training samples is 1.4\%.  (b)~Envelope $\sigma_{x,y}(s)$
along the MEBT with the native element (solid) versus the surrogate
substituted for the shaded buncher (dashed), at zero current:
maximum deviation 0.011\% in $\sigma_{x,y}$ (strip; the
longitudinal moments differ by at most 0.043\%), wall time
$1.35\rightarrow\SI{1.11}{\second}$ ($1.21\times$) with one of four
bunchers surrogated.  (c)~Wall-clock economics (log scale): one probe-tracking
transfer-matrix computation (\SI{336}{\milli\second}) versus one raw
MLP forward pass (\SI{30}{\micro\second}); in registry-engaged
envelope runs with space charge the partial-slice hooks delegate to
the native integrator in the present release (zero neural
evaluations), and the result stays bit-identical.
Apple M3 Max, 14-core CPU.}
\end{figure*}

Field-map elements dominate the cost of an envelope pass: every
evaluation of an element's linearized $6\times6$ transfer matrix
requires tracking twelve probe particles through the tabulated field
with the element's native midpoint kick--drift integrator
(Sec.~\ref{sec:integrators}) and assembling the matrix by central
differences.  HELIX can replace this call with a trained neural
surrogate.  The surrogate class subclasses both the field-map element
and PyTorch's \texttt{nn.Module}~\cite{pytorch}, and implements the
envelope-mode transfer-matrix contract with a compact multilayer
perceptron (three hidden layers of 128 units via the Python API's
default; the training CLI defaults to two layers of 64, the
architecture behind Fig.~\ref{fig:surrogate}; smooth SiLU
activations, double precision) that maps the reference kinematics
$(W_{\mathrm{kin}},\beta,\gamma)$ and the element parameters (field
amplitudes and phase) to the flattened $6\times6$ matrix; only
$W_{\mathrm{kin}}$ is sampled independently, with $\beta$ and $\gamma$
derived from it, so the training set carries no unphysical kinematic
tuples and the two derived inputs are redundant features rather than
free dimensions.  Both the
1-D/2-D and the 3-D field-map classes are surrogatable through the
Python API (RFQ variants
are excluded; the shipped training CLI currently supports the 3-D
class only).  Training data are generated by Latin-hypercube
sampling of this input space, with ground truth from the element's own
numerically differenced transfer matrix (the same kick--drift probe
tracking used in envelope mode); generation parallelizes across worker
processes with bit-identical output.  Trained weights are persisted
with a metadata manifest (training scope, normalization, random
seed, and, when resolvable, code commit and lattice hash) for
reproducibility.  Surrogated cavities are autograd-differentiable,
acting as passive blocks through which the \texttt{gradient} matcher
of Sec.~\ref{sec:matching} can propagate derivatives with respect to
\emph{upstream} knobs; the cavity's own parameters are not optimizer
variables.

Every inference is guarded by the recorded training scope, an
axis-aligned box over the inputs: an input
outside it raises an out-of-scope error (the guard bounds each input
independently and is not a correlated out-of-distribution detector).
How the error is handled depends on the invocation path.  Under
direct element substitution, the configuration demonstrated in
Fig.~\ref{fig:surrogate}, the envelope solver decides the propagation
route before querying the surrogate and fetches the full-element
matrix only where it is consumed: on the zero-current envelope route
an out-of-scope input falls back to the wrapped element's
probe-tracked matrix, forfeiting the speedup rather than the run,
while a nonzero-current run, whose sliced space-charge propagation
delegates its partial slices to the wrapped element's integrator
regardless of scope, never consults the model.  The registry space-charge hooks and the multiparticle
fast path fall back to the wrapped element instead of aborting.  In
the gradient path of Sec.~\ref{sec:diffpic}, where the
wrapped nonlinear element cannot be executed, the error is a hard
stop that names the offending element (an identity map is never
substituted).  An engaged surrogate can therefore degrade speed or
abort
with a diagnostic, but never silently extrapolate beyond its
recorded box.

Engagement takes three forms.  Direct substitution
replaces the lattice element with the trained surrogate and is the
route validated in this work; because the solver queries the neural
model only on the zero-current envelope route that consumes the
full-element matrix, direct substitution's envelope speedup is
confined to zero-current transport.  The multiparticle fast
path below is the only route that accelerates a nonzero-current run,
and it is experimental.
Registry-based engagement
(surrogates are resolved by element name and engaged only after the
registered model is verified to wrap the element being tracked, by
object identity or a structural fingerprint of class, length,
resolved field file, and drive parameters; a same-named element that
does not match is skipped with a warning and tracks natively) is
wired into the
envelope solver's
space-charge partial-slice hooks, which in the present release
delegate to the native integrator, so registry engagement leaves
results bit-identical and invokes no neural evaluations; completing
that wiring is release work.  Multiparticle
engagement requires a separate double opt-in: a bit-identical safe
delegate, plus an experimental linear-map fast path.  Command-line
subcommands (\texttt{train}, \texttt{compare},
\texttt{run-envelope}, and \texttt{register-multi} for sharing one
trained cavity across a same-field-map family) and a dedicated GUI
tab cover training,
registration, comparison, and delegation checks;
the comparison tool runs baseline and registry-enabled passes back to
back and reports per-moment relative differences alongside wall-clock
times.  These
interfaces register surrogates within the running process
(the registry is not yet persistent across processes,
Sec.~\ref{sec:applications}); direct element
substitution currently runs through the Python API.

Figure~\ref{fig:surrogate} shows the workflow end to end on the first
MEBT buncher, trained for this paper at the moderate level.  The
deployed model's held-out test error---a matrix-entry mean absolute
percentage error (MAPE) on 400 samples generated \emph{after}
training at an independent seed---is 1.4\%, the headline
generalization number for everything that follows.  The best
validation error during training is 0.95\% (the same metric, which
overweights small entries), inside the documented interpolation
band, and the persisted model is the checkpoint that achieved it
(training snapshots and restores the best-validation epoch); the
independent test set exists precisely because that validation split
also drives the selection.  Substituting this surrogate for the one element changes the
envelope solution by at most 0.043\% (the longitudinal moments; the
transverse moments agree to $1.1\times10^{-4}$ relative) while
removing about a
sixth of the zero-current wall time, a reminder that a
matrix-entry MAPE does not map one-to-one onto moment error.

\begin{table}[tb]
\caption{\label{tab:surrogate_stats}%
Surrogate accuracy over the 400-sample held-out test set (generated
after training at an independent seed).  Percentiles across the samples
of: the matrix-entry mean absolute percentage error (MAPE); the
whole-matrix Frobenius error relative to the native matrix; the ratio
of the surrogate's $2$-norm condition number to the native element's;
and the relative error in the propagated transverse ($\sigma_x$,
$\sigma_y$) and longitudinal ($\sigma_z$) beam size on the nominal PIP-II input
covariance.  No matrix, native or surrogate, was singular.  The
matrix-level metrics (MAPE, Frobenius, conditioning) are evaluated in
the internal mm/mrad/deg/MeV coordinates and are fixed-convention
measures.}
\begin{ruledtabular}
\begin{tabular}{lrrrr}
 & median & 90th & 99th & max \\
\colrule
matrix MAPE (\%)      & 0.66   & 1.6    & 9.1   & 132   \\
Frobenius error (\%)  & 0.12   & 0.31   & 0.46  & 0.91  \\
cond.\ ratio          & 1.000  & 1.005  & 1.008 & 1.012 \\
$\sigma_x$ error (\%) & 0.0028 & 0.0075 & 0.012  & 0.015 \\
$\sigma_y$ error (\%) & 0.0029 & 0.0075 & 0.0098 & 0.016 \\
$\sigma_z$ error (\%) & 0.012  & 0.026  & 0.045  & 0.059 \\
\end{tabular}
\end{ruledtabular}
\end{table}

Table~\ref{tab:surrogate_stats} makes that gap quantitative across the
whole test set rather than at the single deployed point.  The
matrix-entry MAPE that averages to the quoted $1.4\%$ has a median of
only $0.66\%$; its upper tail (99th percentile $9\%$, maximum $132\%$)
is an artifact of near-zero matrix entries, where a small absolute
error reads as a large relative one.  The whole-matrix Frobenius error,
which weights entries by their magnitude, stays below $1\%$ for every
one of the 400 samples (median $0.12\%$), and the propagated beam
sizes track the native element to better than $0.02\%$ transversely
and $0.06\%$ longitudinally for the nominal input covariance
tested here.  The surrogate also reproduces the native matrix's
$2$-norm condition number to within $1.8\%$ (in the fixed internal
coordinates of the table caption; the ratio spans 0.982--1.013 over
the test set) and produced no singular
matrix.  On this element, then, the matrix-entry MAPE is a conservative
proxy that overstates the impact on beam observables by one to two
orders of magnitude; the surrogate's realized fidelity is set by the
sub-$0.1\%$ observable errors.

As a
structural check, the defect
$\lVert M^{\mathsf{T}} J_0 M - J_0\rVert_F$ of the surrogate's
predicted
matrix at the nominal kinematics, evaluated with the fixed
block-skew matrix $J_0$ of Sec.~\ref{sec:envelope} in the internal
mm/mrad/deg/MeV coordinates, is $0.1327$, against $0.1322$ for the
probe-tracked native matrix it was trained on.  Like the
spectral mode values of Sec.~\ref{sec:physics}, this scalar is
coordinate-convention dependent and is read here only as a
\emph{fidelity} measure---the surrogate reproduces its ground
truth's (nonzero) defect to $0.4\%$ rather than adding structure of
its own---not as a physical symplecticity certificate.

At the
measured net saving of \SI{0.235}{\second} per
zero-current envelope pass (one $\sim$0.34-s probe-tracked matrix
computation replaced per pass), the \SI{264}{\second} training cost
breaks even after roughly 1100 envelope passes.  A single design pass
therefore never repays it; the training amortizes only where one
element is re-evaluated across hundreds of zero-current envelope
passes, as in parameter scans over an unchanged lattice,
envelope-mode matching loops, and tolerance ensembles.  All timings in
this subsection were measured on a 14-core Apple M3 Max CPU.

Performance depends on the quoted granularity.
\emph{Per call}, a surrogate evaluates in 0.15--0.30~ms as documented
(the raw MLP forward pass measured for Fig.~\ref{fig:surrogate} is
\SI{30}{\micro\second}), against roughly 340~ms (336~ms measured
here) for the probe-tracking transfer-matrix computation it
replaces on the MEBT buncher, a factor of $1.1\times10^{4}$ on the
measured pair.
\emph{End to end}, the engine documentation's own MEBT+HWR benchmark
(20 field maps, 1000 macroparticles, 5~mA) reports a factor 0.99 for
the safe delegate (bit-identical physics), 1.22 for the fast path with
smoke-level training (200 samples $\times$ 40 epochs) at the price of a
28.4\% error in $\sigma_z$, and 1.66 with moderate training ($2000
\times 100$ for the cavities), which reduces the $\sigma_z$ error to
4.7\%, with transverse-size errors near 5\% in both cases:
longitudinal accuracy tracks training investment while transverse
accuracy does not improve with it.  The end-to-end gain on this lattice is
limited by eight solenoids that fall back to the native integrator.

\section{Verification, numerical convergence, and cross-code
benchmarking}
\label{sec:validation}

Our verification strategy is hierarchical.  Closed-form and
independently integrated envelope benchmarks gate the envelope
physics; cross-implementation parity
tests pin the redundant implementations of the PIC solver against one
another at floating-point precision; convergence scans quantify the
numerical sensitivities of the reported
quantities; and cross-code benchmarks on PIP-II-family lattices against
TraceWin~\cite{tracewin} (chosen as the reference because it is
among the most widely used and accepted codes for hadron-linac beam
dynamics and is the design code of record for PIP-II) using
envelope or
partran (multiparticle) references, named per comparison, validate
the integrated physics.  The LEBT, MEBT+HWR, and full-machine
TraceWin comparisons
below were all re-run for this paper at the recorded generation
baseline of the pinned code lineage, and verified unchanged through
the audited paper pin to the bounds documented in the paper's
figure-provenance records (the public release postdates that
equivalence audit);
for the full machine the fresh mode-matched comparison covers the
accelerating linac to \SI{186}{\meter}, and the older full-machine
numbers quoted alongside it are the documented cross-mode benchmark
the project maintains as a regression gate, labeled as such where
they appear.  The fresh runs executed the
NumPy PIC kernel path (the compiled C++ kernels had not yet been built
on the benchmark machine; they were built and timed later for
Sec.~\ref{sec:performance}); as shown in Sec.~\ref{sec:scverify}, the
two paths agree to floating-point round-off, so this affects speed
only.

\subsection{Analytic envelope benchmarks}
\label{sec:analytic}

\begin{figure*}[t]
\includegraphics[width=\textwidth]{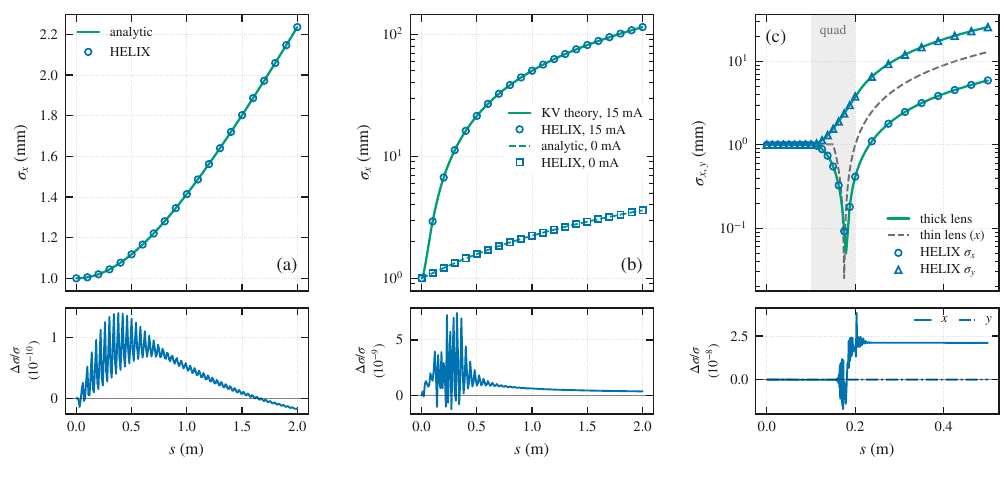}
\caption{\label{fig:analytic}%
Sacherer/KV rms-envelope solver (open markers) against closed-form
and independently integrated references (lines) for an
$\mathrm{H}^-$ beam at \SI{30}{\kilo\electronvolt},
$\varepsilon = \SI{1}{\milli\meter\milli\radian}$; the strip below
each panel shows
the relative deviation.  (a)~Free \SI{2}{\meter} drift from a waist
($\beta_0 = \SI{1}{\meter}$) versus
$\sigma(s) = \sqrt{\beta_0\varepsilon}\sqrt{1+(s/\beta_0)^2}$.
(b)~Space-charge-dominated expansion of a diverging beam at
\SI{15}{\milli\ampere} versus an independently integrated reference:
a fixed-step RK4 integration of the standard rms envelope
equation~\cite{sacherer,wangler},
$\sigma'' = \varepsilon^2/\sigma^3 + K/(4\sigma)$ in the round-beam
limit, with the zero-current Twiss drift shown for scale.  (c)~Thick-lens quadrupole
($G = \SI{10}{\tesla\per\meter}$, $L = \SI{100}{\milli\meter}$,
$\sqrt{k_1}L = 2.0$) versus thick-lens matrix optics in both planes;
the thin-lens equivalent (gray dashed) misplaces and deepens the
focal waist.}
\end{figure*}

Figure~\ref{fig:analytic} compares the standalone Sacherer/KV
rms-envelope integrator (Sec.~\ref{sec:envelope}) with closed-form
references or with independent numerical integration of the same
governing equation, in three
configurations that exercise, respectively, the emittance term, the
space-charge term, and the external-focusing term of the envelope
equation.  For a \SI{2}{\meter} field-free drift from a waist the
solver reproduces the analytic $\sigma(s)$ to a maximum relative
deviation of $1.4\times10^{-10}$ (mean $4.2\times10^{-11}$); the
physics-benchmark regression test for this case gates at $10^{-4}$, six orders of
magnitude looser than the observed deviation.  For the space-charge
test, a diverging \SI{30}{\kilo\electronvolt} $\mathrm{H}^-$ beam at
\SI{15}{\milli\ampere} (generalized perveance $K = 1.88\times10^{-3}$)
enters the drift at $\sigma_x = \SI{1.0}{\milli\meter}$ and exits at
\SI{114.3}{\milli\meter}, thirty-two times the \SI{3.6}{\milli\meter}
zero-current exit size, a strongly space-charge-dominated blowup.
The reference is deliberately independent of \texttt{linac\_gen}'s
solvers: a
fixed-step RK4 integration ($2\times10^4$ steps) of the standard rms
envelope equation~\cite{sacherer,wangler},
$\sigma'' = \varepsilon^2/\sigma^3 + K/[2(\sigma_x+\sigma_y)]$
evaluated in its round-beam limit $K/(4\sigma)$, written directly in
the figure script, with the electromagnetic constants taken from
\texttt{scipy.constants} and only the $\mathrm{H}^-$ rest mass shared
with the code under test.  HELIX tracks the independent
integration to $7.4\times10^{-9}$ maximum relative deviation (mean
$9.7\times10^{-10}$; the zero-current baseline agrees with the Twiss
drift to $2.5\times10^{-10}$), verifying both the space-charge
normalization and its integration.  Finally, a genuinely thick quadrupole
($\sqrt{k_1}L = 2.0$) followed by a drift reproduces thick-lens matrix
optics to $3.8\times10^{-8}$ in the focusing plane and
$6.5\times10^{-11}$ in the defocusing plane, while the thin-lens
equivalent places the waist \SI{4}{\milli\meter} upstream and twice as
deep (0.025 against \SI{0.050}{\milli\meter}) and overestimates the
exit size by a factor 2.2, confirming that the solver integrates the
hard-edge focusing profile through the quadrupole body rather than
applying a thin-lens kick.
These three tests verify the standalone continuous-beam integrator and
the space-charge normalization it shares with the production path in
the nonrelativistic regime where both operate
(Sec.~\ref{sec:envelope}); the
production $6\times6$ covariance solver that carries the
envelope-mode cross-code benchmarks below is verified instead by the
matrix and transport unit tests of the suite and by the TraceWin
comparisons of Sec.~\ref{sec:tracewin}.

\subsection{Cross-implementation PIC parity}
\label{sec:scverify}

\begin{figure*}[t]
\includegraphics[width=\textwidth]{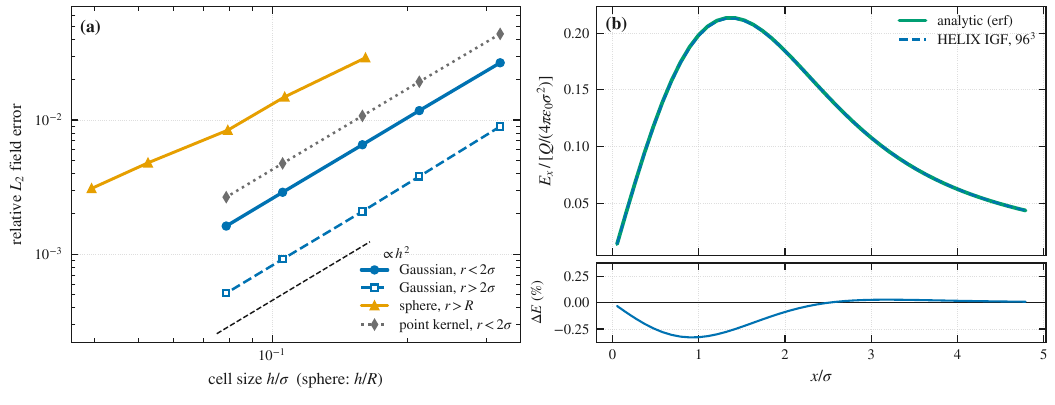}
\caption{\label{fig:fieldacc}%
Space-charge field solver against analytic references.
(a)~Relative $L_2$ error of $\mathbf{E}$ versus cell size for charge
densities assigned directly to the mesh: isotropic Gaussian bunch
(core $r<2\sigma$ and halo $r>2\sigma$, domain $\pm5\sigma$) with the
integrated-Green-function (IGF) kernel and, for contrast, the sampled
point kernel; uniformly charged sphere (exterior).  The smooth cases
follow the $h^2$ guide; the sphere is limited by its staircase surface
representation.  (b)~Gaussian radial field at the production $96^3$
grid along the mesh line nearest the bunch axis, in units of
$Q/(4\pi\varepsilon_0\sigma^2)$, with the residual as a percentage of
the peak field.}
\end{figure*}

The PIC solver exists in several deliberately redundant
implementations---C++ and NumPy deposit/gather kernels, FP64 CPU
and CUDA FFT backends, an FP32 Apple-Metal backend, and the
differentiable PyTorch re-implementation of the full cycle
(Sec.~\ref{sec:diffpic})---and the test suite pins them against one
another.  The C++ cloud-in-cell deposit and gather kernels reproduce
the Python reference to $2$--$4\times10^{-16}$ relative (peak-normalized,
measured over $16^3$--$64^3$ grids: one unit in the last place, the
$10^{-10}$ figure being the conservative regression tolerance), and the
full kick cycle
(deposit, Poisson solve, gather, momentum kick) is pinned to a relative
tolerance of $10^{-10}$ (absolute $10^{-14}$) on every particle
coordinate when the compiled deposit/gather kernels are swapped for
the Python reference, the Poisson stage being common code.  The multithreaded \texttt{scipy.fft}
CPU backend reproduces the serial NumPy reference transform at the
round-off floor of the double-precision transform (the $10^{-12}$
unit-test tolerance is set conservatively).  The CUDA backend matches the CPU field solution to better
than $10^{-9}$ relative; the residual is traced to
$\sim 10^{-15}$ differences between NumPy and CuPy transcendentals,
amplified through the integrated-Green-function (IGF) construction,
FFT, and finite differencing, and the IGF spectrum built natively on
each backend agrees to $10^{-12}$.  The FP32 Metal backend is held to
$5\times10^{-6}$ relative on the solved field by its regression gate,
with $3.4\times10^{-6}$ observed on $E_x$ in the benchmark of
Sec.~\ref{sec:performance}, taken there on a smooth analytic
charge density rather than on a deposited bunch.  That backend is a
performance option and carries no result reported here: every
cross-code benchmark below ran the FP64 CPU path, so its FP32 error
does not enter them.  The PyTorch implementation reproduces the NumPy
deposit/gather at the floating-point round-off floor and the full
space-charge kick to
$\sim 10^{-13}$ relative on the kick increment (the $10^{-12}$ and
$10^{-9}$ unit-test tolerances are set conservatively).  Independently of these pairwise checks, the Poisson
solver (built on the Hockney--Eastwood doubled
grid~\cite{hockney_eastwood} with the IGF Green's
function~\cite{qiang2006,qiang2024igf} as the default) is tested
against its algebraic invariants: deposition conserves charge and the
solved fields scale linearly with the source and superpose, in each
case at the round-off floor of the double-precision solve (the
charge-conservation residual is quantified in
Sec.~\ref{sec:fieldacc}; the linear-rescaling check is bit-exact,
since the test factor is a power of two; the regression gates for
these three checks are set conservatively at $10^{-10}$, $10^{-12}$,
and $10^{-10}$).  These
invariants pin the implementation; the field accuracy itself is
established against analytic references in
Sec.~\ref{sec:fieldacc}, by the convergence scans below, and by the
cross-code benchmarks of Sec.~\ref{sec:tracewin}.

\subsection{Field-solver accuracy, conservation, and numerical
convergence}
\label{sec:fieldacc}

\begin{figure*}[t]
\includegraphics[width=\textwidth]{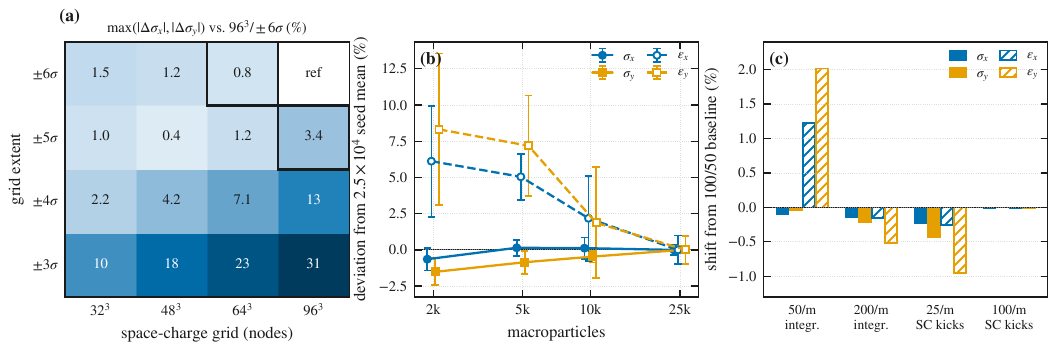}
\caption{\label{fig:convergence}%
PIC numerical-convergence scans for a FODO cell with the PIP-II
$\mathrm{H}^-$ injector beam (\SI{2.123}{\mega\electronvolt},
\SI{5}{\milli\ampere}), re-run for this paper at the pinned code
version ($10^4$ macroparticles, 100/50 integration/space-charge steps
per meter, fixed beam seed, unless the quantity is being scanned).
(a)~Two-dimensional mesh scan varying the grid extent and the node
count independently: each entry shows the larger of the end-of-cell
$|\Delta\sigma_x|$ and $|\Delta\sigma_y|$ relative to the
finest-and-widest corner ($96^3$ nodes, $\pm6\sigma$); boxes mark the
MEBT+HWR multiparticle benchmark mesh ($64^3/\pm6\sigma$, 0.8\%) and
the legacy-compatible default mesh ($96^3/\pm5\sigma$, 3.4\%).
(b)~Macroparticle scan at the $64^3/\pm6\sigma$ mesh: deviation of the
five-seed mean from the $2.5\times10^4$-particle reference; error bars
give $\pm1$ standard deviation over the seeds, whose spread scales as
$N^{-0.49}$ (pooled fit over the four observables).
(c)~Shifts of the same observables when the integration and
space-charge cadences are halved or doubled around the 100/50 per-meter
baseline: at most 0.43\% ($\sigma$) and 2.0\% ($\varepsilon$), the
latter only when the integration cadence is halved.}
\end{figure*}

The tests in this subsection verify the physics of the field solve
itself (IGF Green's function, FFT convolution, gradient, and the
matched CIC deposit/gather pair) against closed-form and
independently integrated references, complementary to the
cross-implementation comparisons above.  Assigning $\rho$ directly to
the mesh isolates the solver from sampling noise: for an isotropic
Gaussian bunch the field error converges at second order over
$32^3$--$128^3$ (fitted orders 1.99 core, 2.03 halo), reaching
$L_2$ errors of $2.9\times10^{-3}$ ($r<2\sigma$) and
$9.2\times10^{-4}$ ($r>2\sigma$) at the production $96^3$ setting,
with $L_\infty = 3.7\times10^{-3}$ [Fig.~\ref{fig:fieldacc}(b);
peak-normalized residual below $0.33\%$].  A uniform sphere, whose
density is discontinuous, converges at the expected reduced order
(1.3--1.6 in $L_2$).  For an anisotropic Gaussian with
$\sigma_z/\sigma_x = 2.5$ the solver matches an independent quadrature
of the ellipsoidal-Gaussian integral (itself verified against adaptive
quadrature to $10^{-12}$) to $2.2\times10^{-3}$ transversely and
$3.0\times10^{-3}$ longitudinally at $96^3$.  With CIC-deposited
macroparticles the error is sampling-dominated, decreasing as
$\sim N^{-1/2}$ from $2.4\times10^{-2}$ at $N=10^5$ toward the
direct-$\rho$ floor.  The discrete scheme also satisfies the expected
conservation laws: deposition conserves charge to $4\times10^{-16}$
(exactly for a bunch straddling the mesh face); the self-field kick of
a single macroparticle is at most $2.3\times10^{-16}$ of the two-particle
kick at one-cell separation (125 in-cell offsets scanned); the net
momentum imparted to a $10^5$-particle bunch by one kick is
$1.2\times10^{-10}$ of the rms single-particle kick when no particle
occupies the outermost grid cell ($7.1\times10^{-4}$ for the full
$5\sigma$ bunch, attributable to a single tail particle in the
one-sided boundary-stencil cell); and shifting the bunch by half a
cell changes the co-moving fields by $0.41\%$ rms.

The integrated PIC convergence scans of Fig.~\ref{fig:convergence},
introduced with the parity tests above, decouple these numerical
knobs.  The two-dimensional mesh scan of panel~(a) varies the grid
extent and the node count
independently: the rms sizes are controlled by the extent, not the
resolution, drifting by 10--31\% at $\pm3\sigma$ without ever
plateauing, whereas at $\pm6\sigma$ the entire $32^3$--$96^3$ span
moves them by at most 1.5\%.  The two
bunched-beam benchmark meshes of Sec.~\ref{sec:tracewin} sit 0.8\%
($64^3/\pm6\sigma$, MEBT+HWR) and 1.2\% (full-machine
$48^3/\pm6\sigma$) from the finest-and-widest corner, the
legacy-compatible default mesh ($96^3/\pm5\sigma$) 3.4\%; adding the $10^4$-particle seed spread (0.4--0.7\% in
$\sigma$, scaling as $N^{-0.49}$ [panel~(b)]), the $\leq0.5$\% shift
of the seed mean out to $2.5\times10^4$ particles, and the
$\leq0.43\%$ cadence sensitivity [panel~(c)] puts the quadrature
scale of these sensitivities at roughly 1.3\%, 1.5\% and 3.6\% for the
three configurations.  Because the mesh and extent shifts are systematic
and correlated rather than independent random errors, that figure is
a sensitivity indicator rather than a confidence interval, with the caveat that the scan is a
FODO/CIC stress test whose numbers transfer to the benchmark
configurations of Sec.~\ref{sec:tracewin} as a guide, not as bounds.  Because the domain extent dominates that
budget (the $\pm5\sigma$ box alone accounts for the 3.4\%), this scan
motivates $\pm6\sigma$ as the preferred default extent, which the
benchmarks of Sec.~\ref{sec:tracewin} use.  The shipped default is
deliberately left at $\pm5\sigma$ so that results reproduce those of
existing studies built on that extent rather than shifting silently
under them; the $3.4\%$ is the deviation of the legacy default mesh
($96^3/\pm5\sigma$) from the reference corner in this space-charge-violent
stress case, the benchmark mesh ($64^3/\pm6\sigma$) sits $0.8\%$ from
it, and the $\pm5\sigma$ sensitivity is quoted wherever it enters a
benchmark here.

The
emittances of this deliberately space-charge-violent stress case are
not mesh-converged at fixed $N$ (in $\varepsilon_x$/$\varepsilon_y$
the MEBT+HWR benchmark mesh sits $-20\%$/$-28\%$ from the corner and
the full-machine mesh $-27\%$/$-41\%$, while the legacy default mesh
sits $+55\%$/$+79\%$), but that mesh dependence is not a
particles-per-cell sampling artifact: raising the count tenfold to
$10^5$ closes only 3.5\% ($\varepsilon_y$) to 8.4\% ($\varepsilon_x$)
of the $64^3$-versus-$96^3$ gap, and at fixed mesh the emittance moves
by less than $0.6\%$ between $2.5$--$5\times10^{4}$ particles and
$10^{5}$.  Emittance
growth here is resolution-limited, so the cross-code benchmarks below
are stated on the rms sizes and the documented production guidance
scales the particle count together with the mesh.

\subsection{Derivative verification}
\label{sec:gradcheck}

\begin{figure*}[t]
\includegraphics[width=\textwidth]{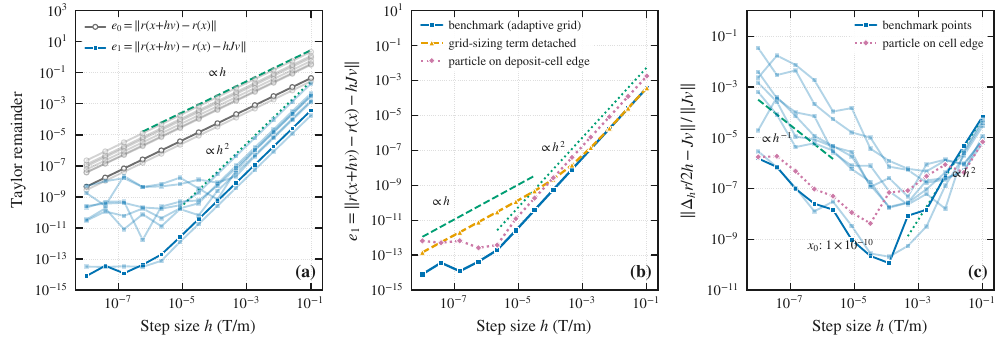}
\caption{\label{fig:gradcheck}%
Direct verification of the autograd Jacobian $J$ of the tracking+PIC
matching residual $r(x)$ of the benchmark problem of
Sec.~\ref{sec:app_gradmatch}, taken in raw native units rather than in the
tolerance-normalized form the benchmark optimizes (four exit
constraints, six quadrupole gradients, 1500 macroparticles, 26
space-charge kicks on a $32^3$ grid).  Diagonal rescaling of $r$ leaves
the Taylor orders unchanged but shifts the quoted norm floors.  (a)~Taylor remainders
$e_0$ and $e_1$ versus step size $h$ at the benchmark start $x_0$
(full color) and at three random in-bounds settings, two random unit
directions each (light); green guides show slopes 1 and 2.
(b)~$e_1$ at $x_0$ for the engine configuration; for a Jacobian
computed with the adaptive grid extents detached from the graph; and
at a point constructed so that one macroparticle sits on a
deposit-cell boundary ($5\times10^{-14}$ cells).  (c)~Relative
disagreement between the central difference $\Delta_h r/2h$ and $Jv$;
the V shape is the $O(h^2)$-truncation versus $O(\epsilon/h)$-roundoff
tradeoff of the finite difference itself, bottoming at its
measurement limit.}
\end{figure*}

Section~\ref{sec:diffpic} calls these Jacobians exact to
floating-point precision for a fixed computational branch; that is a
testable claim, and we test it on the matching residual of
Sec.~\ref{sec:app_gradmatch}.  If $J$ is the true derivative, the
remainders $e_0(h)=\lVert r(x+hv)-r(x)\rVert$ and
$e_1(h)=\lVert r(x+hv)-r(x)-hJv\rVert$ must scale as $h$ and $h^2$.
Over $h\in[10^{-8},10^{-1}]$~T/m at the benchmark start $x_0$ and
three random in-bounds settings, two random unit directions each, the
fitted slopes are $0.999\pm0.002$ for $e_0$ and $2.01\pm0.03$ for
$e_1$ [Fig.~\ref{fig:gradcheck}(a)]; $e_1$ follows $h^2$ down to a
floor of $10^{-14}$--$10^{-11}$ relative to $\lVert r\rVert$, reached
below $h\approx3\times10^{-5}$ (median; $2\times10^{-6}$ at $x_0$).
That floor makes the piecewise-smoothness caveat of
Sec.~\ref{sec:diffpic} quantitative.  The CIC deposit's derivative
jumps when a particle crosses a cell boundary.  With $1.2\times
10^{5}$ deposit coordinates per evaluation, the nearest boundary at
$x_0$ is $8\times10^{-6}$ cells away, so crossings are unavoidable at
any useful $h$.  Yet even at a point engineered from the deposit
indices to hold a macroparticle within $5\times10^{-14}$ cells of a
boundary, the fitted slope remains 2.00 and the floor stays in the
$10^{-14}$ band relative to $\lVert r\rVert$.  This corresponds to a
few $10^{-13}$ in the absolute remainder plotted in
Fig.~\ref{fig:gradcheck}(b).  Central differences
agree with $Jv$ to a best relative error of $1\times10^{-10}$ at
$x_0$ ($2\times10^{-8}$ median, $5\times10^{-7}$ worst case over the
eight scans) on the plateau $h\approx3\times10^{-5}$--$10^{-3}$
[Fig.~\ref{fig:gradcheck}(c)], agreement at the limit of what a
finite difference can resolve.  Detaching the adaptive grid extents
from the graph leaves the forward pass bit-identical but shifts $J$ by
$1.0\times10^{-4}$ in relative Frobenius norm and degrades $e_1$ to
first order at small $h$ [Fig.~\ref{fig:gradcheck}(b)]: differentiating through
the grid geometry is load-bearing, not cosmetic.  The Jacobian is
bit-identical with checkpointing on or off, with vectorized or looped
vector--Jacobian products, and under \texttt{grid\_mode}
\texttt{fixed}/\texttt{adaptive}---an interface-robustness check
rather than a physics toggle, since \texttt{grid\_mode} is honored
only by the numpy solver and the torch kick always refits its grid;
the frozen-grid derivative question is the detach ablation above.
These quantitative derivative checks exercise the CIC deposit used by
the matching residual; the torch TSC deposit shares the same
detached-index, differentiable-weight construction and passes
gradient-flow and forward-parity tests, but its derivatives have not
been subjected to the same Taylor-remainder verification.
Complementary optimizer-level evidence appears in the
benchmark of Sec.~\ref{sec:app_gradmatch}: repeated autograd matches are
bit-identical in their final knobs, residual vector and cost, evaluation
counts, and norm histories, while the finite-difference TRF
baseline tracks the autograd trajectory to within the
finite-difference error itself: residual norms after the first
accepted step within
$2.1\times10^{-8}$ relative, endpoints within
$3.1\times10^{-6}$~T/m on the matched-solution manifold.

\subsection{Field-map transfer-matrix probe robustness}
\label{sec:probesteps}

\begin{figure}[tb]
\includegraphics[width=\linewidth]{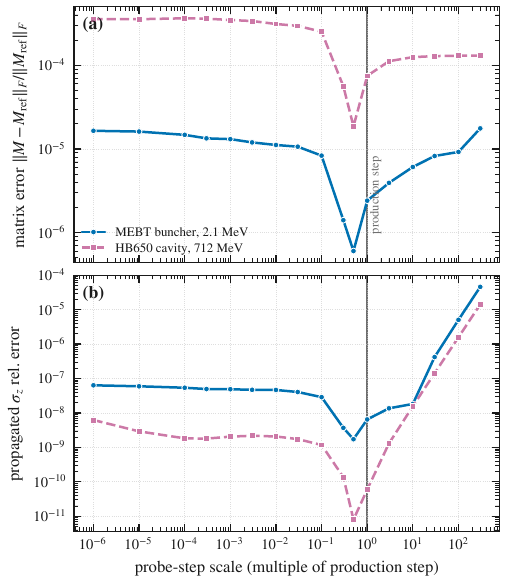}
\caption{\label{fig:probesteps}%
Probe-step robustness of the field-map transfer matrix.  The
central-difference matrix is recomputed at probe displacements scaled
from $10^{-6}$ to $300$ times their production values, for a
\SI{2.1}{\mega\electronvolt} MEBT buncher and a
\SI{712}{\mega\electronvolt} HB650 cavity.  (a)~Whole-matrix Frobenius
error against an $O(h^{4})$ Richardson-extrapolated reference, in the
internal mm/mrad/deg/MeV coordinates (a fixed-convention measure).
(b)~Relative error in the propagated $\sigma_z$ on the nominal PIP-II
input covariance.  The vertical line marks the production step.  The
reference is Richardson-extrapolated from the production and
half-production matrices, so neither of those two steps is independent
of it: the minimum at the half-production step is a construction
artifact, and the small-step level measures the extrapolated
reference's own offset from the converged central-difference matrix
rather than round-off.}
\end{figure}

The envelope solver and the surrogate ground truth of
Sec.~\ref{sec:surrogates} both take each field-map element's $6\times6$
transfer matrix from a central-difference Jacobian: twelve probe
particles are tracked through the tabulated field at fixed coordinate
displacements and differenced.  Those displacements are fixed constants
in the integrator ($10^{-3}$~mm, $10^{-3}$~mrad, $10^{-2}$~deg,
$10^{-4}$~MeV), and a finite-difference derivative is only as
trustworthy as its step: too large and truncation dominates, too small
and floating-point round-off does.  Figure~\ref{fig:probesteps} scans
the step from $10^{-6}$ to $300$ times the production value for two
cavities at opposite ends of the linac---a
\SI{2.1}{\mega\electronvolt} MEBT buncher and a
\SI{712}{\mega\electronvolt} HB650 cavity, entry energies taken from a
zero-current reference pass of the full lattice---and measures the
whole-matrix Frobenius error against an $O(h^{4})$
Richardson-extrapolated reference together with the resulting error in
the propagated beam size.  Reconstructing the matrix at the production
step reproduces the integrator's own result bit for bit, so the scan
isolates the step alone.  Both cavities show the expected truncation rise at
large steps; at small steps the error does not rise but flattens, at a
level set by the extrapolated reference's own offset, so no
floating-point round-off rise appears within the scanned range.  Over
the full eight-decade
scan the transfer matrix moves by at most $2\times10^{-5}$
(\SI{2.1}{\mega\electronvolt}) and $4\times10^{-4}$
(\SI{712}{\mega\electronvolt}) relative to the reference, and the
propagated $\sigma_z$ by at most $5\times10^{-5}$.  What matters for
the production step is that absolute scale rather than its position
along the scan: even the worst step tried
shifts a beam size by less than $10^{-4}$, far below the code-to-code
agreement of Sec.~\ref{sec:tracewin}.  Because the reference is
Richardson-extrapolated from the production and half-production
matrices, the comparison at those two steps is not independent of it,
and no minimum-locating claim is made from this scan.  In the whole-matrix Frobenius error
[Fig.~\ref{fig:probesteps}(a)] the small-step level is about twenty
times higher at the high-energy, high-frequency cavity
($3.6\times10^{-4}$ against $1.6\times10^{-5}$ at the smallest step);
the extrapolated reference's own offset is larger there.
This scan
covers the tight full-element 3-D probe steps; the 1-D/2-D field-map
class, the superposed-cluster class, and the sliced 3-D space-charge
path use larger fixed
displacements ($\pm0.01$ in the first five internal coordinates and
$\pm1$~keV in energy), up to ten times the full-element steps, inside
the multiplier range scanned here, although those map classes were not
independently included in the scan.

\subsection{Comparison with TraceWin on PIP-II lattices}
\label{sec:tracewin}

The anchor validation is against TraceWin on three PIP-II
lattices spanning the code's operating regimes: the PIP-II LEBT
(\SI{30}{\kilo\electronvolt} dc beam, continuous-beam space charge),
the MEBT plus HWR cryomodule (bunched beam, 2.12 to
\SI{10.26}{\mega\electronvolt} over \SI{18.96}{\meter}), and the full
\SI{256.5}{\meter} linac (about 1700 elements, to
$\sim$\SI{821}{\mega\electronvolt})~\cite{pathak_pip2design}; the
TraceWin reference mode, envelope or partran (multiparticle), is
named for each comparison.  All runs use
\SI{5}{\milli\ampere} $\mathrm{H}^-$ with the PIP-II design
Twiss parameters at each entrance, and
Table~\ref{tab:tracewin} summarizes the agreement.  Every benchmark
reported here is a space-charge-on comparison in one of the two
production modes: the envelope solver against TraceWin's envelope
solver, and multiparticle PIC tracking against partran.  On the two
bunched lines the envelope arm uses the three-dimensional linearized
model; on the LEBT the beam is dc, so the longitudinal force vanishes
and both codes use their two-dimensional continuous-beam models
instead.

\begin{figure*}[t]
\includegraphics[width=\textwidth]{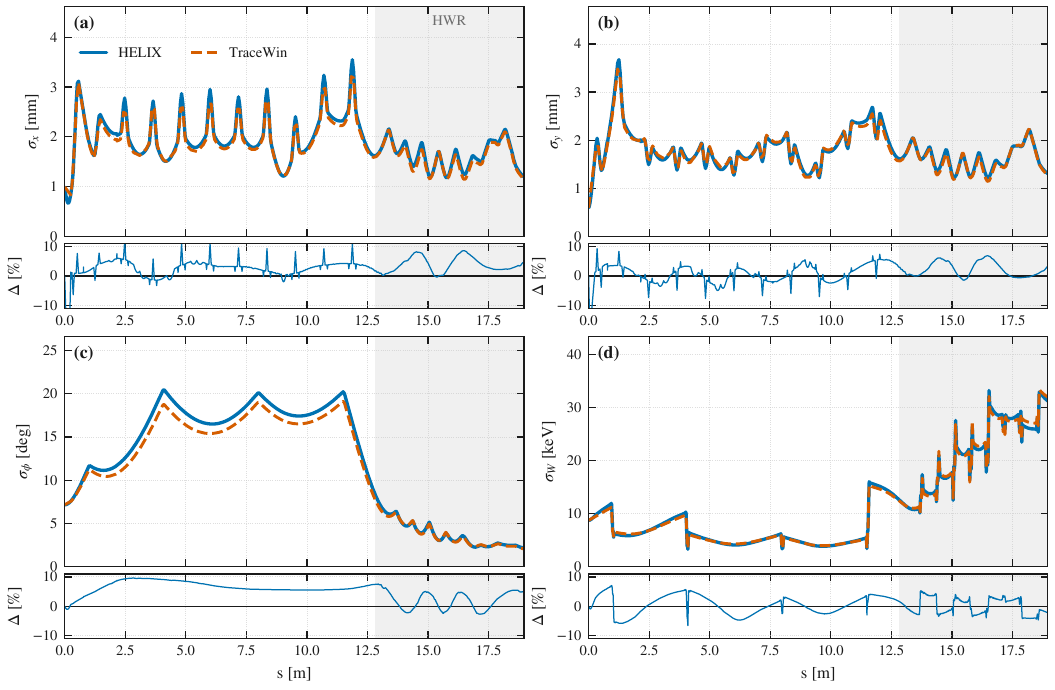}
\caption{\label{fig:tw_mp}%
RMS beam sizes along the PIP-II MEBT and HWR cryomodule in
multiparticle mode ($\mathrm{H}^-$, \SI{2.12}{\mega\electronvolt}
injection, \SI{5}{\milli\ampere}, space charge on): HELIX PIC tracking
($10^4$ macroparticles, $64^3$ grid, $\pm6\sigma$ extent, solid)
versus the TraceWin partran reference with equal particle count
(dashed), an ensemble-level statistical comparison of independently
sampled bunches, not a particle-identical parity test.
(a)~Horizontal and (b)~vertical rms size, (c)~rms phase spread, and
(d)~rms energy spread, each over a strip giving the relative deviation
$(\mathrm{HELIX}-\mathrm{TW})/\mathrm{TW}$; the shaded band marks the
HWR section ($s = 12.8$--\SI{18.96}{\meter}).  HELIX
transmits 97.76\% of the beam.  Residual strips are clipped at
$\pm11\%$: near the entrance waists ($s < \SI{0.15}{\meter}$) the
deviation reaches $-26\%$, where partran's export interpolates
linearly across the waist instead of resolving it.  As a cross-mode
diagnostic, TraceWin's dense envelope export does resolve the waist,
and against it HELIX's multiparticle $\sigma_x$ differs by 0.16\% at
that position.
Beyond that leading drift the maxima are 12.0\% ($\sigma_x$) and
9.4\% ($\sigma_y$); the 99th
percentile of $|\Delta|$ along the line is 8.5\% ($\sigma_x$) and
6.7\% ($\sigma_y$), with means of 2.6--3.9\% across the four
moments.}
\end{figure*}

Three methodological points apply.  First, HELIX follows TraceWin's
documented \texttt{.dat} unit conventions (lengths in
millimeters~\cite{tracewin_manual}), so lattice files drive both
codes within the supported card set; particle interchange uses
TraceWin's binary \texttt{.dst} format, which HELIX reads and writes
in TraceWin's native cm/rad/MeV units.  Second, along-$s$ statistics
must be evaluated at HELIX's recorded $s$-points (the integration
grid: 1586 records at 1074 distinct positions for the MEBT+HWR
envelope run, each element's exit record coinciding with the next
element's entrance record, at a maximum spacing of
\SI{20}{\milli\meter}; coincident records carry identical values and
are counted once) with the dense
TraceWin grid ($28\,947$ points, \SI{0.65}{\milli\meter} spacing)
interpolated onto them, the convention used for the MEBT+HWR rows
(the LEBT statistics use the overlapping $s$-range of the two output
grids).  Interpolating the finer record onto the coarser is the
well-posed direction here: TraceWin's export is some thirty times
denser than HELIX's \SI{20}{\milli\meter} output cadence, so the
reverse asks HELIX for values between its own records.  Transversely
the distinction barely matters, moving the maximum deviation from
0.57\% to 0.64\% ($\sigma_x$) and from 0.61\% to 0.86\%
($\sigma_y$).  The energy spread is the exception: it carries sharp
local minima that the output cadence does not resolve, and
interpolating HELIX across the deepest of them, where TraceWin swings
by 48\% inside a single \SI{19}{\milli\meter} gap, reaches 39\%
against 0.39\% at HELIX's own records.  Third, the
multiparticle rows are ensemble-level statistical benchmarks rather
than particle-identical parity tests: each code draws its own input
bunch from the shared rms Twiss specification (Gaussian in both
codes; truncation and generator details are code-side settings, so
the residuals also include any distribution-shape
difference), and pointwise
residuals carry the shot-noise floor of the finite ensembles
($\sim$1\% at $10^4$ particles).  A separate effect sets the largest
multiparticle residuals: TraceWin's partran export writes at
diagnostic planes and fills the intervals between them linearly, so
wherever one interval spans a waist the exported reference is a chord
across it.  Over the leading \SI{250}{\milli\meter} drift of the
MEBT+HWR line its 409 rows are collinear to within the export's
seven-digit round-off ($10^{-6}$~mm), so the
export does not resolve the waist there, and this sampling effect
dominates the entrance-waist maxima of Fig.~\ref{fig:tw_mp}.  How
much of the residual would survive a densely recorded partran run is
not something the archived exports can settle.
Table~\ref{tab:twconfig} collects the
configurations of the three benchmark lines.

\begin{table*}[t]
\caption{\label{tab:twconfig}%
Configuration of the TraceWin cross-code benchmarks.  Both codes
read the same \texttt{.dat} lattice file; for the multiparticle rows
each code samples its input bunch independently from the same rms
Twiss specification (HELIX: Gaussian with a $4\sigma$ cutoff at a
fixed seed), while the envelope rows propagate the same $\Sigma$
deterministically.  The
TraceWin references are archived exports of the cited
code~\cite{tracewin,tracewin_manual}, run
with its own space-charge solver and mesh settings.  The TraceWin
installation used for this work reports version 2.27.0.0 (64-bit); the
exports themselves carry no version field, so the build that produced
each one is not independently recorded.
In the multiparticle maximum columns of
Table~\ref{tab:tracewin} the parenthetical value excludes the leading
\SI{250}{\milli\meter}, over which the partran export is a chord
between diagnostic planes (footnote~a).  The single-convention statistics of
Table~\ref{tab:tracewin} are evaluated at HELIX's recorded positions
with the TraceWin record interpolated onto them; the per-figure strips
use the same positions and differ only in being unweighted, except for
the LEBT multiparticle pair, whose strip is evaluated on TraceWin's
\num{417}-point export rather than on HELIX's denser substep record.  No deck contains an unrecognized
card that carries physics: the LEBT deck parses warning-free, and each
bunched deck raises one warning about the amplitude convention of the
inert \texttt{ERROR\_*} directive, which nominal runs do not consume.
The full-linac deck additionally reports its 59 colon-terminated
section labels (12 distinct names, e.g.\ \texttt{SSR1 CM:},
\texttt{WPM :}) as unsupported cards; these name positions in the deck
and carry no physics.}
\begin{ruledtabular}
\begin{tabular}{lp{3.7cm}p{4.4cm}p{4.4cm}}
 & PIP-II LEBT & PIP-II MEBT+HWR & PIP-II acc.\ linac (\SI{186}{\meter}) \\
\hline
Lattice file & \texttt{lebt\_pxie.dat} & \texttt{mebt\_plus\_hwr.dat} & \texttt{mebt+hwr+ssr1+ssr2+} \texttt{lb650+hb650.dat} \\
Beam & dc $\mathrm{H}^-$, \SI{30}{\kilo\electronvolt}, \SI{5}{\milli\ampere} & bunched $\mathrm{H}^-$, 2.12--\SI{10.26}{\mega\electronvolt}, \SI{162.5}{\mega\hertz}, \SI{5}{\milli\ampere} & bunched $\mathrm{H}^-$, from \SI{2.12}{\mega\electronvolt}, \SI{162.5}{\mega\hertz}, \SI{5}{\milli\ampere}; installed-voltage rf configuration (full line reaches ${\approx}\SI{821}{\mega\electronvolt}$; nominal delivery \SI{800}{\mega\electronvolt}) \\
Macroparticles (HELIX\,/\,TW) & $2\times10^4$\,/\,not recorded & $10^4$\,/\,$10^4$ & 5000 ($10^4$ check)\,/\,$10^4$ \\
HELIX MP space charge & 2-D Hockney FFT (\texttt{pic2d}) & 3-D FFT PIC (CIC, IGF), $64^3$, $\pm6\sigma$ & 3-D FFT PIC (TSC, IGF), $48^3$, $\pm6\sigma$ \\
HELIX env.\ space charge & 2-D dc model (continuous) & 3-D linearized (bunched) & 3-D linearized (bunched) \\
Recording grids (HELIX\,/\,TW) & env 151\,/\,271 pts; MP 6\,067\,/\,417 pts & env 1586\,/\,28\,947 pts; MP 15\,607\,/\,36\,423 pts & HELIX records to \SI{256.5}{\meter}\,/\,TW exports end at \SI{185.989}{\meter} \\
\texttt{ERROR\_*} directives & none in file & one (\texttt{ERROR\_CAV\_NCPL\_stat}), inert in both codes' nominal runs & same card, inert \\
\end{tabular}
\end{ruledtabular}
\end{table*}

\begin{figure}[t]
\includegraphics[width=\linewidth]{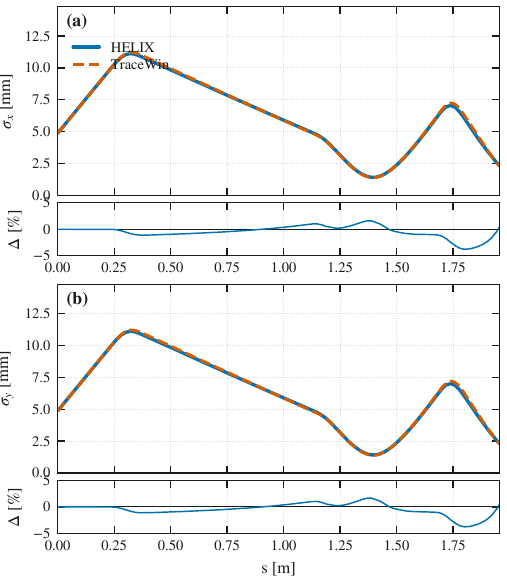}
\caption{\label{fig:tw_lebt}%
Transverse rms envelopes in the \SI{1.96}{\meter} PIP-II LEBT for a
continuous \SI{30}{\kilo\electronvolt} $\mathrm{H}^-$ beam at
\SI{5}{\milli\ampere}: HELIX continuous-beam envelope solver with the
2-D dc space-charge model (solid) versus the TraceWin envelope
reference (dashed), including the solenoid
space-charge-compensation maps.  (a)~Horizontal and (b)~vertical rms
size, each over a strip showing the relative
deviation at HELIX's distinct recorded positions: mean $|\Delta|$ is
0.77\% in both planes, with maxima of 3.79\% and
3.78\% at $s = \SI{1.80}{\meter}$, where the envelope converges most
steeply (the line's only waist, at \SI{1.39}{\meter}, shows 1.6\%),
and exit deviations of $+0.47\%$ and
$+0.51\%$ (the documented statistic on a 60-point
uniform grid, quoted in the text, is 0.90\% mean).}
\end{figure}

\emph{LEBT.}  The low-energy test exercises the dc-beam paths: the
envelope solver in continuous mode and the multiparticle tracker with
the two-dimensional continuous-beam (PICNIC-style) kernel
(Sec.~\ref{sec:spacecharge}).  The deck's
\texttt{SPACE\_CHARGE\_COMP} cards apply sectioned neutralisation
(fully compensated near the source, full space charge downstream),
identically interpreted by both codes.  Against the TraceWin envelope
reference, the HELIX envelope agrees along $s$, on the
arc-length-weighted convention of Table~\ref{tab:tracewin}, to 0.86\%
in both transverse planes with space charge at \SI{5}{\milli\ampere}
(the strips of Fig.~\ref{fig:tw_lebt} give 0.77\% unweighted at
HELIX's distinct recorded positions, and the documented 60-point uniform comparison
grid 0.90\%, with a 3.87\% maximum); at the exit the space-charge envelope gives
$\sigma_x = \SI{2.306}{\milli\meter}$ versus TraceWin's
\SI{2.295}{\milli\meter}.  The multiparticle runs agree to 1.5--1.8\%
mean along $s$ (Table~\ref{tab:tracewin}; 1.6--1.8\% on the
unweighted figure grid).

\begin{figure*}[t]
\includegraphics[width=\textwidth]{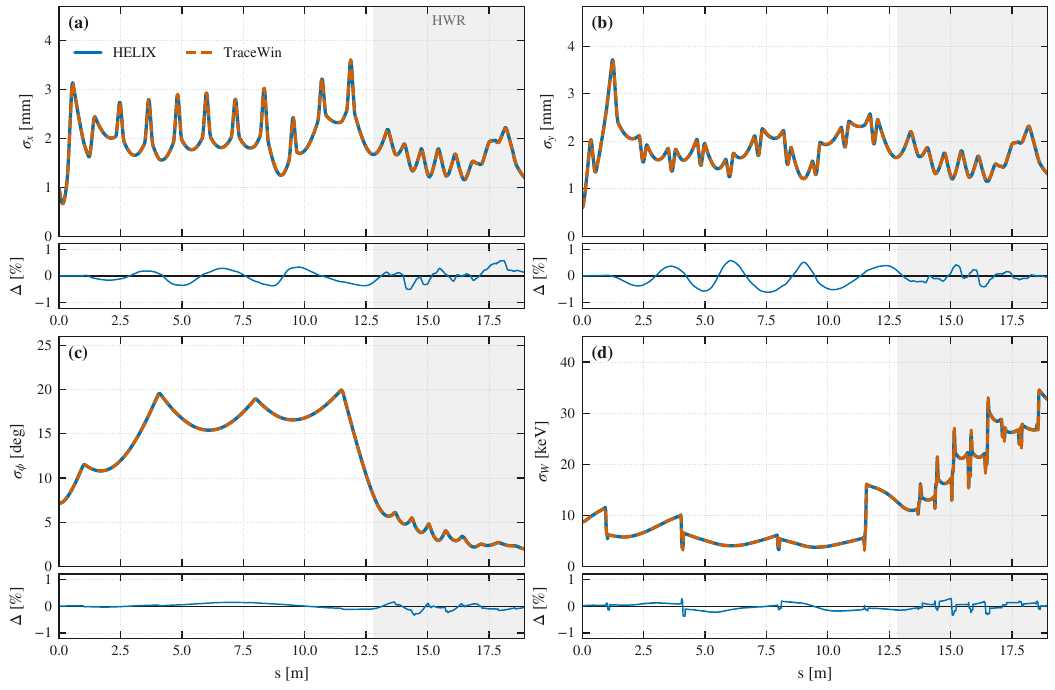}
\caption{\label{fig:tw_env}%
RMS beam sizes along the PIP-II MEBT and HWR cryomodule
($\mathrm{H}^-$, \SI{2.12}{\mega\electronvolt} injection,
\SI{5}{\milli\ampere}, space charge on): HELIX envelope solver
(solid) versus TraceWin envelope mode (dashed).  (a)~Horizontal and
(b)~vertical rms size, (c)~rms phase spread, (d)~rms energy spread;
the shaded band marks the HWR section ($s = 12.8$--\SI{18.96}{\meter})
in which the beam is accelerated to \SI{10.26}{\mega\electronvolt}.
Each strip shows the relative deviation $(\mathrm{HELIX}-\mathrm{TW})
/\mathrm{TW}$ evaluated at HELIX's 1074 distinct recorded positions
with the 28\,947-point TraceWin export interpolated onto them.  Mean
$|\Delta|$ is 0.18\% ($\sigma_x$), 0.24\% ($\sigma_y$), 0.07\%
($\sigma_\phi$), and 0.11\% ($\sigma_W$); exit deviations are
$+0.13\%$, $-0.05\%$, $-0.05\%$, and $+0.03\%$.}
\end{figure*}

\begin{figure*}[t]
\includegraphics[width=\textwidth]{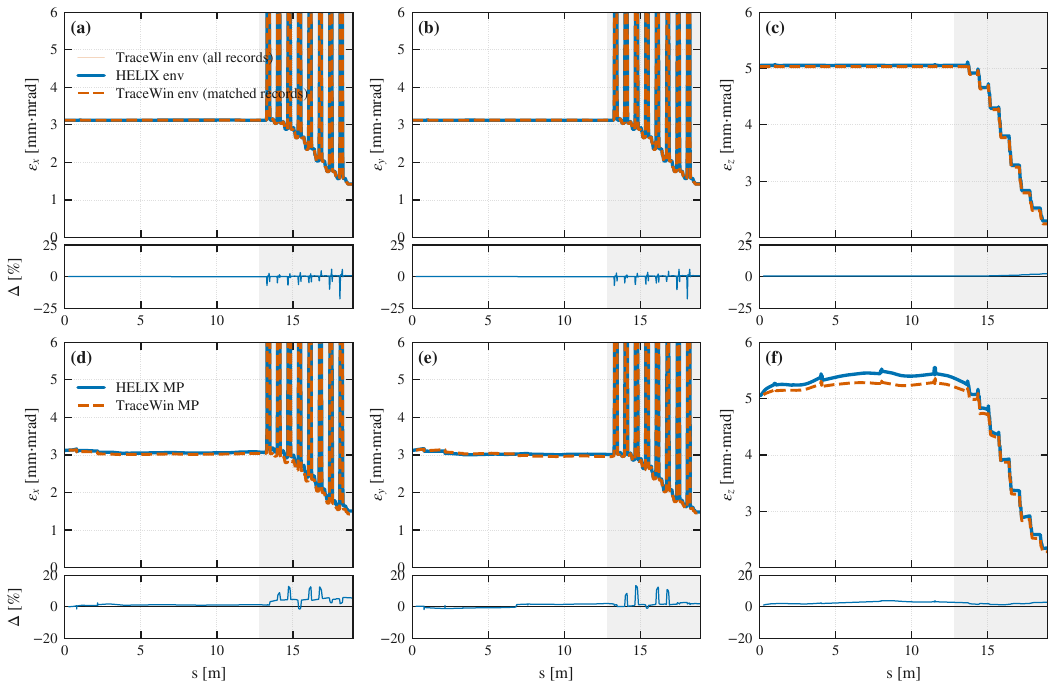}
\caption{\label{fig:tw_emit}%
Geometric rms emittances along the MEBT+HWR line at
\SI{5}{\milli\ampere} with space charge: HELIX (solid) versus
TraceWin (dashed), in envelope mode (a)~$\varepsilon_x$,
(b)~$\varepsilon_y$, and (c)~$\varepsilon_z$ (top row) and in
multiparticle mode (d)~$\varepsilon_x$, (e)~$\varepsilon_y$, and
(f)~$\varepsilon_z$ (bottom row).
The instantaneous transverse geometric emittance is
not an invariant inside the HWR gaps: mid-gap records transiently
spike to ${\approx}21$~mm\,mrad (axes clipped).  Both codes resolve
these transients in both modes, and the spikes themselves overlay.
The envelope strips are set by the sampling of that mid-gap structure
rather than by disagreement where both codes resolve it: at the 248
positions their records share to within \SI{10}{\micro\meter}
(TraceWin's envelope-emittance export resolves 621 such positions,
HELIX 1059), the transverse deviations average 0.06\% and reach
0.64\%, whereas the excursions to
18\% fall only where that sparser export is interpolated across a gap
inside which the instantaneous emittance more than doubles in
\SI{18}{\milli\meter}.  The longitudinal emittance differs more,
averaging 0.54\% at the shared positions.  Exit emittances, at a
position both codes record, agree to $+0.63\%$, $+0.64\%$, and
$+2.25\%$.}
\end{figure*}

\emph{MEBT+HWR.}  This is the primary bunched-beam benchmark, and the
comparison is mode-matched: HELIX envelope versus TraceWin envelope,
and HELIX PIC versus TraceWin partran multiparticle.  With space
charge, the envelope solver agrees with TraceWin along the full
\SI{18.96}{\meter} line to mean absolute deviations of 0.18\%
($\sigma_x$), 0.24\% ($\sigma_y$), 0.07\% ($\sigma_\phi$), and 0.11\%
($\sigma_W$), with a worst single point of 0.61\%
(Fig.~\ref{fig:tw_env}).  The four exit rms moments differ by at most
0.13\%, and the exit geometric emittances agree to
$\varepsilon_x = 1.427$ versus 1.418,
$\varepsilon_y = 1.429$ versus 1.420, and
$\varepsilon_z = 2.294$ versus \SI{2.244}{\milli\meter\milli\radian}
(Fig.~\ref{fig:tw_emit}).
The multiparticle comparison (Fig.~\ref{fig:tw_mp}), evaluated at all
$15\,607$ recorded points, gives mean absolute deviations of 3.87\%,
2.58\%, 3.18\%, and 2.69\% in $\sigma_x$, $\sigma_y$, $\sigma_\phi$,
and $\sigma_W$, exit deltas of $+4.8$\%, $+3.4$\%, $+4.8$\%, and
$-2.3$\%, and 97.76\% transmission.  The maxima of 26\% ($\sigma_x$)
and 15\% ($\sigma_y$) are localized at $s < \SI{0.2}{\meter}$, inside
the leading drift, where partran's export runs straight between
diagnostic planes and so chords across the curvature there.
TraceWin's envelope export does resolve it; as a cross-mode
check, HELIX's multiparticle sizes sit within 0.16\% ($\sigma_x$)
and 0.04\% ($\sigma_y$) of it at those positions, which bounds how
much of the apparent maximum the export sampling accounts for rather
than establishing partran parity.  Excluding the leading
\SI{250}{\milli\meter} the maxima are 12.0\% and 9.4\%, and the
99th-percentile deviations are below 9.2\%.  (The benchmark lattice file carries \texttt{ERROR\_*}
directives, but these define statistical error \emph{studies} and are
inert in the nominal runs of both codes: TraceWin applies them only
in error runs, and HELIX's nominal solvers likewise only store them
for its Monte Carlo machinery.  No random machine errors therefore enter
this comparison, and both codes sample their input bunches from the same
rms Twiss parameters.)  The remaining few-percent multiparticle offset (e.g.,
$\varepsilon_x = 1.506$ versus partran's 1.427 at the exit) is consistent with
the documented $\sim$3\% unresolved systematic difference between the
two codes' space-charge treatments, a stable
overshoot of HELIX relative to partran in the code's documented
section-level benchmarks at \SI{5}{\milli\ampere}.  The recorded
$\sigma_x$/$\sigma_y$ rms residuals are below 3\% for the MEBT alone
and near 3\% for both MEBT+HWR and MEBT+HWR+SSR1+SSR2, with matching
transmission at every stage.

To test whether that residual is instead an artifact of the two codes
sampling their own bunches, the multiparticle run was repeated on
TraceWin's own input distribution, read from the binary \texttt{.dst}
that seeded the reference.  The file is authoritative for energy,
frequency, current, emittance and Twiss, and its rms parameters
reproduce the specification of Table~\ref{tab:twconfig} to
$3.4\times10^{-6}$ or better transversely and $2.4\times10^{-5}$
longitudinally, so the two runs start from the same rms beam.  The
generators differ in shape: TraceWin's bunch is an untruncated
Gaussian and carries the expected $6\times10^{-5}$ population beyond
$4\sigma$ that HELIX's cutoff removes.  Tracking it changes the
along-$s$ agreement very little.  Using its first $10^4$ particles the
mean absolute deviations are 3.39\%, 2.24\%, 2.66\% and 1.96\% in
$\sigma_x$, $\sigma_y$, $\sigma_\phi$ and $\sigma_W$; using all
$10^5$ they are 3.23\%, 2.69\%, 3.14\% and 2.89\%, against 3.87\%,
2.58\%, 3.18\% and 2.69\% for the independently sampled run.  No
moment moves by more than 0.8 percentage points, and the shifts are
not systematically toward closer agreement: changing the particle
count within the same file moves them as much as changing the
generator does.  Transmission is 97.84\% and 97.56\% against 97.76\%,
and the entrance maximum is unmoved at 26.3--26.4\%, as expected for a
feature set by the partran export's sampling rather than by either
input bunch.  The few-percent multiparticle residual is therefore not
an artifact of HELIX's input sampling or of its $4\sigma$ truncation.

The exit reference energies, by contrast, agree to better than
$10^{-4}$ relative in both modes (10.2614 and
\SI{10.2625}{\mega\electronvolt} for envelope and multiparticle
against TraceWin's 10.2622 and \SI{10.2624}{\mega\electronvolt}),
confirming that the rf energy-gain model is essentially exact; the
space-charge treatments, together with the codes' different meshes
and the $\sim$1\% shot-noise floor
of $10^4$ macroparticles, remain the leading candidates for
the transverse residuals above, though the cause has not been
isolated.

\begin{figure*}[t]
\includegraphics[width=\textwidth]{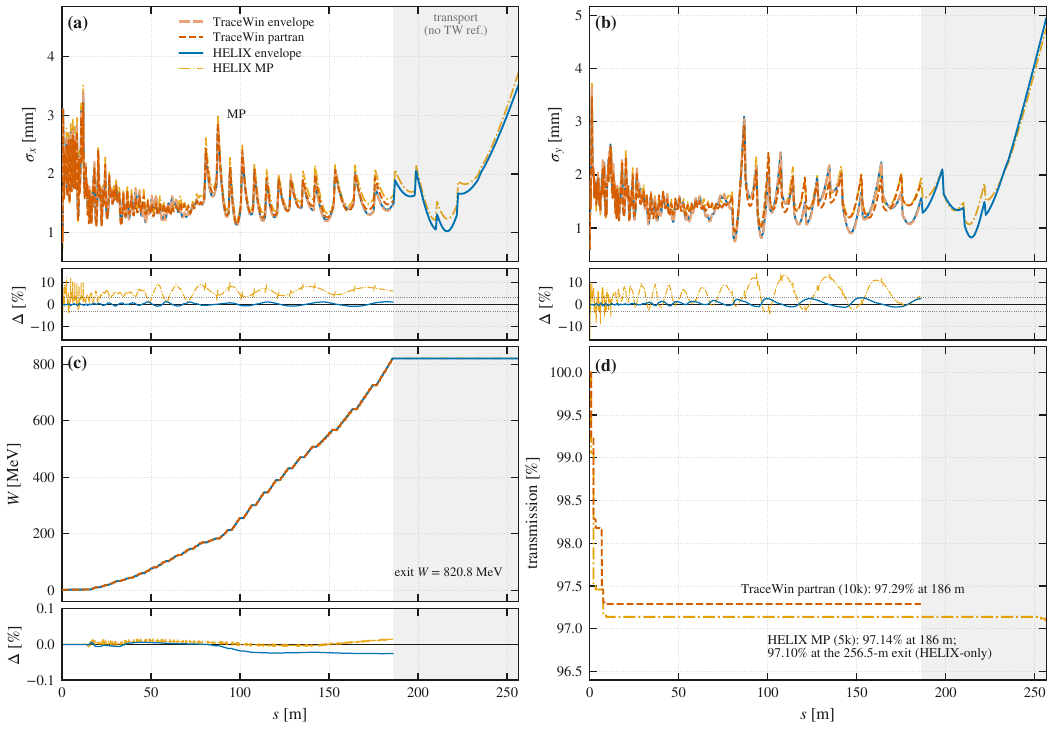}
\caption{\label{fig:pipii}%
PIP-II accelerating linac and modeled downstream transport
(\SI{256.5}{\meter} in total, $\mathrm{H}^-$ from 2.12 to
\SI{821}{\mega\electronvolt}, \SI{5}{\milli\ampere}, space charge on),
re-run for this paper: (a)~horizontal and (b)~vertical rms beam size
for the HELIX envelope solver (solid), HELIX multiparticle mode
(dash-dotted; 5000 macroparticles, TSC deposition, $48^3$ PIC grid),
and both TraceWin references, the mode-matched envelope solution
(long-dashed, lighter) and partran (short-dashed, $10^4$ particles).
Strips show the relative residual
$(\mathrm{HELIX}-\mathrm{TW})/\mathrm{TW}$, each mode against its own
plotted reference: the envelope residual is taken with respect to the
TraceWin envelope solution and the multiparticle residual with respect
to partran, with the $\pm3\%$ band dotted.  Rms residuals over the
accelerating linac (first and last meter trimmed) are 0.5\%/1.0\%
(envelope) and 5.2\%/5.3\%
(multiparticle) in $x$/$y$.  (c)~Kinetic energy along the machine;
HELIX matches TraceWin to within 0.03\% everywhere.
(d)~Transmission: at the common \SI{186}{\meter} endpoint HELIX MP
transmits 97.14\% versus partran's 97.29\%---a difference of 0.15
percentage points against a combined binomial uncertainty of 0.29
points for the two finite ensembles (5000 and $10^4$ particles); the HELIX curve continues to 97.10\%
at the \SI{256.5}{\meter} exit.  The
available mode-matched TraceWin references track the accelerating
linac ($s < \SI{186}{\meter}$); the shaded downstream transport
carries HELIX curves only.}
\end{figure*}

\emph{Full machine.}  The end-to-end benchmark tracks the beam through
the complete MEBT--HWR--SSR1--SSR2--LB650--HB650 sequence, including
the frequency jumps from 162.5 to 325 to \SI{650}{\mega\hertz}
(Fig.~\ref{fig:pipii}).  The superconducting linac accelerates to a
full \SI{821}{\mega\electronvolt}; the final
$\sim$\SI{21}{\mega\electronvolt} is operational energy overhead, and
the machine is run to deliver the nominal \SI{800}{\mega\electronvolt}
of the intro, the energy for which the downstream beam-transfer line
and the stripper-foil study of Sec.~\ref{sec:applications} are
designed and tuned.  The benchmark proper is the mode-matched
comparison over the accelerating linac, $0 \leq s \leq
\SI{186}{\meter}$, the range covered by both fresh TraceWin
references (lattice file
\texttt{mebt+hwr+ssr1+ssr2+lb650+hb650.dat}, tracked in the paper's
figure-provenance records): HELIX envelope against the TraceWin envelope solution
and HELIX multiparticle ($5000$ particles, TSC kernel, $48^3$ grid)
against partran.  Over that range the per-step rms residuals are
0.51\%/0.98\% (envelope, $\sigma_x$/$\sigma_y$) and 5.22\%/5.31\%
(multiparticle), both evaluated on the common grid over
$1 \leq s \leq \SI{184.99}{\meter}$ (one meter trimmed at each end),
with output energy matching TraceWin to within
0.03\% everywhere and transmission at the common \SI{186}{\meter}
endpoint of 97.14\% versus partran's 97.29\% (at its own
\SI{256.5}{\meter} exit, downstream of any reference, HELIX
transmits 97.10\%); the weighted single-convention
statistics of Table~\ref{tab:tracewin} give the same picture.  The
nominal ${\sim}2.8\%$ interception is a design feature, not a
numerical artifact: in both codes the losses occur almost entirely at
the two MEBT collimators (aperture cards at $s = 0.78$ and
\SI{2.16}{\meter}), which scrape the sampled Gaussian tails at
identical locations, and the superconducting linac downstream of
$s \approx \SI{9}{\meter}$ is loss-free in HELIX to 0.04 percentage
points.
Beyond \SI{186}{\meter} the curves are HELIX-only modeled downstream
transport with no TraceWin reference (shaded in
Fig.~\ref{fig:pipii}).  For continuity with the code's documentation,
the documented full-machine \emph{regression record} (a cross-mode
comparison of both HELIX modes against the same partran run over the
full \SI{256.5}{\meter}, predating the mode-matched
references) lists per-step rms residuals of
1.90\%/2.29\% (envelope) and 8.56\%/13.31\% (multiparticle);
re-running it for this
paper at the release version gives
envelope exit sizes of 3.521 and \SI{4.948}{\milli\meter}
($-0.9$\%/$+0.2$\% versus partran) and the
multiparticle
signature within
seed-level statistics.  Re-running
the multiparticle benchmark at TraceWin's own $10^4$ particle count
moves the agreement slightly toward the reference (per-step rms
$5.22\rightarrow4.50\%$ in $\sigma_x$; transmission 97.30\% versus
partran's 97.29\% at the common \SI{186}{\meter} endpoint, 97.28\% at
the \SI{256.5}{\meter} exit), with all shifts inside the $\sim$1\% statistical
floor---the 5000-particle benchmark is not particle-count limited.  The larger multiparticle
residuals reflect the $\sim$1\% statistical floor at 5000 particles,
the unresolved systematic difference noted above, and the fact that
partran applies its
own space-charge algorithm with independent mesh choices.  The
longitudinal comparison requires care with conversion
conventions: HELIX reports $\sigma_\phi$ in degrees at the local RF
frequency (\SI{650}{\mega\hertz} at the linac exit), TraceWin at the
\SI{162.5}{\mega\hertz} bunch frequency, so the codes are compared in
physical bunch length,
$\sigma_z = \sigma_\phi\beta\lambda/360^\circ$, with each code's own
wavelength ($\beta = 0.846$ there).  At the \SI{186}{\meter} exit of
the accelerating linac, the last point covered by the mode-matched
reference, this gives $\sigma_z = 1.073$ versus
\SI{1.080}{\milli\meter} ($-0.6\%$) in envelope mode and 0.895 versus
\SI{0.903}{\milli\meter} ($-0.9\%$) in multiparticle mode.  Applying
the same single convention to the documented full-machine benchmark
table, whose entries mix the two conventions, turns its apparent
$+22\%$ entry into 6.64 versus \SI{6.44}{\milli\meter} ($+3.2\%$) at
the \SI{256.5}{\meter} exit, after the bunch has lengthened by
${\sim}6\times$ in the rf-free downstream transport, in line with
the few-percent agreement of the other full-machine quantities rather
than an anomaly.

\emph{Validated scope.}  This section establishes the following, at
the stated levels: analytic envelope benchmarks close at the
$10^{-8}$ level and cross-implementation PIC parity at
$10^{-9}$--$10^{-13}$ across the FP64 implementations, with the FP32
Metal backend held to $5\times10^{-6}$ on the solved field
(Sec.~\ref{sec:scverify}); against TraceWin,
rms beam \emph{sizes} are validated to sub-percent mean deviation in
envelope mode (worst recorded point 3.8\%, in the LEBT's steepest
converging section) and to
the few-percent level in multiparticle mode (worst localized 26\%
where the partran reference is interpolated across an entrance waist,
14\% beyond that drift; Table~\ref{tab:tracewin}), output
\emph{energies} to within 0.03\%, \emph{transmission} to
${\sim}0.1$ percentage point at the common endpoint, and $\sigma_z$
at the accelerating-linac exit to ${\sim}1\%$.  Geometric \emph{emittances} carry
weaker claims: the envelope pair agrees to ${\sim}0.6\%$
transversely at the MEBT+HWR exit but to only $2.2\%$
longitudinally, the multiparticle pair differs by the unresolved
few-percent systematic, and the convergence stress case shows
emittances not mesh-converged at fixed particle count---rms-size,
energy, and transmission agreement, not emittance agreement, is the
validated claim of this section.

\begin{table*}[t]
\caption{\label{tab:tracewin}%
Agreement between HELIX and TraceWin on the three PIP-II benchmark
lattices at \SI{5}{\milli\ampere} with space charge on.
For each row the relative deviation
$e(s)=[\sigma_{\mathrm{HELIX}}(s)-\sigma_{\mathrm{TW}}(s)]/
\sigma_{\mathrm{TW}}(s)$ is evaluated at HELIX's recorded $s$-points
over the $s$-range common to both codes' outputs, with the TraceWin
record interpolated onto them; the rms and mean columns are
arc-length-weighted (trapezoidal weights, so substep-dense regions
are not over-counted), the max column is the largest $|e|$ over
points where the reference exceeds 5\% of its median (a near-zero
guard that in practice excludes no points), and the exit column is
the signed $e$ at the physical common endpoint of the two records,
which can differ from each
code's final logged record.  These single-convention table values can
differ slightly from the per-figure strip statistics, which are
unweighted.  Every row is mode-matched: envelope comparisons use
TraceWin envelope references and multiparticle (MP) comparisons use
TraceWin partran references.}
\begin{ruledtabular}
\begin{tabular}{llcccccccc}
& & \multicolumn{3}{c}{$\Delta\sigma_x$ along $s$ (\%)} &
\multicolumn{3}{c}{$\Delta\sigma_y$ along $s$ (\%)} &
\multicolumn{2}{c}{Exit $\Delta$ (\%)} \\
Lattice & Mode & rms & mean & max & rms & mean & max &
$\sigma_x$ & $\sigma_y$ \\
\colrule
LEBT             & Envelope & 1.23 & 0.86 & 3.79 & 1.23 & 0.86 & 3.78 & $+0.47$ & $+0.51$ \\
                 & MP       & 2.11 & 1.48 & 6.78 & 2.45 & 1.83 & 6.93 & $+5.76$ & $+5.72$ \\
MEBT+HWR         & Envelope & 0.22 & 0.18 & 0.57 & 0.29 & 0.24 & 0.61 & $+0.13$ & $-0.05$ \\
                 & MP       & 4.55 & 3.55 & 26.3 (12.0)\footnotemark[1] & 3.38 & 2.63 & 15.2 (9.4)\footnotemark[1] & $+4.84$ & $+3.42$ \\
PIP-II acc.\ linac & Envelope\footnotemark[3] & 0.60\footnotemark[2] & 0.50\footnotemark[2] & 1.23\footnotemark[2] & 1.21\footnotemark[2] & 0.94\footnotemark[2] & 2.99\footnotemark[2] & $+1.04$\footnotemark[2] & $+2.57$\footnotemark[2] \\
(\SI{186}{\meter}) & MP     & 5.81\footnotemark[2] & 5.45\footnotemark[2] & 26.4 (12.1)\footnotemark[1]\textsuperscript{,}\footnotemark[2] & 6.09\footnotemark[2] & 4.81\footnotemark[2] & 15.3 (13.8)\footnotemark[1]\textsuperscript{,}\footnotemark[2] & $+5.99$\footnotemark[2] & $+3.98$\footnotemark[2] \\
\end{tabular}
\end{ruledtabular}
\footnotetext[1]{Maximum localized at $s < \SI{0.2}{\meter}$, in the
leading drift, where TraceWin's partran export interpolates linearly
between diagnostic planes and chords across the envelope curvature
there (over that drift its rows are collinear to within the export's
seven-digit round-off), so the tabulated maximum is set by the export cadence
rather than by a resolved disagreement.  Excluding the leading
\SI{250}{\milli\meter}, the
MEBT+HWR maxima fall to 12.0\% and 9.4\% and the full-linac maxima
to 12.1\% ($\sigma_x$) and 13.8\% ($\sigma_y$), the latter at
$s = \SI{93.7}{\meter}$ and so unrelated to the entrance sampling.  For MEBT+HWR the 99th-percentile
deviations are 8.5\% ($\sigma_x$) and 6.7\% ($\sigma_y$).}
\footnotetext[2]{From this paper's re-run at the documented benchmark
configuration; the common $s$-range ends at the TraceWin export's end,
$s = \SI{186.0}{\meter}$ (the accelerating-linac exit), so the exit
column is the linac exit.  The documented full-machine
\emph{regression record} (cross-mode against partran over
\SI{256.5}{\meter}: envelope 1.90\%/2.29\%, MP 8.56\%/13.31\% over
$s\in[\SI{1}{\meter}, L-\SI{1}{\meter}]$) is quoted, so labeled, in
the text.}
\footnotetext[3]{Mode-matched: this row compares the HELIX envelope
against the fresh TraceWin \emph{envelope} reference plotted in
Fig.~\ref{fig:pipii} (the earlier documented full-machine comparison,
which predated that reference, was cross-mode against partran and is
quoted separately in the text).}
\end{table*}

\clearpage
\section{Applications}
\label{sec:applications}

Four studies show the validated physics of
Sec.~\ref{sec:validation} and the optimization machinery of
Sec.~\ref{sec:matching} doing design work: the envelope-first loop on
the PIP-II front end, gradient-based matching through the
differentiable space-charge model, misalignment Monte Carlo with orbit
correction, and stripper-foil transport in the Booster transfer line.
They differ in evidentiary weight, and the distinction is worth
stating at the outset.  The matching study of
Sec.~\ref{sec:app_gradmatch} and the foil study of
Sec.~\ref{sec:app_foil} are quantitative results, the first an
optimizer comparison with measured cost scaling and the second a
per-particle verification of the implemented interaction models.  The
workflow of Sec.~\ref{sec:app_frontend} demonstrates infrastructure
rather than a converged design study, and is labeled as such; the
error study of Sec.~\ref{sec:app_errors} ships and audits a
full-machine template and quantifies the corrected workflow in a
25-seed paired corrected--uncorrected ensemble on the front end, short
of a tolerance study but no longer a bare demonstration.  A closing subsection surveys
the remaining design-study tools (parameter scans, failure studies,
the error-directive surface, Pareto exploration, and the surrogate
workflow) on the same shipped lattices.

\subsection{Envelope-first workflow and card-driven matching}
\label{sec:app_frontend}

\begin{figure}[t]
\includegraphics[width=\linewidth]{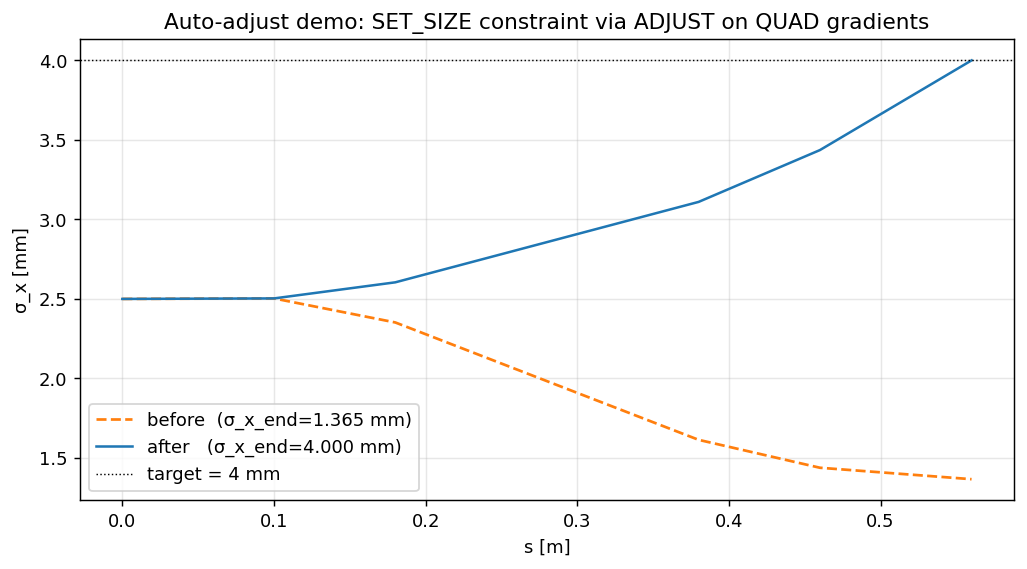}
\caption{\label{fig:matchdemo}%
Card-driven matching demo shipped with the code: $\sigma_x(s)$ before
(dashed) and after (solid) matching on a \SI{0.56}{\meter} two-quadrupole
cell at zero current, whose lattice file carries two link-grouped
\texttt{ADJUST} cards
on the quadrupole gradients and one \texttt{SET\_SIZE} card requesting
$\sigma_x = \SI{4}{\milli\meter}$ at the exit (dotted line).  The
default least-squares matcher moves the exit size from 1.365 to
\SI{4.000}{\milli\meter}.}
\end{figure}

The reference line for this workflow is the MEBT+HWR line of
Sec.~\ref{sec:tracewin} (\SI{18.96}{\meter} from 2.12 to
\SI{10.26}{\mega\electronvolt} at \SI{5}{\milli\ampere}
$\mathrm{H}^-$), distributed with the code as a TraceWin-format lattice
file containing the quadrupole doublets and triplets, buncher and HWR
field-map cavities and solenoids, aperture cards, and BPM markers of
the machine.  The intended
workflow is envelope-first: design iterations run against the rms
envelope solver with space charge (Sec.~\ref{sec:envelope}), which on
this line costs \SI{12.7}{\second} against \SI{109}{\second} for a
$10^4$-macroparticle PIC pass at the benchmark's $64^3$ mesh, a factor
of 8.5 that widens with macroparticle count (a separate
$10^5$-particle run of the same line, tracking TraceWin's own input
distribution, costs 5.3 times its own $10^4$ pass in the same process,
499 against \SI{94}{\second}), and which
Sec.~\ref{sec:tracewin} showed follows TraceWin along this line to
0.18\% ($\sigma_x$) and 0.24\% ($\sigma_y$) mean absolute deviation.
Decisions made at envelope fidelity are then confirmed at PIC fidelity,
where the same lattice reproduces TraceWin partran to a few percent
with 97.76\% transmission, so the expensive mode is a confirmation
step, not the inner loop.

Matching problems travel with the lattice as \texttt{SET}/\texttt{ADJUST}
cards (Sec.~\ref{sec:matching}), and two shipped demos exercise this
path end to end on compact demo cells.  The forward demo
(Fig.~\ref{fig:matchdemo}) loads a \SI{0.56}{\meter} two-quadrupole
cell (\SI{3}{\mega\electronvolt} proton beam, zero current) whose file carries two
\texttt{ADJUST} cards on the quadrupole gradients sharing a link
group (so the pair moves as a single ganged variable) plus one
\texttt{SET\_SIZE} constraint; a single \texttt{match()} call drives
the exit $\sigma_x$ from 1.365 to \SI{4.000}{\milli\meter} against a
\SI{4}{\milli\meter} target.  The inverse demo declares the
\emph{input} beam as the unknown: an \texttt{ADJUST\_BEAM\_TWISS} card
exposes the injected $\alpha_{x,y}$, $\beta_{x,y}$ and a
\texttt{SET\_TWISS} card pins the exit optics to $\alpha=0$,
$\beta=1.5$ in both planes, recovering the injection optics that
produce a prescribed downstream condition.  The demo lattices carry
these cards, and the benchmark file of Sec.~\ref{sec:tracewin} does
not; equipping a design lattice is scripted rather than manual, and
the repository ships this line already equipped (sixty-three
\texttt{ADJUST} cards covering every quadrupole, cavity and solenoid,
closed by a
four-dimensional emittance-growth objective and an exit-energy
floor), together with the utility that generates it from a parsed
deck.  Those two closing cards are envelope-mode matcher constraints,
outside the differentiable path of Table~\ref{tab:gradscope}.  The card syntax
and the engine are the same at either fidelity.  What these two demos
establish is the mechanism, not a front-end design decision:
Sec.~\ref{sec:app_gradmatch} takes the same engine to a matching
problem where the result, rather than the plumbing, carries the
argument.

\subsection{Gradient-based matching through space charge}
\label{sec:app_gradmatch}

\begin{figure*}[t]
\includegraphics[width=\textwidth]{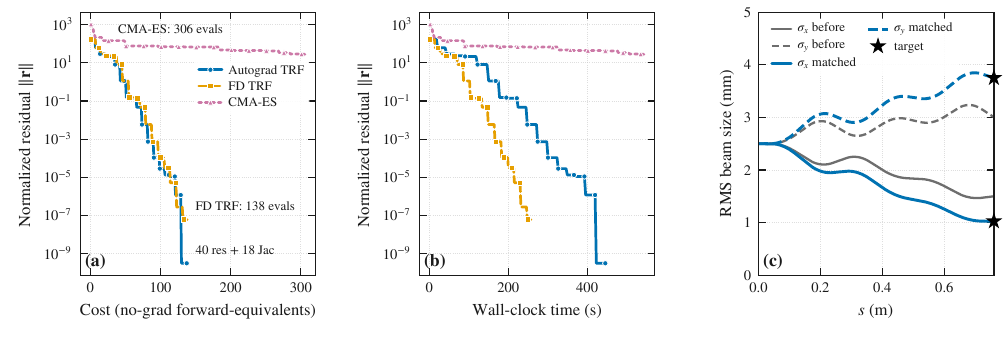}
\caption{\label{fig:gradmatch}%
Matching through the nonlinear PIC space-charge model: exact-Jacobian
trust-region-reflective least squares (autograd TRF,
Sec.~\ref{sec:diffpic}) versus a
finite-difference TRF and a CMA-ES baseline, all minimizing the
identical nondimensionalized residual
$r_i=(q_i-q_i^\star)/\tau_i$ with $\tau_\alpha=0.01$ and
$\tau_\sigma=\SI{10}{\micro\meter}$ (the convention of
Fig.~\ref{fig:scaling}), from the same bunch, seed, starting point,
and bounds.
(a)~Best-so-far residual norm versus cumulative cost in no-gradient
forward-equivalents: every finite-difference and CMA-ES evaluation is
one forward pass, while each autograd plain residual and each autograd
Jacobian is charged its measured cost in units of the same run's
no-gradient forward pass (1.0 and 5.5 here), so one Jacobian advances
the autograd curve by 5.5 units.  (b)~The
same versus wall-clock time.  (c)~rms beam sizes $\sigma_x$,
$\sigma_y$ along the section before (gray) and after (blue) matching,
with the target exit sizes starred.}
\end{figure*}

The second case study is matching \emph{through} the nonlinear
space-charge dynamics rather than around them.  The benchmark problem
is a \SI{0.76}{\meter}, six-quadrupole MEBT-style matching section
(\SI{80}{\milli\meter} quadrupoles, \SI{40}{\milli\meter} drifts)
carrying a \SI{5}{\milli\ampere}, \SI{3}{\mega\electronvolt} proton
bunch at \SI{352.21}{\mega\hertz}.  Six \texttt{ADJUST} cards expose
the six gradients (bounds $\pm\SI{15}{\tesla\per\meter}$), starting
from a setting detuned by up to 1.7~\si{\tesla\per\meter} in both
directions; \texttt{SET\_TWISS} and \texttt{SET\_SIZE} cards impose
four exit constraints ($\alpha_x$, $\alpha_y$, $\sigma_x$,
$\sigma_y$).  The benchmark is a manufactured-solution verification
test: the targets are the exit optics that the same model produces at
a generating setting, so a zero-residual solution exists by
construction and the optimizers can be compared on how completely
they find one.  With four constraints on six knobs the zero-residual set is
generically underdetermined, and the optimizer's endpoint
indeed differs from the generating setting (final gradients
$(6.35, -8.21, 7.96, -5.53, 5.27, -5.56)$ against generating
$(6.5, -7.5, 7.0, -6.8, 7.2, -6.0)$~\si{\tesla\per\meter}), so the
test verifies the optimizer and the differentiable model on a
problem the model itself generated; how the Jacobian-cost advantage
scales with the knob count is taken up below.
The forward model is exactly the one
\texttt{match(algorithm="gradient", space\_charge=True)} evaluates
internally: a fixed-seed 1500-macroparticle bunch tracked through the
differentiable stepwise tracker with the $32^3$ PyTorch PIC solver
(adaptive grid extent, CIC deposit, IGF Green function) applied inside
every element.  The benchmark wraps that engine rather than calling it
bare: the production matcher minimizes card residuals in native units
times their card weights, so a thin wrapper divides each component by
$\tau_i$ before the engine's loop sees it, putting the autograd,
finite-difference and CMA-ES runs on one common objective.  The
comparison below is therefore between three optimizers on an identical
residual, not a demonstration of the default weighting.

The gradient path (Sec.~\ref{sec:diffpic}; scope in
Table~\ref{tab:gradscope}) wraps this residual in a
bound-constrained trust-region-reflective least-squares loop whose
Jacobian is computed
by reverse-mode automatic differentiation through the entire
cycle (deposit, FFT Poisson solve, gather, kick, and transport), with
checkpointing to bound memory.  Measured in-process against the same
run's no-gradient forward pass, one Jacobian costs 5.5
forward-equivalents (Sec.~\ref{sec:diffcost}), set by the
four residual components rather than by the number of knobs; a
finite-difference Jacobian instead costs one forward pass per
knob.

From an initial normalized residual norm of 170, the autograd TRF
finds a matched solution: it converges on its step-size
criterion to a norm of $3.2\times10^{-10}$ in 40 residual plus 18
Jacobian evaluations, at a
mean of \SI{3.2}{\second} per residual evaluation (about
seven and a half
minutes end to end on the Apple M3 Max laptop CPU of
Sec.~\ref{sec:performance}), and a repeated run reproduces the final
knobs, the residual vector and cost, the evaluation counts, and the
residual- and Jacobian-norm histories bit for bit, so matches are
exactly reproducible on fixed hardware, library versions, thread
count, and backend.  The nondimensionalization makes the norm
invariant under unit changes: a tolerance on it bounds each $\alpha$
mismatch in units of $0.01$ and each size mismatch in units of
\SI{10}{\micro\meter}.  Physically, the converged
solution reproduces the targets to $\Delta\alpha_x =
-1.2\times10^{-12}$ and $\Delta\alpha_y = -9.7\times10^{-14}$, with
size errors $\Delta\sigma_x = 2.7\times10^{-9}$ and
$\Delta\sigma_y = 1.0\times10^{-9}$~\si{\micro\meter}.  A finite-difference least-squares baseline on
the identical residual, start, bounds, and tolerances stops on its
step-size criterion at a norm of $6.2\times10^{-8}$, a factor of
about 200 short of the autograd solution, after 138
forward
evaluations (36 main iterations plus 17 six-point Jacobians): the two
trust-region runs track each other closely while both proceed, their
residual norms after the first accepted step agreeing to
$2.1\times10^{-8}$ relative, and the finite-difference run then stops
one accepted iteration earlier.  Both report the same step-size
termination condition; the archive records residual norms and final
knobs rather than per-iterate steps, so it does not resolve why that
condition is met sooner.  On the forward-equivalent axis of
Fig.~\ref{fig:gradmatch}(a) the two terminate at essentially the same
budget (138.3 versus 138 forward-equivalents), each counted in units
of its own forward pass, so at six knobs the autograd advantage is
\emph{depth at an equal forward-equivalent budget} (from many fewer
top-level model evaluations, 58 against 138, but a costlier one each),
not a wall-clock win.
Those two units are not the same size, for a reason outside the
physics: a no-gradient forward timed in the gradient run's own process
costs \SI{3.25}{\second}, while the finite-difference run evaluates the
identical residual at \SI{1.87}{\second} per call.  Graph recording is
not the driver: in the gradient run the optimizer's forward and a
\texttt{torch.no\_grad} forward of the same residual agree to
0.35\%, and that run's own first forward, taken before any Jacobian,
cost \SI{1.97}{\second}.  The inflation is consistent with allocator and
cache state left behind by the checkpointed Jacobian builds
(Sec.~\ref{sec:diffcost}), and it is what the wall clock below
reflects.  In wall clock
the FD baseline finished about 42\% sooner (259
versus \SI{449}{\second}; Fig.~\ref{fig:gradmatch}(b)).  Both
endpoints sit far inside the matching tolerances, the
finite-difference norm of $6.2\times10^{-8}$ being a mismatch some
seven orders of magnitude below $\tau_\alpha$ and $\tau_\sigma$, so the
extra depth measures the optimizer rather than deciding whether the
match is usable.  The two properties that carry engineering weight
appear elsewhere and are taken up below: a per-Jacobian cost that does
not grow with the knob count, and a converged setting whose remaining
error is set by the model rather than by the optimizer.  The derivative-free CMA-ES baseline (the engine's option set,
population 9, without its usual least-squares polish) is not
competitive on this smooth landscape: stopped by its evaluation cap
after 306 residual evaluations, its best normalized norm is 27.9,
eleven orders of magnitude above the autograd solution.  It had not
converged and was still improving when the cap ended it (its best
moved from 29.0 to 27.9 over the final 21 evaluations), so 27.9 is
where a capped run reached, not a floor: the comparison shows how far
derivative-free search remains after a fixed budget, and does not
measure its asymptote.  Its role in the matching suite is the
multimodal and plateaued cost landscapes of Sec.~\ref{sec:matching},
not deep convergence on smooth landscapes.

\begin{figure*}[t]
\includegraphics[width=\textwidth]{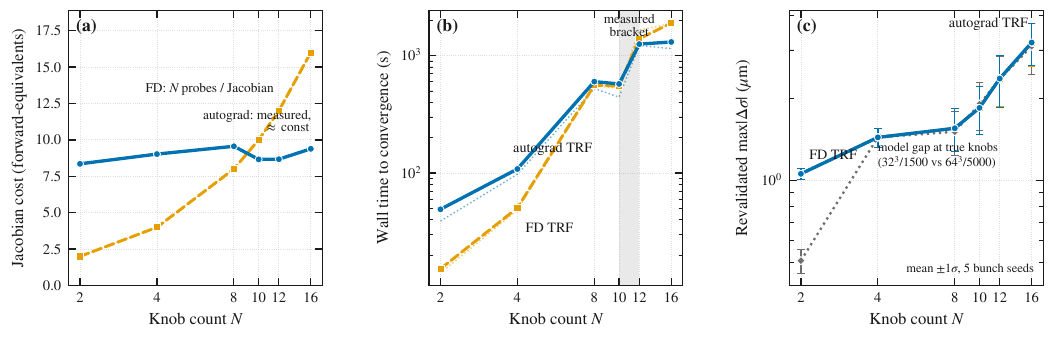}
\caption{\label{fig:scaling}%
End-to-end matching cost across a knob-count benchmark family.  The
matching benchmark of Fig.~\ref{fig:gradmatch} is repeated
on $N=2,4,8,10,12,16$ independent quadrupole-gradient knobs (same
\SI{5}{\milli\ampere}, \SI{3}{\mega\electronvolt} bunch and $32^3$ PIC
model, four fixed exit constraints throughout), with the targets here
generated by a finer $64^3$/5000-particle model at an independent seed,
so no zero-residual solution exists (see text), comparing autograd
trust-region-reflective least squares (blue) against a
finite-difference TRF (orange)
run on the identical differentiable residual from identical starts.
With only the exit optics constrained the problem is under-determined
for $N>4$, so this isolates Jacobian cost and convergence, not a
unique solution.  (a)~Jacobian cost in forward-equivalents: the
reverse-mode Jacobian holds near-constant (\numrange{8.4}{9.6}, set by
the four residual components) while the two-point finite difference
costs exactly $N$ forward probes.  (b)~Wall time to convergence,
measured directly in idle, load-gated runs of the identical
deterministic optimizations (the $N=8$, $N=10$, and $N=12$ points
are means of two repetitions whose spreads do not exceed 6.1\%; log
axes); the finite-difference route wins the clock through the
measured $N=10$ and the reverse-mode route from $N=12$, and at both
bracketing points the two repetitions of the faster method lie
entirely below those of the slower, so the crossover is bracketed
between $N=10$ and $N=12$
(log-log interpolation of the central values gives $N\simeq10.5$;
$1.45\times$ at $N=16$).  The shaded band marks the measured
bracket.  Faint dotted curves: the per-call reconstruction,
which under-predicts the autograd walls by 3--23\% and tracks the
finite-difference walls to within 8\% except at $N=12$, where it
overshoots by 19\%; it misorders the two closest pairs, at $N=8$ and
$N=10$, so it is a sanity check rather than a substitute for
the stopwatch.  (c)~Both solutions re-evaluated against an
independent higher-fidelity model ($64^3$ grid, 5000 macroparticles),
mean $\pm1\sigma$ over five bunch seeds: the worst revalidated size
mismatch $\max|\Delta\sigma|$ tracks the coarse-to-fine model gap
(gray), not the optimizer; the two methods are statistically
indistinguishable, and the mismatch exceeds the revalidation's own
seed-noise floor at every $N$.  All four constrained quantities
($\alpha_x$, $\alpha_y$, $\sigma_x$, $\sigma_y$) were revalidated;
the $\alpha$ mismatches follow the same pattern.}
\end{figure*}

The six-quadrupole benchmark fixes the knob count; the operational
value of the exact Jacobian is how its cost behaves as that count
grows.  Figure~\ref{fig:scaling} repeats the match at
$N=2,4,8,10,12,16$,
holding the beam, the PIC model, and the four exit constraints fixed
and running both methods on the identical differentiable
residual (nondimensionalized as $r_i=(q_i-q_i^\star)/\tau_i$ with
$\tau_\alpha=10^{-2}$ and $\tau_\sigma=\SI{10}{\micro\meter}$) from
identical starts.  Each added knob inserts a quadrupole and a drift, so
the lattice, and with it the absolute per-evaluation cost, grows with
$N$; the forward-equivalent ratio below factors that out.  The targets
are the exit optics the higher-fidelity model ($64^3$ grid, 5000
macroparticles) produces at a chosen generating setting, deliberately
evaluated at a finer grid, more particles, and an independent seed from
the $32^3$/1500 optimizer model so the match is not an inverse crime.  The generating
knobs are therefore \emph{not} the solution (they leave the
coarse-to-fine model gap), and how far the residual is driven varies
with the knob count relative to the four constraints, across that
gap, though each $N$ also changes the lattice, targets and
conditioning, so the knob count is not isolated as the cause.  Two knobs do not steer the four coarse-model outputs onto the
four higher-fidelity targets and floor at a normalized norm of $0.13$; four
knobs (a square system) reproduce the targets to $5.9\times10^{-13}$,
the recovered gradients absorbing the model gap; and beyond four knobs
the match is under-determined and the optimizer stops on its
step-size tolerance short of machine zero ($3.0\times10^{-8}$ at $N=8$,
$9.9\times10^{-7}$ at $N=16$; finite differences reach
$8.4\times10^{-7}$ at $N=16$).

None of this bears on the Jacobian-cost scaling, which is the point
[Fig.~\ref{fig:scaling}(a)]: one reverse-mode Jacobian costs a
near-constant \numrange{8.4}{9.6} forward-pass equivalents across the
whole range (fixed by the four residual components, not the knob
count), whereas a two-point finite difference costs exactly $N$ forward
probes (measured: 2, 4, 8, 10, 12, and 16).  These per-stage ratios are
measured in a fresh process each, normalized by the graph-recording
forward (the scaling harness runs its plain residuals under
\texttt{no\_grad}, so these are timings of that harness rather than of
the production matcher; the harness's own graph-recording and
no-gradient forwards differ by under 2\%); normalized instead by a
no-gradient forward they read \numrange{8.5}{9.7}.  The six-quad
demonstration's self-contained in-process measurement is 5.5 under
either normalization, its two forwards differing by 0.3\%, the
remaining gap to the scan being process state rather than physics
(Sec.~\ref{sec:diffcost}); under either convention the Jacobian cost
is set by the four residual components and is flat in $N$.  In
directly measured wall time [Fig.~\ref{fig:scaling}(b)] (idle,
load-gated runs; at the four knob counts that had been run before, these
were reruns of the identical deterministic optimizations and
reproduced every archived residual norm, evaluation count and final
knob set bit for bit, so there the stopwatch was the only new
information; $N=8$, $N=10$ and $N=12$ were each measured twice, the
second an idle-gated repetition that reproduced the first's
optimization bit for bit), the measurements bracket the crossover
between ten and twelve knobs.  Finite differences win the clock at
$N=8$ (\SI{600}{\second} versus \SI{560}{\second}, ahead by 7\%) and
still at $N=10$ (\SI{574}{\second} versus \SI{549}{\second}, ahead by
4\%), each the mean of two repetitions, as they do at the six knobs
of Fig.~\ref{fig:gradmatch}; reverse mode wins from $N=12$
(\SI{1253}{\second} versus \SI{1393}{\second}, means of two) and by
$1.45\times$ at $N=16$ (\SI{1306}{\second} versus
\SI{1892}{\second}).  The
individual
wall times are not monotone in $N$, because the number of iterations
the optimizer happens to need varies from problem to problem, while
the autograd-to-finite-difference wall ratio falls monotonically across all six sampled knob counts,
crossing unity once.  At both bracketing
points the two repetitions of the faster method lie entirely below
the two of the slower ($N{=}10$: finite differences 537.6 and
\SI{560.4}{\second} against 573.0 and \SI{574.1}{\second}; $N{=}12$:
reverse mode 1214.8 and \SI{1290.4}{\second} against 1365.7 and
\SI{1419.7}{\second}), so the repeats bracket the crossover between
ten and twelve knobs rather than merely suggesting it.  The
memory cost of the exact Jacobian stays bounded: the
checkpointed reverse-mode matching run peaks at
\SIrange{2.6}{3.4}{\giga\byte} of resident memory across the
archived repetitions, with run-to-run variation at fixed $N$
(repeated stages differ by as much as \SI{0.67}{\giga\byte}) as large as
the variation across $N$,
against \SIrange{0.44}{0.50}{\giga\byte} for the
finite-difference passes, so reverse-mode differentiation buys the
$N$-independent Jacobian at a bounded, laptop-scale memory overhead
rather than one that grows with problem size.  Re-validating both
solutions against the higher-fidelity model over five independent
bunch seeds [Fig.~\ref{fig:scaling}(c)] leaves the two optimizers at
the same physical mismatch of \numrange{1.1}{3.2}~\si{\micro\meter}
(means over seeds; statistically indistinguishable, differing by
$\leq\SI{0.02}{\micro\meter}$ at every $N$), exceeding the
revalidation's own seed-noise floor by factors of 2.6--13, so the
mismatch is real, and it is the compound gap between the optimizer's
model ($32^3$, 1500 particles, seed~42) and the higher-fidelity model
($64^3$, 5000).  Its grid and particle-count changes were not
separated from each other; the higher-fidelity seed was sampled on its
own (five seeds), and that sample is what the floor above
measures (the mean worst-case per-seed mismatch): for
$N\geq4$ the solution mismatch equals the gap within one standard
deviation with component signs anticorrelated seed by seed (the optimizer
drives the coarse-model residual to its optimum, and the finer-model
revalidation exposes the negative of the coarse-model bias), while the
over-determined $N=2$ case, unable to absorb the gap with two knobs,
sits about twice above it.  The $\alpha$ mismatches, revalidated
alongside the sizes, follow the same model-gap-dominated pattern.  In
units of the matching tolerances the finer-model mismatch is
$|\Delta\sigma| \leq 0.32\,\tau_\sigma$ at every $N$ and
$|\Delta\alpha| \leq 0.26\,\tau_\alpha$ for $N \leq 4$, but
$\Delta\alpha_x$ reaches $1.6\,\tau_\alpha$ ($0.0158 \pm 0.0006$) at
$N=16$: the finer model exposes the coarse-model bias, and at the
largest knob count that bias sits modestly outside the $\alpha$
tolerance, a statement about the coarse optimization model's
fidelity, not about either optimizer.

\subsection{Misalignment Monte Carlo and orbit correction}
\label{sec:app_errors}

\begin{figure}[t]
\includegraphics[width=\linewidth]{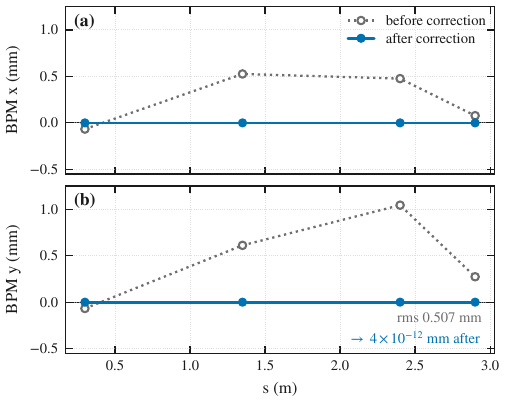}
\caption{\label{fig:correction}%
Orbit-correction demo: (a)~horizontal and (b)~vertical BPM centroid
readings versus $s$ before (gray)
and after (blue) \texttt{ADJUST\_STEERER}-driven correction on a
six-cell FODO (\SI{5}{\mega\electronvolt} protons) with
\SI{0.2}{\milli\meter} rms transverse misalignments planted on every
quadrupole.  The one-to-one steerer/BPM solve drops the rms reading
from \SI{0.507}{\milli\meter} to $4\times10^{-12}$~mm in one
iteration.}
\end{figure}

\begin{figure*}[t]
\includegraphics[width=\textwidth]{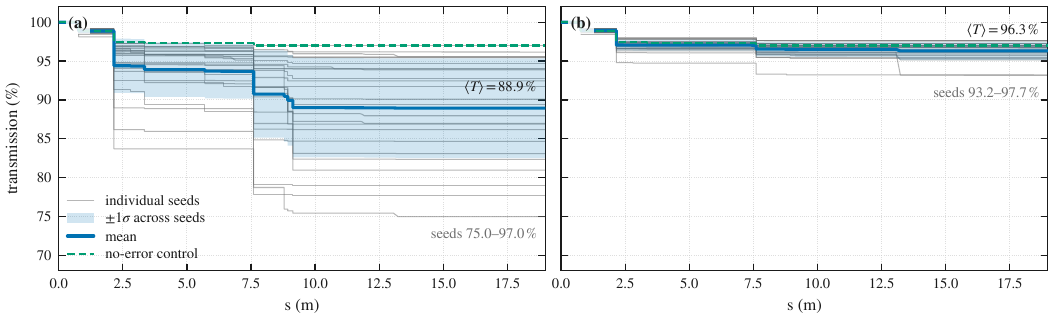}
\caption{\label{fig:errors}%
Misalignment Monte Carlo with orbit correction: transmission versus
$s$ for 25 paired seeds on the leading \SI{18.99}{\meter} of the
PIP-II error-study deck (MEBT and HWR span, 12 steerer cards onto 11
of 19 BPM markers, truncated-SVD solve),
each seed's error draw applied identically to (a)~an uncorrected and
(b)~a corrected pass (single correction iteration).  Individual seeds
(gray), cross-seed mean $\pm1\sigma$ (blue), nominal no-error
reference (green
dashed; a single design-lattice, design-beam run with all error draws
disabled and no correction, not seed-matched to the 25 realizations;
the in-plot legend labels it ``no-error control'').  Correction narrows
the final-transmission spread from
75.0--97.0\% (mean 88.9\%) to 93.2--97.7\% (mean 96.3\%) against the
97.0\% reference.}
\end{figure*}

Error studies accept two input routes into the same error model:
TraceWin \texttt{ERROR\_*} directives parsed from the \texttt{.dat},
or programmatic \texttt{ErrorDef}/\texttt{BeamErrorDef} specs (a glob
pattern over element names times one parameter: $dx$, $dy$, roll,
relative gradient or voltage, phase, steerer field; input-beam
centroid, emittance scale, Twiss mismatch, current); either populates
an \texttt{ErrorStudy}, the card route covering a subset of the
programmatic surface.  The campaign reported below uses the
programmatic specs.  Per seed the engine draws truncated Gaussians in
TraceWin's redraw convention, applies them to a copy of the lattice
and beam, optionally corrects the orbit, tracks, and records; the
results object reduces the ensemble to means, standard deviations,
percentiles, and transmission statistics.  The correction driver,
\texttt{run\_correction\_from\_lattice}, mirrors TraceWin's
\texttt{ADJUST\_STEERER} semantics: each card is paired with a BPM
marker and steerer, a finite-difference response matrix is measured,
and the kicks are solved either one-to-one or, when the system is not
square, by a truncated-SVD pseudoinverse, iterated with per-card
$v_\mathrm{max}$ clips and optional BPM noise.  The shipped demo
(Fig.~\ref{fig:correction}), a noiseless square-system consistency
test and a unit test of the response-matrix machinery rather than an
operational correction study, plants \SI{0.2}{\milli\meter} rms
quadrupole offsets on a six-cell FODO with four BPM--steerer pairs and
reduces the rms BPM reading from \SI{0.507}{\milli\meter} to
$4\times10^{-12}$~mm in a single iteration (the demo's acceptance
gate is 1\%); with four steerers and four BPMs per plane the system
is square and the card/BPM pairing is clean, so the solve is
one-to-one; the SVD branch handles the remaining cases, and either
route can be selected explicitly.

The production-scale template is the PIP-II misalignment study
distributed with the code: the full \SI{256}{\meter} lattice, which the
parser resolves into 1791 objects (the benchmark deck's 1772 plus the
nineteen \texttt{ADJUST\_STEERER} pairing cards the study's deck
generator inserts), under a
PIP-II-derived error budget of
quadrupoles at \SI{0.10}{\milli\meter} offsets,
\SI{1}{\milli\radian} rolls and 0.5\% gradient error; solenoids at
\SI{0.20}{\milli\meter}, \SI{1}{\milli\radian} and 0.5\% field;
cavities at \SI{0.30}{\milli\meter},
\SI{1}{\milli\radian}, 1\% voltage and $1^\circ$ phase (0.5\% and
$0.5^\circ$ for the 650-MHz sections); \SI{e-4}{\tesla\meter}
per-plane steerer-field jitter; and input-beam jitter, with
5000 macroparticles per seed.  Its element selectors gate a name band
on the resolved element kind, so a cavity cannot take a solenoid draw
or vice versa; a pre-run audit prints the matched, applied, skipped
and multiply-selected counts for every error class and by default
refuses to run on any anomaly; \texttt{ADJUST\_STEERER} cards generated with the
deck pair each steerer with a BPM; and each archive records the
resolved specifications, the per-seed draws and correction kicks, the
particle count, correction state and code commit, alongside a
nominal no-error reference run: a single pass of the design lattice
with the design beam, all element- and input-beam error draws
disabled and no correction applied.  It is one run, not seed-matched
to the error realizations.

Figure~\ref{fig:errors} reports a paired campaign on the leading
\SI{18.99}{\meter} of that separately generated deck, covering the
same MEBT and HWR span as the \SI{18.96}{\meter} benchmark file above
but cut at a different drift, retaining 31
quadrupoles, 8 solenoids, 12 cavities and 12 steerer cards paired
onto 11 of the deck's 19 BPM markers (two co-located steerers share
one, and twelve cards cannot pair one-to-one onto nineteen
markers).  The front-end solve therefore takes the truncated-SVD route
rather
than a one-to-one solve, fitting twelve kicks per plane to all
nineteen readings.  The budget is restricted to the error classes
present: 25 seeds, each
drawn once and tracked twice, without and with the correction pass (a
single iteration, with \SI{0.05}{\milli\meter} BPM noise; the solver's
\SI{0.05}{\milli\meter} rms convergence tolerance cannot bind at one
iteration).  Against the 97.0\%
nominal no-error reference the uncorrected ensemble transmits
$88.9\% \pm 6.4\%$ (cross-seed sample standard deviations, in
percentage points; range 75.0--97.0\%, twelve of the 25 seeds below
90\%); correction leaves every seed above 93\%, at
$96.3\% \pm 1.2\%$ (range 93.2--97.7\%).  The paired recovery
averages $+7.4$ percentage points, with a sample standard deviation of
6.2 points and a two-sided Student-$t$ 95\% confidence interval for the
paired mean of $+4.8$ to $+9.9$ points, improving 24 of the 25 seeds:
the worst seed climbs from 75.0\% to 93.2\%, the largest single
recovery is $+19.9$ points, and one seed near the top of the
uncorrected range (94.7\%) emerges 1.5 points lower.  A signed mean deficit
of 0.7 points against the reference remains (3.8 at worst, while seven
corrected seeds end above it); that deficit is a $2.1\sigma$ effect
against the quadrature of the reference run's own binomial spread
(0.24 points at 5000 macroparticles) and the cross-seed standard error
of the corrected mean (0.23 points), so it is marginal rather than
firmly resolved.  Twenty-five paired seeds quantify the corrected workflow.  One
methodological limitation remains: the correction pass measures its
response matrix and BPM readings without space charge, and each
corrected seed is then tracked with the full PIC configuration.
Tolerance statements would need hundreds of seeds, and a corrected
ensemble on the full \SI{256}{\meter} deck is deferred on cost, since
the response matrix re-tracks the line once per steerer plane and
pass, so its per-seed price grows with the product of steerer count
and line length.

\subsection{Stripper-foil transport in the Booster transfer line}
\label{sec:app_foil}

\begin{figure*}[t]
\includegraphics[width=\textwidth]{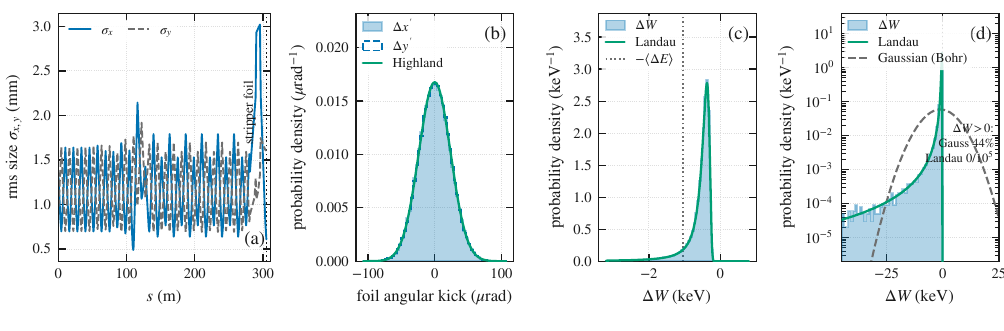}
\caption{\label{fig:foil}%
PIP-II Booster-transfer-line stripper foil (carbon,
\SI{600}{\micro\gram\per\square\centi\meter}, \SI{800}{\mega\electronvolt}
$\mathrm{H}^-$).  (a)~Envelope-solver rms sizes along the
\SI{305}{\meter} line at \SI{4.84}{\milli\ampere} for a demo beam
specified by transverse Twiss only, with no initial longitudinal
spread, so this trace carries no dispersive contribution through the
line's 36 dipoles; panels (b)--(d) use the full six-dimensional beam.
The foil location is marked; at the foil the envelope solver applies the
covariance diffusion $\Sigma \to \Sigma + D$ (adding the Highland
$\theta_0^2$ to the divergence moments, and to the energy moment a
Gaussian straggling $\sigma_E^2$: Bohr's thin-absorber form without
its relativistic $(1-\beta^2/2)/(1-\beta^2)$ factor) and the minimum-ionizing mean loss to the reference
energy; the position traces are unchanged because the foil
terminates the line.  (b)~Per-particle angular kicks
$\Delta x'$, $\Delta y'$ isolated by paired $10^5$-particle runs with
and without the foil ($I=0$), against the Highland Gaussian.
(c)~Per-particle energy change $\Delta W$ against the
mean-pinned truncated-Landau model (solid) and the model mean
loss (dotted).  (d)~The same on a log scale against the previous
Gaussian straggling model (dashed): the Landau sample assigns
a net energy \emph{gain} to 0 of $10^5$ particles, where the Gaussian
assigned it to 44\%.}
\end{figure*}

The last study exercises an interaction element rather than optics:
the carbon stripper foil
(\SI{600}{\micro\gram\per\square\centi\meter}) terminating the
\SI{305}{\meter} PIP-II Booster transfer line at
\SI{800}{\mega\electronvolt}.  The foil element implements three interaction
models: Highland
multiple Coulomb scattering~\cite{highland} on $x'$, $y'$; a
minimum-ionizing mean
ionization loss~\cite{pdg}; and
energy straggling whose form is
selected by the thin-absorber parameter $\kappa = \xi/T_\text{max}$.
Each carries a caveat.  At
$x/X_0 \simeq 1.4\times10^{-5}$ this foil lies far below the
$10^{-3} \lesssim x/X_0$ range over which the Highland fit was
established~\cite{pdg}, so the angular model is an extrapolation, and
Fig.~\ref{fig:foil}(b) verifies faithful sampling of it rather than the physical
single-scattering tail of so thin a target.  The mean loss is a fixed
tabulated mass stopping power,
in \si{\mega\electronvolt\square\centi\meter\per\gram},
multiplied by the areal density; the $\beta$-dependent
Bethe rise is not modeled, so the fixed value is representative near
the stopping-power minimum and increasingly understates the mean
loss toward low energies.  And
this foil sits deep in the Landau regime ($\xi =
\SI{0.065}{\kilo\electronvolt}$, $T_\text{max} =
\SI{2.48}{\mega\electronvolt}$, $\kappa = 2.6\times10^{-5}$), so the
per-particle loss is drawn from a mean-pinned, kinematically
truncated Landau distribution~\cite{vavilov,pdg}, bounded
by $T_\text{max}$ and shifted so its expectation equals the
model value; a Gaussian draw is retained for thick absorbers
($\kappa > 10$).  It does
\emph{not} model charge-state fractions: the $\mathrm{H}^-$ to proton
conversion is represented by the element's position at the
line--Booster boundary, not by stripping-fraction dynamics.  In
envelope mode the foil is no longer invisible: the solver applies the
second-moment update $\Sigma \to \Sigma + D$, adding $\theta_0^2$ to
the divergence moments ($\sigma_{x'}$ 0.2363 to
\SI{0.2375}{\milli\radian}, $\sigma_{y'}$ 0.0867 to
\SI{0.0899}{\milli\radian} at the foil), the Gaussian straggling
$\sigma_E^2$ to the
energy moment ($\sigma_W$ 0 to \SI{6.78}{\kilo\electronvolt} for this
initially monoenergetic demo beam; the Landau sample rms is
tail-dominated at \SI{11.9}{\kilo\electronvolt}, so the envelope
carries the Gaussian value rather than the sample rms as its
second-moment bookkeeping), and the mean loss to the
reference energy.  The kicks are weak
against the beam itself: the \SI{23.8}{\micro\radian} scattering
angle is 10.1\% of the horizontal beam divergence
(\SI{236.3}{\micro\radian}) and 27.5\% of the vertical
(\SI{86.7}{\micro\radian}) at the foil.  The study therefore
uses paired $10^5$-particle runs, identical except for the foil and
bit-identical at the foil entrance, whose per-particle exit
differences isolate the foil physics exactly.  Both runs transmit
100\%.  The extracted kick statistics reproduce the underlying models:
$\mathrm{std}(\Delta x') = \SI{23.88}{\micro\radian}$ and
$\mathrm{std}(\Delta y') = \SI{23.83}{\micro\radian}$ against the
Highland $\theta_\mathrm{rms} = \SI{23.811}{\micro\radian}$ (within
0.3\%), and the sampled loss distribution shows the Landau
signature [Fig.~\ref{fig:foil}(c,d)]: a most probable loss of
\SI{0.375}{\kilo\electronvolt} well below the
\SI{1.047}{\kilo\electronvolt} model mean (sample mean
\SI{1.013}{\kilo\electronvolt}; the sampler's ensemble mean is
pinned to the model value, so the \SI{0.03}{\kilo\electronvolt}
deficit is finite-sampling scatter, one standard error at $10^5$
particles given the tail-dominated \SI{11.9}{\kilo\electronvolt}
sample rms), a core of full width at half maximum
\SI{0.26}{\kilo\electronvolt} ($4.02\,\xi$, the Landau value), a long
high-loss tail (0.13\% of particles lose more than
\SI{45}{\kilo\electronvolt}; largest single loss
\SI{2.3}{\mega\electronvolt}), and \emph{no} net energy
gain in $10^5$ particles---where the previous Gaussian model, with
$\sigma_E = \SI{6.8}{\kilo\electronvolt}$ against the
\SI{1.05}{\kilo\electronvolt} mean, unphysically assigned a gain to
44\% of them.  For ring
injection this quantifies the foil's contribution per traversal: an
angular kick of 10\% (horizontal) to 28\% (vertical) of the local
beam divergence (rms divergence growth in quadrature of 0.5\% and
3.7\% respectively), a core energy smear of a few tenths
of a keV, and a rare-event tail that drives the sample rms to
\SI{11.9}{\kilo\electronvolt}, \SI{2.8}{\percent} of the
\SI{420.8}{\kilo\electronvolt} rms energy spread of the transfer-line
beam itself.  The remaining model limitation
is the intermediate Vavilov regime ($0.01 \lesssim \kappa \lesssim
10$), which this release approximates by the Gaussian branch; the
foils of interest here sit far from it.

\subsection{Design-study toolset}
\label{sec:app_toolset}

\begin{figure}[t]
\includegraphics[width=\linewidth]{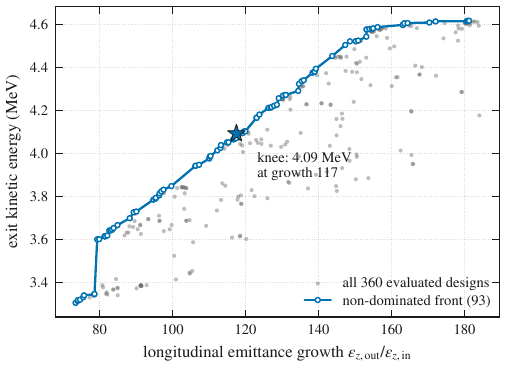}
\caption{\label{fig:pareto}%
Multi-objective optimization interface demonstration on the shipped
six-knob demo cell (two
solenoid fields plus the voltage and phase of a buncher and an
accelerating cavity; \SI{2.5}{\mega\electronvolt},
\SI{5}{\milli\ampere} proton beam, \SI{0.74}{\meter}; space charge off,
so the declared current does not enter the forward pass), posed
at multiparticle fidelity: NSGA-II~\cite{pymoo} (population 24, 15
generations, seed 0, 360 evaluations, about two seconds) minimizing
longitudinal emittance growth while maximizing exit kinetic energy
over a 1000-particle fixed-seed forward pass.  All 360 evaluated
designs (gray), the 93 non-dominated designs (blue, connected), and
the knee-point pick (star): growth factors 73--181 against exit
energies from 3.30 to \SI{4.62}{\mega\electronvolt}, knee at
\SI{4.09}{\mega\electronvolt}.  The knee is selected by an explicit
rule, the same one the graphical workbench applies: the front point of
smallest Euclidean distance to the ideal corner after each objective is
min--max normalized over the front.  The mismatched, longitudinally hot
demo beam (8\% rms energy spread) also scrapes on the cell's
\SI{20}{\milli\meter} apertures (final transmission 66--83\% across
all 360 evaluations; 70--83\% on the front), so the emittance
objective is computed from surviving
particles.  A repeated seeded run returns a bit-identical front.}
\end{figure}

The four studies above each drove one capability hard; we close with
the rest of the design-study surface, each tool tied to an example
distributed with the code.

\emph{Parameter scans.}  The \texttt{scan} subcommand
(Sec.~\ref{sec:interfaces}) sweeps any beam-configuration scalar, the
grid size, extent and per-meter step counts, or any single element
parameter addressed by name or index; repeated range arguments form a
Cartesian product,
and each point runs an independent envelope or multiparticle
simulation whose end-of-lattice metrics (rms sizes, emittances,
transmission, reference energy, peak excursions, wall time) become one
CSV row.  Points distribute over a process pool whose workers default
their FFT threading to one thread, existing environment settings
taking precedence, so the parallelism comes from the pool alone
(Sec.~\ref{sec:performance}).  The shipped two-dimensional example
sweeps beam current against a quadrupole gradient over nine points and
four workers; the same command accepts any lattice input, including
the PIP-II files of Sec.~\ref{sec:validation}.

\emph{Failure studies.}  The \texttt{failures} engine (CLI, GUI
Failure Study tab, or Python API) asks which element failures hurt
the beam most and whether the machine can recover.  Failures are
injected through the elements' relative error slots: \texttt{off}
zeroes an element's strength, \texttt{partial} scales a magnet to a
fraction of nominal, and \texttt{detune} applies a cavity amplitude
scale and/or additive RF-phase offset---swept singly, as all pairs
(an $N \times N$ criticality matrix), or as named groups.  Each
scenario is tracked and scored against the nominal baseline by a
criticality that weights fractional transmission loss and exit-energy
deviation above normalized-emittance growth; an optional compensation
pass then re-tunes neighboring elements (k-out-of-n,
neighboring-period, or manual selection) through the matcher of
Sec.~\ref{sec:matching} to recover the design exit energy, following
the local-compensation scheme of the MYRRHA fault-tolerance studies
and the LightWin tool~\cite{lightwin}.  The shipped
demo is a compact PIP-II-like superconducting section of four cavities
and three solenoids, rerun for this paper (the sweep and recovery
complete in about half a second).  On it the single-element \texttt{off}
sweep ranks the last cavity most critical
(\SI{1.13}{\mega\electronvolt} of the \SI{6.22}{\mega\electronvolt}
baseline exit energy lost; the cavities order by energy contribution
and the solenoids score zero on this loss-free envelope run), and
compensation by the three surviving cavities restores the exit energy
to the baseline value of \SI{6.2196}{\mega\electronvolt}.

\emph{Element and beam error classes.}  Section~\ref{sec:app_errors}
exercised the error engine on a full-machine template and in a
paired corrected--uncorrected front-end campaign; its file-level
compatibility surface is TraceWin's tolerance-study
language~\cite{tracewin}: \texttt{ERROR\_QUAD\_NCPL\_STAT},
\texttt{ERROR\_CAV\_NCPL\_STAT}, and \texttt{ERROR\_BEND\_NCPL\_STAT}
cards apply alignment offsets, rolls, and relative
gradient/field/voltage plus RF-phase errors to the next $N$ matching
elements; \texttt{ERROR\_BEAM\_STAT} declares input-beam jitter; and
\texttt{ERROR\_GAUSSIAN\_CUT\_OFF} sets the global truncation.  The
beam-error class is what the thirteen jitter specs of the PIP-II
study instantiate: centroid offsets in all six coordinates ($x$,
$x'$, $y$, $y'$, $\Delta\phi$, $\Delta W$), per-plane emittance
scaling, per-plane Twiss mismatch, and current jitter, drawn fresh
per seed and applied to the input beam before generation.  The
support status is explicit in the documentation: draws are
truncated Gaussian (redraw convention) or uniform, with TraceWin's
constant mode approximated by a Gaussian under a parse warning;
pitch/yaw rotations and cavity/bend longitudinal offsets are parsed
but inert (of the rotations, only the roll is applied by the
tracker); and the dynamic
(\texttt{ERROR\_*\_DYN}) and coupled (\texttt{\_CPL}) variants are
absorbed as static, uncoupled per-seed errors; their time-varying
and coupled-group semantics are documented as deferred.

\emph{Multi-objective exploration.}  Where the matcher collapses
\texttt{SET}/\texttt{ADJUST} cards into one scalar residual, the
\texttt{mo} subcommand (and the Matching tab's Pareto dialog)
explores the trade-off surface between competing objectives over the
same \texttt{ADJUST} knobs.  The objective library covers per-plane
normalized emittance growth, four-dimensional transverse emittance
growth, transmission loss (multiparticle mode), negative exit energy,
and exit or peak rms sizes; the engines are NSGA-II~\cite{pymoo} and,
for expensive forward passes, the sample-efficient Bayesian qNEHVI~\cite{qnehvi} of
BoTorch~\cite{botorch}, for which the shipped documentation reports
reaching a front comparable to NSGA-II's with roughly an order of
magnitude fewer evaluations on the envelope-mode demo (a
documentation-level illustration with no front-distance metric
defined; not re-measured here).  The shipped demo poses longitudinal emittance growth against
exit energy on a six-knob cell; run as shipped, with the envelope
forward pass, the emittance objective is degenerate on this
cell: linear transport conserves rms emittance, so all 360 evaluated
designs return a growth of 1.0 to within $5\times10^{-12}$ and the
front spans only exit energy.  Figure~\ref{fig:pareto} therefore
poses the same problem at multiparticle fidelity
(\texttt{cost\_solver="mp"}, 1000-particle fixed-seed bunch, no space
charge), where each particle samples the cavity kick at its own RF
phase and the two objectives trade off: the seeded NSGA-II
search (population 24, 15 generations, 360 evaluations, about two
seconds) returns 93 non-dominated designs whose exit energy rises
monotonically from 3.30 to \SI{4.62}{\mega\electronvolt} as the
tolerated emittance growth rises from 73 to 181, with a
min--max-normalized nearest-to-ideal knee at
\SI{4.09}{\mega\electronvolt}, and a repeated seeded run reproduces
the front bit for bit; Pareto exploration inherits the determinism
of the underlying solvers.  Two qualifications keep this example in
its place, and together they mean the demo does not separate the
phase-spread mechanism from aperture scraping: the emittance
objective is evaluated on surviving
particles and the shipped demo imposes no transmission floor, so the
designs on this front transmit only 70--83\% (66--83\% across all
360 evaluations) and the optimizer is
free to improve its objective by scraping (the loss-gaming loophole
of Sec.~\ref{sec:matching}), and the front is a single-seed result.
The figure is therefore an interface-and-determinism demonstration
rather than a machine-design tradeoff; a design study would add the
\texttt{MIN\_TRANSMISSION} card or the transmission-loss objective
(the objective is available on this path; the card is not) and repeat
over seeds.

\emph{Surrogate workflow.}  The surrogate machinery of
Sec.~\ref{sec:surrogates} ships as a three-step CLI sequence,
exercised here on the PIP-II MEBT: \texttt{train} fits one field-map
element and writes its weights and metadata manifest (the moderate
level, the run behind Fig.~\ref{fig:surrogate}), \texttt{compare}
runs baseline and registry-enabled envelope passes back to back and
prints per-moment differences and wall times, and
\texttt{run-envelope} registers the trained surrogates for a
production pass.  Two limits surface at the application level.  At
space-charge current the partial-slice hooks delegate to the native
integrator, so \texttt{compare} at \SI{5}{\milli\ampere} verifies
bit-identical fallback with zero neural evaluations rather than
surrogate accuracy.  And one trained cavity surrogate can be shared
across the eight HWR cavities that use the same field map by
registering it under each name, provided the trained sweep spans every
cavity's drive setting and reference energy---on the shipped deck the
eight settings differ by a factor of 6.5, well outside the default
$\pm20\%$ sweep, so this is a training-scope decision and not free
reuse.  \texttt{register-multi} scripts the registration and prints the
trained scope against each element's current setting, but the
runtime registry is an in-process singleton and that command's process
exits before any tracking: as a standalone command it only reports
that comparison, and an actual shared-surrogate run requires registering
from the Python process that does the tracking.
The accuracy--speed economics are quantified in
Sec.~\ref{sec:surrogates}.

\section{Performance and scaling}
\label{sec:performance}

Two kinds of timing numbers appear below and are labeled as such:
\emph{documented} reference benchmarks that ship with the code as
maintained runtime expectations, and \emph{measured} numbers taken for
this paper on an Apple M3~Max laptop (14-core CPU).  The
figure-generation runs of Secs.~\ref{sec:validation}
and~\ref{sec:applications} used the NumPy PIC kernels on a host
shared with other workloads---order-of-magnitude scales, not
controlled benchmarks.  For this revision the C++/OpenMP
deposit/gather extension was additionally built on the same machine
(Apple clang with Homebrew \texttt{libomp}), and the timings reported
below as controlled follow a fixed protocol: medians of repeated runs
are quoted (with min--max spreads for the end-to-end comparison), and
the hardware, library versions, and observed load state are archived
with the timing data.  Idle gating differs by suite: each end-to-end
repetition starts only when the one-minute load average is below six
(recorded per run), the kernel microbenchmark applies that gate once
at launch, and the Poisson-solve benchmark records the load state
with its data.  Every measured number names its hardware; documented
numbers name it where the documentation records it.

\subsection{Solver-mode cost hierarchy}
\label{sec:costhierarchy}

The three solver modes span the orders of magnitude anticipated in
Sec.~\ref{sec:solvers}.  Matrix composition, once per-element
matrices are available, is a chain of matrix products with no
particle tracking or space-charge kicks, and is the cheapest mode by
construction.
For envelope-mode work the documented runtime expectation on the full
\SI{256}{\meter} PIP-II lattice is about
\SI{340}{\second} for the first evaluation and about \SI{3}{\second}
for every subsequent one against the same lattice and beam
configuration (host unspecified in the documentation); both figures
describe the opt-in per-element transfer-matrix cache, exercised
through the GUI's phase-advance computation; the underlying development-log benchmark is
\SI{339}{\second} on a cold cache versus \SI{2.67}{\second} warm, a
factor of 127, with bit-identical results, and the 127 is a property
of that cache path, not of the envelope solver itself.  The envelope
solver does not consume this cache: the two archived executions of
the full-machine envelope+space-charge benchmark behind
Fig.~\ref{fig:pipii} took \SI{185}{\second} (current run of record)
and \SI{472}{\second} (original execution, which its provenance
notes began with a cold field-map cache) for the envelope pass.
The two come from different revisions of that figure's history and
are not a controlled pair (both on this paper's M3~Max, per the
figure's provenance), but each sits far above the cached
seconds-scale figure, which belongs to the transfer-matrix cache
alone.
That warm-cache regime is what the GUI's phase-advance panels
exercise; the matching engine re-evaluates its physics model at every
iteration without this cache, so an envelope-mode matching evaluation
pays the envelope cost, not the cached figure.  For the multiparticle tracker the documented
reference on the \SI{18.96}{\meter} MEBT+HWR benchmark of
Sec.~\ref{sec:tracewin} ($10^4$ macroparticles, $64^3$ grid,
\SI{5}{\milli\ampere}) is \SI{97.2}{\second} on an Apple M3~Max with
the CPU FFT backend.  Under the controlled protocol above (five
repetitions per kernel mode, each begun only below a one-minute load
average of six), the complete MEBT+HWR space-charge comparison
(parsing both TraceWin~\cite{tracewin} reference exports, one
envelope+space-charge pass, one PIC pass at the same $10^4$/$64^3$
settings, and four comparison figures) ran in a median
\SI{139}{\second} with the NumPy deposit/gather kernels and
\SI{147}{\second} with the C++/OpenMP kernels, consistent with the
documented \SI{97.2}{\second} PIC pass dominating the total.  The
compiled build was slower in every one of the five paired
repetitions, by about \SI{9}{\second} (6\%) at the median, though the
run-to-run spreads overlap (\SIrange{134}{157}{\second} and
\SIrange{146}{165}{\second}): at these settings the compiled kernels
buy nothing end to end.  An earlier uncontrolled run on the shared
host, timed from the log including figure generation, gave the same
$\sim$\SI{144}{\second} scale.  That the compiled kernels fail to
pay off at $10^4$ macroparticles is consistent with per-element
field-map transport, rather than deposit/gather, dominating the step:
in the repository's $N{=}2000$ profile of the same line, field-map
transport accounts for 56\% of cumulative run time (largest
exclusive share: field sampling at 22\%) against 7.5\% for the CIC
deposit, and the isolated kernel timings of
Sec.~\ref{sec:picbackends} put the NumPy gather at only
$\sim$\SI{2}{\second} of this \SI{139}{\second} run.  Per matching
evaluation the engine documentation quotes the multiparticle solver
at 50--100 times the envelope solver, but on the MEBT+HWR line the
measured ratio is $1.9\times$ at the matcher's default $10^{3}$
macroparticles and $8.5\times$ at $10^{4}$ (Sec.~\ref{sec:matching});
envelope-mode matching is the documented interactive default, with
multiparticle fidelity reserved for constraints, such as
transmission, that only the tracker can evaluate.

Two further mechanisms address this hierarchy.  The machine-learned
surrogates of Sec.~\ref{sec:surrogates} convert
\SI{336}{\milli\second} field-map transfer-matrix computations
into sub-millisecond calls, but the engine manual's own MEBT+HWR
benchmark (20 field maps, 1000 particles at \SI{5}{\milli\ampere},
14-core M3~Max) reports end-to-end multiparticle speedups of only
$1.22\times$ and $1.66\times$ with the experimental double-opt-in
fast path engaged at two training depths ($0.99\times$ with
surrogates registered but the fast path off), at a documented
longitudinal accuracy cost that shrinks with training depth
($\sigma_z$ error 28.4\% to 4.7\%) while transverse-size errors stay near
5\%; the paper's own retained surrogate measurement (the campaign
behind Fig.~\ref{fig:surrogate}) is a zero-current envelope-mode
drop-in at $1.21\times$ (1.35 to \SI{1.11}{\second}).  And because scan points are independent
simulations, the parameter-scan machinery
(Sec.~\ref{sec:interfaces}) distributes them over a process pool: each
worker defaults its internal FFT threading to a single thread
(existing environment settings take precedence) so that
parallelism comes from the pool rather than oversubscribed threading,
GPU backends are forced to the CPU path when more than one worker is
requested, and results are collected asynchronously.  The
documentation states the design intent (throughput scaling with core
count across points) without publishing a scan-scaling curve.

\subsection{PIC backends and GPU offload}
\label{sec:picbackends}

Within a single multiparticle run the acceleration levers are those of
Sec.~\ref{sec:acceleration}: C++/OpenMP deposit and gather kernels
(CIC only; the TSC pair runs in Python), and FFT backend dispatch for
the Poisson solve.  Measured in isolation on the $64^3$ grid with medians of nine timed
calls after three warm-ups (C++ and Python outputs agree to a maximum
grid-point difference below $2\times10^{-15}$ of the array peak), the
compiled kernels accelerate the CIC gather by
\numrange{11}{25}$\times$ at $10^4$ particles (rising to $41\times$
on eight threads at $10^5$), while the CIC deposit is no faster at
$10^4$ (\numrange{0.7}{1.0}$\times$, i.e., up to 44\% more wall time)
and pulls ahead (\numrange{4.9}{8.7}$\times$) at $10^5$.  The deficit
at $10^4$ is consistent with fixed threading and with the
per-thread-buffer reduction the kernel uses as a deterministic
alternative to atomics, an overhead that amortizes only at larger
$N$.  At the
$10^4$-macroparticle MEBT+HWR scale, then, only the gather is
materially faster, and at \SI{1.4}{\milli\second} per NumPy call over
the line's 1296 space-charge kicks it accounts for only
$\sim$\SI{2}{\second} of the \SI{139}{\second} end-to-end run,
consistent with the profiling shares and end-to-end times of
Sec.~\ref{sec:costhierarchy}; the compiled path earns its keep at
higher particle counts.  Table~\ref{tab:gpubench} reproduces the documented
benchmark of one Hockney--Eastwood Poisson solve comparing the
CuPy/CUDA backend on an NVIDIA RTX~2000 Ada laptop GPU against a
16-thread \texttt{scipy.fft} CPU path on the same machine, across a
WSL2-hosted PCIe host--device link.  On this laptop-class hardware the
GPU is faster only at small and medium grids ($1.6\times$ at $48^3$ and
$1.5\times$ at $64^3$), while the CPU path wins at $96^3$ and $128^3$,
i.e., at and above the documented $96^3$ shipped default.  The
documentation attributes the crossover to host--device transfer cost
rather than FFT throughput and recommends the CPU path for
$\geq 96^3$ workloads on such systems; \texttt{auto} backend
selection remains overridable.

\begin{table}[b]
\caption{\label{tab:gpubench}%
Documented wall time of one Hockney--Eastwood Poisson solve versus
grid size: CuPy/CUDA backend on an NVIDIA RTX~2000 Ada laptop GPU
(FP64, WSL2-hosted PCIe host--device link) against a 16-thread
\texttt{scipy.fft} CPU path.  On this laptop-class GPU the offload
is faster only below the $96^3$ shipped default; the documentation
attributes the crossover to host--device transfer cost, not FFT
throughput.}
\begin{ruledtabular}
\begin{tabular}{lccc}
Grid & CPU (ms) & GPU (ms) & GPU speedup \\
\colrule
$48^3$  & 7.6   & 4.9   & $1.6\times$ \\
$64^3$  & 17.5  & 11.7  & $1.5\times$ \\
$96^3$  & 41.7  & 55.5  & $0.8\times$ (CPU wins) \\
$128^3$ & 100.6 & 128.7 & $0.8\times$ (CPU wins) \\
\end{tabular}
\end{ruledtabular}
\end{table}

\begin{figure}[t]
\includegraphics[width=\linewidth]{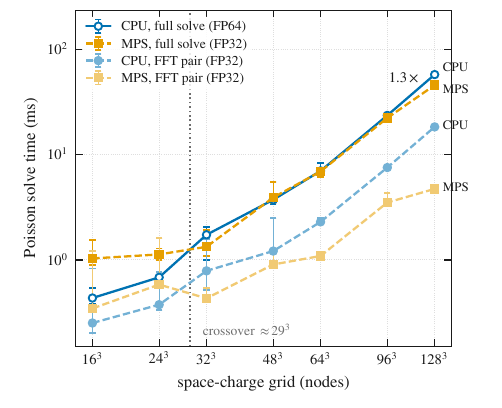}
\caption{\label{fig:gpubench}%
Measured wall time of one Hockney--Eastwood Poisson solve versus grid
size on an Apple M3~Max (median of nine repetitions, error bars
min--max): 14-thread CPU path at FP64 versus the Metal (MPS) backend
at FP32, each at its production precision (solid), and the
forward/inverse real-FFT pair alone at equal FP32 precision (dashed).
On unified-memory Apple silicon the full solve is roughly
backend-neutral across the measured grid sizes; the Metal median leads
at $32^3$ (within overlapping min--max spreads) and by
$\approx$$1.3\times$ at the largest measured grid, $128^3$.}
\end{figure}

On Apple silicon the trade-offs invert: unified memory removes the
PCIe transfer, though staging, precision conversion, and
synchronization overheads remain, and the Metal backend runs in FP32
(Sec.~\ref{sec:acceleration}).  Warmed measurements on an Apple M3~Max
with PyTorch~2.10~\cite{pytorch} put the forward/inverse real-FFT pair
on the padded $256^3$ Hockney grid (the doubled transform grid of a
$128^3$ physical grid) several times faster on MPS than on
the multithreaded CPU path: 4.7--5.7 versus
16.6--\SI{18.3}{\milli\second} at FP32, across this paper's benchmark
(Fig.~\ref{fig:gpubench}) and the repository's profiling notes; an
older documented figure that showed the inverse transform far slower
on MPS was a cold-start artifact, as recorded in those profiling
notes.  End to end, the documented MEBT+HWR multiparticle run ($10^4$
macroparticles, $64^3$ grid) improves only from \SI{97.2}{\second} on
the M3~Max CPU path to \SI{90.1}{\second} with MPS, because
the non-FFT costs (per-element transport and deposit/gather)
dominate; the documentation
projects larger gains for lattices with proportionally more PIC kicks.
The measurement taken for this paper on the M3~Max
(Fig.~\ref{fig:gpubench}; median of nine repetitions per grid, load
state recorded with the data) resolves this at the level of the
Poisson solve itself: comparing each backend at its production
precision (FP64 CPU, FP32 MPS), the two are within a few percent of
each other from $48^3$ through $96^3$, with the Metal path ahead by
about $1.3\times$ at $128^3$ and (by medians that sit within
overlapping min--max spreads) at $32^3$, and behind below $32^3$,
where the solve takes under a millisecond and fixed per-call
dispatch overhead is proportionally largest.  An FFT-pair-only
comparison at equal FP32 precision (dashed) shows the transforms
themselves run up to $3.9\times$ faster on Metal at $128^3$; the
remainder of the solve (real-to-complex packing, spectral multiply,
and field finite differences) absorbs most of that advantage at
the larger measured grid sizes.  An independent idle re-run during revision
reproduces the pattern (FFT pair $3.8\times$ at $128^3$, Metal ahead
$1.2\times$ on the full solve there), with the full-solve medians at
the measured grids ($48^3$--$128^3$) shifting by up to $\sim$15\%
between runs (comparable to the min--max bars) and by more at the
smallest measured grids and on the FFT-pair curves; the archived
run remains the run of record.

\subsection{The cost of differentiability}
\label{sec:diffcost}

The differentiable PIC of Sec.~\ref{sec:diffpic} buys gradients at a
documented price: it runs in \texttt{torch.float64} on the CPU only
(FP64 is required for parity with the production path, and Apple MPS
is FP32-only), and a plain forward run is documented as roughly five
times slower than the production PIC on a small MEBT lattice.  It is
worthwhile only where gradients change the optimization economics.  A
bound-constrained trust-region-reflective (\texttt{trf}) least-squares
step with a two-point finite-difference (FD)
Jacobian costs one baseline residual plus $N$ probe passes for $N$
knobs (the baseline is shared with the main loop, so the Jacobian's
marginal cost is $N$ forwards), whereas
reverse-mode autograd assembles the exact Jacobian from one
graph-recording forward pass plus reverse sweeps whose cost grows far
more slowly with $N$.  The documented microbenchmarks on the
matrix-tracking (no space charge) residual give per-Jacobian costs of
0.4 versus \SI{1.0}{\milli\second} at $N{=}1$ (FD wins), 2.7 versus
\SI{3.3}{\milli\second} at $N{=}4$, 10 versus \SI{7}{\milli\second} at
$N{=}8$ ($1.4\times$), and 37 versus \SI{13}{\milli\second} at
$N{=}16$ ($2.8\times$), with the crossover near five knobs; with space
charge the documented crossover sits in the same range and the
autograd Jacobian wins by about $2.25\times$ at $N{=}8$.  The exact
Jacobian is also documented to roughly halve the iteration count of
the least-squares loop (12 to 6 on the single-knob benchmark); the
six-knob measurement below shows no such reduction (40 residual and
18 Jacobian evaluations for autograd against 36 and 17 for FD), so at
that scale the demonstrated autograd benefit is convergence depth,
not iteration count---and not per-Jacobian cost either, which at six
knobs still favors FD.

The measured anchors from the six-knob matching demonstration of
Sec.~\ref{sec:applications} [Fig.~\ref{fig:gradmatch};
\SI{5}{\milli\ampere} at \SI{3}{\mega\electronvolt}, on the shared
M3~Max] quantify this through the full nonlinear PIC residual (1500
macroparticles, $32^3$ grid).  These are single archived runs rather
than repetitions under the controlled protocol; every ratio quoted
from them compares calls within one process or between the two
matched optimization processes of the same session, which is where
single runs are informative.  One autograd Jacobian costs
\SI{17.8}{\second}; a no-gradient forward pass timed in the same
process costs \SI{3.25}{\second}, so the Jacobian is $5.5$
forward-equivalents, set by the four residual components rather than
the knob count, whereas a finite-difference Jacobian costs $N$ such
probes.  Two subtleties in that measurement deserve note.  First,
graph recording is not the cost driver: the autograd loop's
graph-building residual and the \texttt{torch.no\_grad} forward in
the same process agree to within 0.35\% (3.236 versus
\SI{3.247}{\second}).  Second, per-forward cost is process-state dependent: the
finite-difference run, in its own process, evaluates the identical
residual at \SI{1.87}{\second} per call, so the autograd process's
forward cost is $1.7\times$ higher, consistent with allocator and
cache state left behind by Jacobian builds, whose peak resident
memory runs several times the finite-difference path's
(Sec.~\ref{sec:app_gradmatch}).  Against the
finite-difference probe's own cost the Jacobian is $\sim$9.5
forward-equivalents, in line with the fresh-process per-stage ratios
of Fig.~\ref{fig:scaling} (8.4--9.6 normalized by the
graph-recording forward, 8.5--9.7 by a no-gradient one); under
either convention the ratio is flat in the knob count.  The comparison is
not one-sided: at \SI{1.87}{\second} per probe an FD six-point
Jacobian costs about \SI{11.2}{\second} against the
\SI{17.8}{\second} autograd figure.  On this problem the two
trust-region runs track each other while both proceed, but the FD
run stops one accepted iteration earlier (36 main-loop residual and 17
Jacobian evaluations versus 40 and 18), and the two reach the same
knob values within $3.1\times10^{-6}$~\si{\tesla\per\meter}.  The FD
run finished in \SI{259}{\second} of wall clock
against \SI{449}{\second} for the autograd run.  At six knobs, in
other words, FD won the wall clock by a factor of 1.74 while the
autograd run converged a factor of about 200 deeper (final
normalized residual norm $3.2\times10^{-10}$ versus
$6.2\times10^{-8}$); the
per-Jacobian ratio moves in autograd's favor as the knob count grows,
the FD cost scaling linearly in $N$ while the autograd cost stays
flat.
Memory is the remaining cost axis: the space-charge residual builder
enables gradient checkpointing in the step tracker automatically, and
the documentation notes that without checkpointing and vectorized
Jacobian evaluation the autograd path memory-thrashes on lattices with
more than about four space-charge-coupled knobs.  With both enabled,
the knob-count scaling study of Sec.~\ref{sec:app_gradmatch}
(Fig.~\ref{fig:scaling}) confirms the intended behavior through the
complete differentiable nonlinear PIC residual: the
reverse-mode Jacobian cost holds flat in the knob count (8.4--9.6
fresh-process forward-equivalents across the scan; 5.5 in-process at
the six-knob anchor) and its peak resident memory stays within
\SIrange{2.6}{3.4}{\giga\byte} across the archived runs from two to
sixteen knobs, with run-to-run variation at fixed knob count as
large as the variation across knob counts and no growth with $N$, so
that the measured wall-clock advantage over finite differences
changes sign between ten and twelve knobs (bracketed there by two
non-overlapping repetitions on each side) and widens to $1.45\times$
at sixteen.

\section{Discussion}
\label{sec:discussion}

\subsection{Relation to existing codes}
\label{sec:related}

Table~\ref{tab:codes} places HELIX alongside the codes closest to it in
purpose, restricted to capabilities we can document from the cited
references.  Two families frame the comparison.  The established
hadron-linac workhorses (TraceWin~\cite{tracewin},
IMPACT-Z~\cite{impactz}, and TRACK~\cite{track_code}) cover the physics
scope HELIX targets: TraceWin pairs an rms-envelope mode with
multiparticle space-charge tracking behind a graphical interface and
defines the \texttt{.dat} lattice language that HELIX adopts, but it is
distributed as precompiled binaries; IMPACT-Z and TRACK are
multiparticle codes with three-dimensional space charge developed for
linear-accelerator design.  AVAS~\cite{avas} is the most recent
addition to this group and, among the non-TraceWin codes surveyed
here, the closest to the workflow HELIX targets:
an MPI-parallel C++ code carrying both envelope and multiparticle
modes, a symmetry-reduced PICNIC space-charge solver (S-PICNIC,
reported at ${\approx}4\times$ the standard algorithm and applied to
${\sim}10^8$ macroparticles), an element-dependent switch of independent
variable, built-in parameter matching and error analysis, and a PyQt
graphical interface.  AVAS has been applied to presetting the
operating parameters of the CAFe front-end demonstrator to 0.5\% in
energy, and cross-checked against TraceWin on the \SI{202.75}{\meter}
CiADS superconducting section and against Impact-T on a separate
high-current benchmark.  The newer general
frameworks (Bmad~\cite{bmad}, Xsuite~\cite{xsuite},
ImpactX~\cite{impactx}, and Cheetah~\cite{cheetah}, a PyTorch package
built for machine-learning applications around fast linear beam
dynamics) are open source and architecturally modern but
are not organized around the envelope-first, TraceWin-style linac
workflow.  HELIX
deliberately occupies TraceWin's niche (hybrid
envelope-plus-multiparticle hadron-linac design driven by a single
lattice file) while adopting a Python-first, extensible architecture
with optional C++/OpenMP and GPU acceleration, in the spirit of the
newer frameworks (Sec.~\ref{sec:methods}).

\begin{table*}[t]
\caption{\label{tab:codes}%
Capability comparison between HELIX and representative beam-dynamics
codes.  Env.: deterministic rms-envelope (moment) mode with space
charge.  PIC SC: three-dimensional particle-in-cell space charge
(lower dimensionality noted where documented).  Diff.\ SC:
differentiable space-charge model.  Grad.\ match: demonstrated
gradient-based matching or optimization through a space-charge model.
Match.\ suite: built-in matching/optimization engine.  TW input: reads
TraceWin-format (\texttt{.dat}) lattice files.  Entries reflect
capabilities as documented in the cited sources as of July 2026 and,
for availability, the codes' public distribution channels; a dash
indicates a capability that is absent from, or not documented in,
those sources, not a definitive claim of absence.}
\begin{ruledtabular}
\footnotesize
\begin{tabular}{llcccccccc}
Code & Scope & Env. & PIC SC & Diff.\ SC & Grad.\ match &
Match.\ suite & TW input & GUI & Availability \\
\colrule
\textbf{HELIX} (this work) & hadron linacs, lines & yes & yes &
  yes (PIC) & yes (PIC)\footnote{Exact-Jacobian matching through the
  differentiable PIC (Secs.~\ref{sec:diffpic} and~\ref{sec:matching}),
  demonstrated on a compact MEBT-style matching section
  (Sec.~\ref{sec:applications}).} &
  7 alg.\ + Pareto & yes\footnote{HELIX follows the unit conventions
  of the TraceWin \texttt{.dat} language, including card lengths in
  millimeters~\cite{tracewin_manual} (Sec.~\ref{sec:lattice}).  Import
  is limited by front-end coverage rather than units, and export is
  not yet lossless; both are discussed in
  Sec.~\ref{sec:limitations}.} &
  yes & open source \\
TraceWin~\cite{tracewin} & linacs, lines & yes & yes & --- & --- &
  yes & native & yes & binary\footnote{Precompiled executables
  distributed under a CEA license; the source code is not public.} \\
IMPACT-Z~\cite{impactz} & linacs & --- & yes & ---\footnote{%
  Differentiable space-charge simulation has been demonstrated by the
  same group in separate research
  codes~\cite{qiang2023differentiable,qiang2025admodule}.} & --- &
  --- & --- & --- & open source \\
TRACK~\cite{track_code} & proton, ion linacs & --- & yes & --- & --- &
  --- & --- & --- & binary\footnote{Precompiled executables distributed
  by Argonne National Laboratory.} \\
AVAS~\cite{avas} & high-intensity linacs &
  ---\footnote{An envelope mode \emph{is} documented ($\sigma$-matrix
  phase-ellipse transport, used to match beam and element parameters
  during design), but the reference specifies the element transport
  without a space-charge term, so the dash records the absence of the
  space-charge coupling this column requires, not the absence of an
  envelope mode~\cite{avas}.} & yes\footnote{Two three-dimensional
  solvers: an FFT Poisson solve (FFTW, periodic boundaries) and the
  symmetry-reduced S-PICNIC algorithm, reported at
  ${\approx}4\times$ the standard algorithm and applied to
  ${\sim}10^8$ macroparticles~\cite{avas}.} & --- & --- &
  yes\footnote{The documented function list comprises ``beam and
  element parameters matching, particle tracking, error analysis, and
  preset operating parameters''; the Python control level integrates
  the optimization algorithms~\cite{avas}.} & --- &
  yes\footnote{``The user level includes a graphical interface
  developed using pyQt''~\cite{avas}.} & not stated \\
Bmad~\cite{bmad} & general (rings, linacs) & --- &
  yes\footnote{Three-dimensional integrated-Green-function space
  charge on an FFT grid via the companion OpenSC
  library~\cite{mayes2018opensc}.} & --- & --- &
  yes\footnote{Via the companion Tao design and analysis program,
  which provides nonlinear optimization on Bmad
  lattices~\cite{tao}.} & --- & --- & open
  source \\
Xsuite~\cite{xsuite} & general (rings, lines) & --- &
  yes\footnote{Frozen, quasi-frozen, and particle-in-cell
  space-charge models are provided; the PIC element accepts both a
  full 3-D FFT solver (\texttt{FFTSolver3D}) and 2.5-D solvers
  (Xfields documentation and source)~\cite{xsuite_docs}.} & --- & --- &
  yes\footnote{Built-in matching through
  \texttt{Line.match()}, which drives a numerical optimizer over
  \texttt{Vary}/\texttt{Target} specifications~\cite{xsuite_docs}.} &
  --- & --- & open source \\
Cheetah~\cite{cheetah} & beamlines (ML-oriented) & --- &
  yes\footnote{Three-dimensional integrated-Green-function space
  charge added in a 2026 differentiable extension~\cite{dhamrait2026};
  the base code implements linear beam dynamics~\cite{cheetah}.} &
  yes~\cite{dhamrait2026} & --- & --- & --- & --- & open source \\
ImpactX~\cite{impactx} & linacs, beamlines & yes & yes &
  envelope\footnote{Compiler-level automatic differentiation (Enzyme)
  of the envelope model including space charge, demonstrated on
  gradient-based FODO Twiss matching~\cite{huebl2025impactx}.} &
  yes (envelope) & --- & --- & yes\footnote{The code's documentation
  lists three interfaces, the third a graphical user interface
  (\texttt{impactx-dashboard})~\cite{impactx_docs}.} & open source \\
\end{tabular}
\end{ruledtabular}
\end{table*}

The differentiable ingredients of HELIX likewise have clear published
antecedents, and we position them as engineering integration rather
than primacy.  Differentiable self-consistent space-charge simulation
was established by Qiang~\cite{qiang2023differentiable}, who
demonstrated gradient-based accelerator design optimization through a
self-consistent space-charge PIC model; the same group has since
described a multi-language automatic-differentiation module for a
parallel MPI PIC code and used it for gradient-based (BFGS) Twiss
rematching of a four-quadrupole matching section with space
charge~\cite{qiang2025admodule}, and applied automatic
differentiation to a fully coupled three-dimensional envelope model
with space charge, uncovering a new instability
stopband~\cite{qiang2026envelope}.  JuTrack~\cite{jutrack} brings
Enzyme-based automatic differentiation to accelerator tracking in
Julia, including sensitivities of space-charge-driven emittance
growth, and Huhn and Velotti~\cite{huhn2025} benchmark automatic
differentiation against finite differences, and gradient-based against
gradient-free optimization, in a purpose-built differentiable tracker
including a space-charge FODO cell.  Closest to the HELIX combination,
Cheetah has been extended with a reverse-mode differentiable
integrated-Green-function space-charge model whose memory cost scales
linearly with the numbers of macroparticles, grid cells, and
space-charge kicks~\cite{dhamrait2026}, and ImpactX has demonstrated
compiler-level (Enzyme) differentiation of its envelope model
including space charge, matching the Twiss parameters of a 0.5-A,
6.7-MeV proton beam through a FODO cell in 37 simulations where
Nelder--Mead required 167~\cite{huebl2025impactx}; the two codes share
the same three-dimensional open-boundary integrated-Green-function
space-charge model~\cite{huebl2025impactx,dhamrait2026}.

Four things distinguish the HELIX combination.  First, the division of
labor between the two modes is not itself new: TraceWin has long
paired them, and AVAS documents the same split, envelope transport to
match parameters during design and multiparticle simulation for design
evaluation~\cite{avas}.  HELIX's claim here is not the pairing but
what it is built on: an open, Python-native implementation of that
workflow reading TraceWin's own \texttt{.dat} language, so one file
and the same \texttt{SET}/\texttt{ADJUST} cards drive
second-scale envelope design iteration and full-PIC confirmation
(Secs.~\ref{sec:lattice} and~\ref{sec:solvers}) in a code that can be
read, extended, and differentiated---which the binary-distributed
incumbent cannot be.  That openness also closes the front end: the
established RFQ codes are separately licensed or export-controlled and
the openly available ones compute space charge on an $r$--$z$ mesh
(Sec.~\ref{sec:lattice}), so, to our knowledge, no other openly
licensed code covers a linac start to end---dc low-energy transport,
RFQ, and superconducting linac---in one lattice description.  Second,
gradient-based matching through self-consistent \emph{nonlinear PIC}
space charge is delivered inside the code's matching
engine: the exact-Jacobian \texttt{gradient} algorithm
is dispatched through the same interface as the other six
single-objective optimizers, driven by the
same TraceWin-style constraint cards as the derivative-free methods,
within its documented element and constraint scope
(Sec.~\ref{sec:matching}); the Pareto explorer is a separate entry
point, but it ranges over the same \texttt{ADJUST} decision variables
and is reachable from the same three interfaces.  The
nearest published demonstrations optimize through space
charge: gradient-based design optimization through a
differentiable PIC~\cite{qiang2023differentiable}, BFGS rematching of
a four-quadrupole section through an automatically differentiated
parallel PIC~\cite{qiang2025admodule}, and envelope-level
gradient Twiss matching of a FODO cell in
ImpactX~\cite{huebl2025impactx}.  HELIX, by contrast, applies PIC-level
gradient matching from its standard constraint cards, demonstrated on a
compact MEBT-style matching section (Sec.~\ref{sec:applications}),
with the underlying physics cross-validated against TraceWin on
PIP-II-family lattices in Sec.~\ref{sec:validation}: sub-percent
\emph{mean} envelope deviations on the LEBT and MEBT+HWR benchmarks
(0.18--0.86\%, with a 3.8\% worst single point in the LEBT's steepest
converging section and every MEBT+HWR point below 0.7\% with space
charge),
multiparticle rms sizes at the
${\sim}3\%$ level on MEBT+HWR, and, on the \SI{186}{\meter} PIP-II
accelerating linac, mode-matched envelope per-step rms residuals of
0.51\% and 0.98\%,
few-percent multiparticle agreement, and
longitudinal $\sigma_z$ agreement within 0.9\% at the linac exit
($-0.6\%$ envelope, $-0.9\%$ multiparticle at
\SI{186}{\meter}).  Third, the machine-learned element
surrogates of Sec.~\ref{sec:surrogates} are scope-guarded,
reproducible, and autograd-transparent, so they compose with the
gradient matcher instead of sitting beside it: an out-of-scope input
in the gradient path raises a hard error that names the offending
element rather than silently substituting an identity map, so
surrogate-in-the-loop matching is confined by construction to the
recorded training scope.  Fourth, these capabilities are
reachable from the graphical workbench and the
batch command line alike (Sec.~\ref{sec:interfaces}), with two
exceptions, both on the surrogate side: direct element substitution
runs through the Python API, and the shipped surrogate-training CLI
covers only the 3-D field-map class, so training a 1-D/2-D field-map
surrogate requires the workbench or the API
(Sec.~\ref{sec:surrogates}).  Among the
sources surveyed here as of July 2026, no single tool documents the
combination of a TraceWin-format front
end, hybrid envelope/multiparticle solvers, a seven-algorithm matching
suite with gradient matching through nonlinear PIC space charge, and a
GUI.  Each is represented by one or more columns of
Table~\ref{tab:codes}.  Machine-learned
element surrogates are not a column of that table, so the further
claim that none of these tools adds them rests on the works cited
above rather than on the table.  Each ingredient, taken
alone, has an antecedent in Table~\ref{tab:codes} or those works.

\subsection{Limitations and future work}
\label{sec:limitations}

HELIX's collective-effects scope is bounded.  Aside from the 1-D
steady-state CSR wake of Sec.~\ref{sec:physics}, no impedance-based
wakefield models are implemented (structure, multibunch, or
resistive-wall), and the
PIC solve is strictly open-boundary, so the beam pipe acts only as an
aperture cut and never as a boundary condition for the Poisson
equation; image-charge effects are therefore absent, and
synchrotron-radiation damping is likewise not modeled.  The
benchmark geometries here are the favorable case for that omission:
the chambers are circular and the beams round and close to centered,
the configuration in which the leading image term is smallest.  We
make no quantitative claim about its size, and none of the results
reported here depends on one.  Studies whose chambers are
non-circular, or whose orbits are displaced far enough for coherent
image forces to matter, fall outside what this paper establishes.  For
high-power accumulator
rings or high-energy
electron machines HELIX is not the right tool.

The TraceWin front end is broad but not complete, and the gaps are
graded.  Within the quadrupole, cavity, bend, and beam error families
that HELIX implements, dynamic directives
(\texttt{ERROR\_*\_DYN}) are absorbed as ordinary static
per-seed draws rather than per-step jitter, and coupled error groups
(\texttt{ERROR\_*\_CPL\_*}) as independent per-element draws; multi-seed
ensembles and manually scripted group draws are the documented
workarounds.  Outside those families the degradation is sharper: the
\texttt{ERROR\_*\_FILE} and \texttt{ERROR\_RFQ\_*} directives are
recognized but apply no error at all, \texttt{ERROR\_SET\_RATIO} is
stored and never consumed, TraceWin's constant-amplitude
($r=0$) draws are approximated as Gaussian, and random-subset
(\texttt{Nb}) selection is ignored so errors reach every matched
element.  Longitudinal-alignment offsets are stored but inert; pitch
and yaw are read only far enough to warn, and are not retained in the
alignment model when they arrive on \texttt{ERROR\_*} cards (the
programmatic error API does store them, equally inert).  Superposed
field-map clusters carry their own downgrades: mixed-frequency
clusters fall back to sequential maps, which lengthens the line and
shifts every downstream element, and curved-reference
(\texttt{SUPERPOSE\_MAP\_OUT}) clusters are unsupported.  Strict
mode refuses both downgrades outright rather than returning a
mislengthened lattice, while permissive mode accepts the fallback
with a warning.
Every case above raises a parse-time warning, and the code's
known-limitations page carries the fuller list.
Finally, HELIX's \texttt{.dat}
dialect follows the unit conventions documented for the TraceWin
language itself (card lengths in millimeters and field-map files in
meters~\cite{tracewin_manual}), so supported cards carry the same
units in both programs; import is limited by the coverage
gaps above rather than by units.  Export is not yet lossless in the
other direction: the \texttt{.dat} writer serializes no
\texttt{ERROR\_*} cards, so saving a lattice that carries an error
study discards it, which the writer warns about.  Systematic round-trip testing
against the TraceWin executable has not been performed.  Particle
interchange through the \texttt{.dst} format follows TraceWin's
on-disk cm/rad/MeV units (Sec.~\ref{sec:io}).

The differentiable path trades speed for gradients.  The PyTorch PIC
runs in double precision on the CPU only, re-fits its grid extent to
the bunch at every kick, and is markedly slower than the default
compiled PIC for a plain forward run (about a factor of five on a
small MEBT lattice), so it is engaged only when gradients are
required.  Its exactness is also conditional: the Jacobian is exact to
floating-point precision for a fixed computational branch, and the CIC
deposit's derivative jumps whenever a macroparticle crosses a cell
boundary, which sets the remainder floors measured in
Sec.~\ref{sec:gradcheck}; those Taylor-remainder checks cover the CIC
deposit that the matching residual uses, and the torch TSC deposit,
while it passes forward-parity and gradient-flow tests, has not been
verified to the same standard.  The gradient matcher's
through-space-charge residual is evaluated by default on a $32^3$ node
count with an adaptive $\pm4\sigma$ extent and a 1500-particle bunch,
against the $96^3$/$\pm5\sigma$ default of the compiled solver.  All
three are overridable: through the multiparticle configuration from
the Python API and the graphical workbench, and through
\texttt{--mp-grid}, \texttt{--mp-extent} and \texttt{--mp-n-particles}
on the batch command line.  Of the two mesh knobs,
the node count is the cost
concession: $32^3$ against $96^3$ is a factor of 27 in cells, carried
through a recorded autograd graph.  The extent is not a cost knob at
all: at a fixed node count a $\pm4\sigma$ box costs the same solve as
$\pm6\sigma$ and simply trades tail capture for finer cells in the
bunch core.  Both defaults are deliberate (every gradient-matching
result reported here was obtained at them, so changing either would
move those results rather than merely refine them), and what has been
costed is the node count only in compound with the particle count,
never either mesh knob alone.  The
revalidation of Sec.~\ref{sec:app_gradmatch} re-evaluates each matched
point on a $64^3$, 5000-particle model over five independent bunch
seeds,
and the mismatch it exposes is set by that coarse-model bias rather
than by the optimizer, reaching modestly outside the $\alpha$ matching
tolerance in $\alpha_x$ at sixteen knobs; the seed contribution is
separated out as the five-seed noise floor that the mismatch exceeds
by 2.6--13$\times$, but the node-count and particle-count changes
remain compounded, as that section states.  The extent
is not in that budget at all: the optimizer model and its revalidation
both sit at $\pm4\sigma$, so the revalidation cannot see the extent
choice, and the only evidence bearing on it is the mesh scan of
Sec.~\ref{sec:fieldacc}---which does find the rms sizes controlled by
the grid extent rather than the resolution, but on a different lattice
and as a deliberate stress test.  Isolating those two contributions,
and re-costing the matched points against a wider extent, is therefore
the open item that would most sharpen this section.  In the acceleration layers, the compiled C++
kernels cover the CIC shape function only (the TSC deposit and gather
pair exists only in the Python implementation), and the Apple-MPS GPU
path runs in FP32; it is excluded from the reported cross-code
baselines but is covered by its own CPU-parity regression tests
(Secs.~\ref{sec:acceleration} and~\ref{sec:scverify}).
Cross-code validation is also concentrated: the quantitative
benchmarks of Sec.~\ref{sec:validation} are against TraceWin on
PIP-II-family lattices, and systematic comparisons against other
space-charge codes and on other machine classes remain to be done.

These limitations set the near-term program.  The documented roadmap
lists per-step dynamic error directives, wakefield infrastructure
(longitudinal first), and pitch/yaw misalignment support as items
under consideration.  On the optimization side, the natural next steps
are a GPU-resident differentiable PIC, which would lift the CPU-only
restriction of the gradient path and make matching at the benchmark
mesh and extent affordable rather than merely configurable, and
broader cross-code benchmarks
beyond the TraceWin/PIP-II axis.

\section{Conclusion}
\label{sec:conclusion}

HELIX integrates a differentiable tracking-plus-PIC forward map,
covering a documented fixed-energy element subset, into a card-driven
matching engine, and assembles around it, in one Python tool, the
pieces of the hadron-linac design workflow that have historically
lived in separate codes.  A TraceWin-format lattice front end feeds
rms-envelope, multiparticle-PIC, and matrix solvers over one data
model.  The same tool carries a seven-algorithm matching engine with
TraceWin-style constraint cards and multi-objective Pareto
exploration, machine-learned element surrogates, element- and
beam-error Monte Carlo with orbit correction, failure-criticality and
compensation analysis, and GUI and batch interfaces.

Its verification is hierarchical and quantified
(Sec.~\ref{sec:validation}).  Analytic envelope benchmarks agree at
the $10^{-8}$ level and below.  Cross-implementation PIC parity runs
from $10^{-9}$ to $10^{-13}$ across the C++, Python, CUDA, and PyTorch
backends.  The FP32 Metal backend is held instead to
$5\times10^{-6}$ on the solved field, with $3.4\times10^{-6}$
measured, and no cross-code benchmark reported here depends on it.
Mode-matched cross-code benchmarks against TraceWin use PIP-II-family
lattices: along the 18.96-m MEBT+HWR line, envelope rms moments with
space charge agree to better than 0.7\% at every recorded point and to
below 0.15\% at exit, and multiparticle rms sizes agree at the
few-percent level.  Agreement on the PIP-II LEBT and the 186-m PIP-II
accelerating linac is quantified in the same mode-matched terms, and
the downstream transport to \SI{256.5}{\meter} is modeled without a
TraceWin reference.

That integration is what differentiates HELIX from both the
established accelerator-design codes and the recent
differentiable-simulation research codes (Sec.~\ref{sec:related}).
Reverse-mode Jacobians of the discretized forward map, exact to
floating-point precision for a fixed computational branch, flow
through the self-consistent space-charge solve.  The resulting
gradient matcher drove a tolerance-normalized six-quadrupole matching
residual through nonlinear space charge to $3.2\times10^{-10}$,
bit-identically across repeated runs, at essentially the same
cumulative forward-equivalent budget at which an otherwise identical
finite-difference solver stopped a factor of about 200 short.  A
derivative-free baseline remained eleven orders above when its
evaluation cap stopped it.  At this knob count the finite-difference
run won the wall clock by $1.74\times$.  Because the Jacobian cost
scales with the number of residual components rather than the number
of knobs, that balance reverses as knobs are added.  Measured end to
end at six knob counts from two to sixteen, the crossover falls
between the ten- and twelve-knob measurements, a bracket set by two
non-overlapping repetitions at each point
(Sec.~\ref{sec:app_gradmatch}).  At sixteen knobs reverse mode holds a
$1.45\times$ advantage, and its peak memory stays within
\SIrange{2.6}{3.4}{\giga\byte} throughout.

What the paper establishes is therefore not any solver in isolation,
but a reproducible path from a TraceWin-format lattice through
self-consistent tracking and PIC to card-defined, gradient-based
matching within one data model.  The version of the code used here,
its input lattices, and the scripts behind every figure are recorded
in the paper's provenance records and are available
as described in the data availability statement.

The present release has quantified limits.  It models no
impedance-based wakefields beyond the 1-D CSR model and no image
charge.  The error families it implements carry static uncoupled
semantics, and several it only recognizes apply no effect at all.  The
differentiable path is CPU-only float64, and its matching residual
defaults to a coarse $32^3$/$\pm4\sigma$ mesh.  Cross-code
benchmarking is concentrated on PIP-II-family lattices.  These limits
set the near-term development agenda: broader cross-code benchmarks
and GPU-resident differentiable matching.

\begin{acknowledgments}
This work was produced by FermiForward Discovery Group, LLC under Contract
No.~89243024CSC000002 with the U.S. Department of Energy, Office of Science,
Office of High Energy Physics.
The cross-code comparisons reported here were carried out using TraceWin
under a CEA research license.  TraceWin is authored by the Commissariat
\`a l'\'Energie Atomique et aux \'Energies Alternatives (CEA), which holds
the associated intellectual property rights.
\end{acknowledgments}

The author declares no competing interests.

\section*{Data availability}
\label{sec:availability}
HELIX is released as open-source software under the GNU General Public
License v3.0 at \url{https://github.com/Accel-Toolkit/HELIX}~\cite{helix_code}, with the
release approved by Fermi National Accelerator Laboratory and the
U.S.\ Department of Energy; the code this paper describes is tagged
\texttt{v1.0.11}, which reproduces every figure in this paper from the
archived inputs.  The figures were themselves generated during
pre-release development, and the provenance of each is recorded in the
paper's figure-provenance records.  Those
records, the input lattices, the TraceWin reference exports used in
Sec.~\ref{sec:validation}, and the scripts and run records behind
every figure are available from the author upon reasonable request,
excluding third-party field-map data governed by their originating
institutions.

\bibliography{helix_prab}

\end{document}